\documentclass[amssymb,11pt]{article}
\usepackage{amsfonts}
\usepackage[dvipdfm]{graphicx}
\usepackage{amsthm}
\usepackage{amsmath}
\usepackage{multirow}
\usepackage{indentfirst}
\usepackage{booktabs}
\usepackage{bm}
\usepackage{subfigure}
\usepackage{algorithm,algorithmic}
\usepackage[numbers,sort&compress]{natbib}

\date{}
\textwidth=16cm \textheight=25.8cm \vspace{1.4cm}

\usepackage[colorlinks,linkcolor=blue,citecolor=cyan,urlcolor=black,hyperindex,CJKbookmarks]{hyperref}
\usepackage{hyperref}

\begin{document}

\title{\Large Hyperspectral Image Fusion via Logarithmic Low-rank Tensor Ring Decomposition}
%\begin{upshape}
%\author{\ Jun Zhang$^{1,2}$, Lipeng Zhu$^3$, Chao Wang$^4$, Shutao Li$^{1,\ast}$}
%\end{upshape}\maketitle

\begin{upshape}
\author{\ Jun Zhang$^{1,2}$, Lipeng Zhu$^3$, Chao Wang$^4$, Shutao Li$^{1,\ast}$}
\end{upshape}\maketitle

\footnotetext{
$^{\ast}$\scriptsize Corresponding author.\\
$^1$College of Electrical and Information Engineering, Hunan University, Changsha 410082, Hunan, China\\
$^2$College of Science, Nanchang Institute of Technology, Nanchang 330099, Jiangxi, China\\
$^3$Jiangxi Province Key Laboratory of Water Information Cooperative Sensing and Intelligent Processing, Nanchang
Institute of Technology, Nanchang 330099, Jiangxi, China\\
$^4$Department of Statistics and Data Science, Southern University of Science and Technology, Shenzhen 518055, Guangdong, China\\
\textsl{Email addresses:} junzhang0805@126.com (J.~Zhang), lipengzhu0619@163.com (L.~Zhu), wangc6@sustech.edu.cn (C.~Wang),
shutao\_li@hnu.edu.cn (S.~Li).
\indent\indent\small}
%%%%%%%%%%%%%%%%%%%%%%%%%%%%%%%%%%%%%%%%%%%%%%%%%%%%%%%%%%%%%

\textbf{Abstract.}
% Fusing a low-spatial-resolution hyperspectral image (LR-HSI) with a high-spatial-resolution multispectral image (HR-MSI)
Integrating a low-spatial-resolution hyperspectral image (LR-HSI) with a high-spatial-resolution multispectral image (HR-MSI)
is  recognized as a valid method for acquiring HR-HSI.
Among the current fusion approaches, the tensor ring (TR) decomposition-based method has received growing attention owing to its superior performance on preserving the spatial-spectral correlation.
Furthermore,  the low-rank property in some TR factors has been exploited via the matrix nuclear norm regularization along mode-2.
On the other hand,
the tensor nuclear norm (TNN)-based approaches have recently demonstrated to be more efficient on keeping high-dimensional low-rank structures in tensor recovery. Here, we study the low-rankness of TR factors from the TNN perspective and  consider  the mode-2 logarithmic TNN (LTNN) on each TR factor.  A novel fusion model is proposed by incorporating this LTNN regularization and the weighted total variation which is to promote the continuity of HR-HSI in the spatial-spectral domain.
Meanwhile,
we have devised a highly efficient proximal alternating minimization algorithm to solve the proposed model.
The experimental results indicate that our method improves the visual quality and exceeds the existing state-of-the-art fusion approaches with respect to various quantitative metrics.

\begin{flushleft}
\end{flushleft}

%%%%%%%%%%%%%%%%%%%%%%%%%%%%%%%%%%%%%%%%%%%%%%%%%%%%%%%%%%%%%%%%%%%
\section{Introduction}
Hyperspectral imaging technology emerges from the fusion of imaging and spectral technologies. In contrast to conventional imaging methods, hyperspectral imaging captures numerous minuscule spectral regions across the electromagnetic spectrum, offering a wealth of spectral information \cite{S1,S2}. Consequently, the exploitation of hyperspectral data proves particularly advantageous for precise feature classification \cite{class1,class2}, denoising \cite{noise}, unmixing \cite{unmix}, and more. Unfortunately, due to limitations inherent in the imaging techniques employed, hyperspectral images often suffer from low spatial resolution, constraining their practical utility \cite{KCI,NLSTF}. On the other hand, multispectral imaging sensors yield higher spatial resolution images \cite{WQ}. Hence, a viable strategy for achieving hyperspectral super-resolution involves fusing a low-resolution HSI (LR-HSI) with a high-resolution MSI (HR-MSI) of the same scene.

In recent years, significant efforts have been devoted to exploring HSI-MSI fusion techniques. These methods can be systematically classified into three overarching categories: deep learning-based  \cite{LJD,DSQ,RR}, matrix decomposition-based  \cite{TV1,LRSR,TMD,SLRTV1,WNNSTV}, and tensor decomposition-based methods \cite{CP1,CSTF,TT1,CTRF}.
Matrix decomposition-based methods operate under the assumption that each spectral vector within HR-HSI can be formulated as a linear combination of distinct spectral features.
Consequently, estimating the desired HR-HSI involves the computation of both the spectral basis and its corresponding coefficient matrix.
Particularly, by imposing prior constraints on the coefficient matrix, some effective fusion approaches have been proposed \cite{TV1,LRSR,SLRTV1,WNNSTV}.
In \cite{TV1}, the spatial smoothness of the fused image was improved by applying the total variation (TV) constraint to the coefficient matrix.
In pursuit of promoting local low-rank characteristics within images, Dian et al. \cite{LRSR} introduced an HSI super-resolution technique utilizing nuclear norm regularization.
On this foundation, to better retain crucial information in the image, Zhang et al. \cite{WNNSTV} innovatively integrated a weighted vector with the nuclear norm regularization.
Notably, the TV regularization was incorporated to effectively preserve the piecewise smoothness in the spatial domain.
Although the vectorization of individual bands in these methods does not alter the values within the original 3D data, it disrupts the inherent structure of HSI, resulting in a deterioration of fusion performance.

Tensor representation can effectively preserve the inherent structural information of HSI.
Nowadays, tensor factorization, which includes Canonical polyadic (CP) decomposition \cite{CP1,CP2}, Tucker decomposition \cite{CSTF,UTV,NLTSU}, tensor-train (TT) decomposition \cite{TT1,TT2}, and tensor ring (TR) decomposition \cite{TRD,TRC}, has become a popular method to deal with the HSI-MSI fusion problem.
As presented in \cite{CP1},
the CP decomposition was employed to partition the HR-HSI into three parts, followed by iterative updates for each part. Nonetheless, using the CP decomposition presents challenges in locating the optimal latent tensors.
To overcome this problem, based on the Tucker decomposition, Li et al. \cite{CSTF} presented a coupled sparse tensor factorization approach for HSI-MSI fusion.
Specifically, they postulated that the HR-HSI could be formulated as a core tensor multiplied by the dictionaries corresponding to the three modes.
On this basis, Xu et al. \cite{UTV} introduced a unidirectional TV-based approach.
Especially, they applied the $\ell_1$-norm to the core tensor so as to capture sparsity, while employing unidirectional TV on three dictionaries to characterize piecewise smoothness.
In \cite{NLTSU}, Wang et al. proposed a novel HSI-MSI fusion approach using spectral unmixing and nonlocal low-rank tensor decomposition.
They harnessed nonlocal Tucker decomposition to capture both spatial-spectral correlations and spatial self-similarity within HSI, augmenting it with spectral regularization technique to mitigate spectral distortions.
Although the aforementioned methods based on Tucker decomposition excel in preserving the spectral-spatial correlation of HSI, they face the challenge of dimensionality catastrophe.

Fortunately, compared to CP decomposition and Tucker decomposition, tensor network-based approaches such as TT decomposition \cite{TT1,TT2} and TR decomposition \cite{TRD,TRC,TRTV,TRLRF,HCTR,LRTRTNN,CTRF,FSTRD} have a stronger ability to mine the intrinsic structure of data.
In \cite{TT1}, Dian et al. utilized non-local similarity to group similar patches within an image, leading to the creation of multiple 4D tensors. They further applied a low TT rank constraint to these 4D tensors, and thus acquired better fusion performance.
However, the TT decomposition requires the imposition of rank-1 constraint on the border factors, resulting in limitations in its expressiveness and flexibility.
To address the problem caused by the TT decomposition, the TR decomposition model was initially proposed for color image and video recovery \cite{TRD}.
TR can be seen as the linear combination of a set of TT representations \cite{TRC}.
It inherits the main properties of TT.
Furthermore, owing to the rotation invariance property of latent factors, TR exhibits greater flexibility in its model constraints compared to TT.
In \cite{TRTV}, a TV-regularized TR completion model was proposed for HSI reconstruction.
In this work, the TR decomposition was employed to explore low-rank tensor information, and TV was utilized to promote spatial smoothness.
In addition, TR decomposition was successfully applied to HSI denoising \cite{TRHSId} and neural network compression \cite{TRN}.
Recently, TR decomposition has also been developed for HSI-MSI fusion \cite{HCTR,LRTRTNN,CTRF,FSTRD}.
He et al. \cite{CTRF} proposed a coupled TR decomposition approach, which simultaneously learns the TR factors of the HR-HSI among a pair of HSI and MSI.
This approach is very different from traditional methods since the aim of fusion is transformed into estimating the TR factors.
Based on this TR framework,
Chen et al. \cite{FSTRD} further imposed weighted TV regularization to constrain the TR factors so as to capture its spatial-spectral continuity.
Although these approaches in the TR decomposition framework achieve relatively satisfactory fusion results, they don't take into account the low-rank characteristics of HR-HSI.
In other words, the low-rank property of TR factors is ignored in the TR decomposition framework.

Owing to the low-rank nature of HSIs, the nuclear norm regularization has been widely used in HSI denoising and fusion \cite{NLRD,HID,LRSR,WNNSTV}.
In \cite{TRLRF}, Yuan et al. applied the matrix nuclear norm to the rank-mode unfoldings of the TR factors and proposed a TR low-rank factors model for tensor completion.
In \cite{CTRF}, a coupled TR factorization model with nuclear norm regularization was proposed for HSI-MSI fusion.
Specifically, to characterize the low-rank property of HSI along the spectral dimension, the authors utilized matrix nuclear norm regularization to the third core tensor along mode-2.
Although these two approaches acquired acceptable performance, the core tensors were all unfolded to mode-n matrices, neglecting the correlation between different modes.
In contrast, the tensor nuclear norm (TNN) based on tensor-singular value decomposition (t-SVD) can effectively maintain the inherent low-rank structure of the tensor \cite{TNN,LRTRTNN,TRPCA}.
By introducing TNN into all 3D TR factors, Xu et al. \cite{LRTRTNN} presented a TR and subspace decompositions-based model for HSI-MSI fusion to preserve the internal tensor structure naturally.
However, TNN treats each singular value equally, ignoring the physical significance of the different singular values and generating a biased approximation of tensor low-rank.
To address this problem, the use of non-convex functions instead of the $\ell_1$-norm to approximate the rank function can more accurately extract the low-rank components from the degenerate HSI \cite{logTR,LTMR}.
In this way, it allows larger singular values to be penalized less, effectively preserving the main information of the image data.
In approximating the nuclear norm, the log-sum of singular values has demonstrated superior performance when compared to the sum of singular values.
Thus, in \cite{LTMR}, Dian et al. introduced an HSI-MSI fusion technique that employs subspace-based logarithmic TNN (LTNN) regularization.

Inspired by the success of LTNN in subspace-based HSI-MSI fusion, in this study, we expand its applicability to fusion based on the TR decomposition framework.
Specifically, we develop a novel model for HSI-MSI fusion by
combining LTNN with weighted TV regularization.
To exploit the intrinsic low-rank structure of HSIs, we
directly apply the LTNN regularization to each of the TR factors in the framework of TR decomposition.
In addition, we introduce a weighted TV regularization to preserve the continuity of the fused image in the spatial-spectral domain.
We devise a proximal alternating minimization algorithm to address the proposed model.
Our proposed approach exceeds the current state-of-the-art fusion approach by increasing the PSNR value by over 0.69 and reducing the ERGAS value by 6.9$\%$.

The main contributions can be summarized as follows:
\begin{itemize}
    \item We propose a novel fusion model under TR decomposition and utilize TNN for TR factors. To the best of our knowledge, this is the first work in HSI-MSI fusion that investigates the low-rankness of TR factors from the perspective of TNN.
\item 
We address the skew effect of TNN and introduce the LTNN method to capture the low-rankness among these TR factors. Additionally, we incorporate the weighted TV to ensure the continuity of the fused image.
\end{itemize}

The remainder of the paper is structured as follows.
In Section \ref{Preliminaries}, we introduce the related notations and the fusion framework based on the TR decomposition, and two previous fusion models.
In Section \ref{method}, we propose a novel HSI-MSI fusion model by combining logarithmic tensor nuclear norm with the weighted TV and present its optimization algorithm.
Experimental results and convergence analysis are shown in Section \ref{Experiments}.
Finally, some conclusions and outlook of this paper are given in Section \ref{Conclusion}.

%%%%%%%%%%%%%%%%%%%%%%%%%%%%%%%%%%%%%%%%%%%%%%%%%%%%%%%%%%%%%%%%%%%%%
\setcounter{equation}{0}
\section{Preliminaries}\label{Preliminaries}
In this section, we will first introduce concepts and notations related to tensor, and the definition of tensor ring (TR) decomposition.
Subsequently, the problem formulation concerning the fusion of HSI and MSI is outlined.
Finally, we will present existing fusion models related to TR decomposition.

\subsection{Notations}
In this paper, we primarily adopt tensor notations consistent with those found in existing literature \cite{LTMR,FSTRD}.
Lowercase and uppercase letters are used for denoting scalars, i.e., $n$, $N\in\mathbb{R}$.
Bold lowercase is adopted to indicate vectors, i.e., $\mathbf{x}\in\mathbb{R}{^N}$. Correspondingly,
 bold uppercase letters denote matrices, i.e., ${\mathbf{X}}\in\mathbb{R}{^{M\times N}}$.
The calligraphic letters are employed to represent tensors of order $N\ (N\ge3)$, i.e.,
$\mathcal{X} \in {\mathbb{R}^{{I_1} \times {I_2} \times  \cdots  \times {I_N}}}$, where ${I_i}$ is the dimension of the $i$th mode.
Especially, scalars, vectors, and matrices can be seen as the zero-order, first-order, and second-order tensors, respectively.
$\mathcal{X}({i_1},{i_2}, \cdots,{i_N})$ or ${x_{{i_1},{i_2}, \cdots,{i_N}}}$ stands for the elemental value of $\mathcal{X}$ at position $({i_1},{i_2}, \cdots ,{i_N})$.
${\mathbf{X}_{(n)}}\in\mathbb{R}{^{{I_n} \times {I_1}{I_2} \cdots {I_{n-1}}{I_{n+1}}\cdots {I_N}}}$ and ${\mathbf{X}}_{<n>}\in\mathbb{R}^{{I_n} \times {I_{n+1}} \cdots {I_N}{I_1} \cdots {I_{n-1}}}$  represent the first and second mode-$n$ expansion matrices of tensor $\mathcal{X}$, respectively.
${\mathrm{Fold}}_{n}(\cdot)$ is the inverse operator of mode-$n$ matricization.
${\left\| \mathcal{X} \right\|_F} = \sqrt {\left\langle {\mathcal{X},\mathcal{X}} \right\rangle } $ denotes the Frobenius-norm of tensor $\mathcal{X}$,
while its $\ell_1$-norm is defined by
${\left\| \mathcal{X} \right\|_1} = \sum\nolimits_{{i_1},{i_2}, \cdots ,{i_N}} {\left| {x_{{i_1},{i_2}, \cdots ,{i_N}}} \right|}$.
The mode-$n$  product of the matrix ${\mathbf{D}}\in\mathbb{R}{^{N\times {I_n}}}$ and  $\mathcal{X}$ is defined as $\mathcal{Y} = \mathcal{X} \times_n{\mathbf{D}}$, where $\mathcal{Y} \in \mathbb{R}{^{{I_1}  \times \cdots\times {I_{n-1}}\times N \times{I_{n+1}}\times \cdots \times {I_N}}}$.
% The fast Fourier transform on the tensor $\mathcal{X}$ along the $n$th dimension and its inverse transform are denoted as
% $\overline{\mathcal{X}}=\mathrm{fft}\left( \mathcal{X},[\enspace],n \right)$ and $\mathcal{X}=\mathrm{ifft}\left( \overline{\mathcal{X}},[\enspace],n \right)$, respectively.

% The definition of TR decomposition used in this paper is given below.
% The purpose of TR decomposition is to represent a high-dimensional tensor by circular multilinear products on a series of third-order factor tensors (called TR factors)
The objective of TR decomposition is to express a high-dimensional tensor using circular multilinear products involving a sequence of third-order factor tensors, commonly referred to as TR factors \cite{TRD}.
The TR decomposition of $\mathcal{X}$ is expressed as a search for
$N$ latent third-order core tensors
$\mathcal{G}=\left\{ \mathcal{G}^{(1)},\mathcal{G}^{(2)},\cdots,\mathcal{G}^{(N)} \right\}$, where
$\mathcal{G}^{(n)} \in {\mathbb{R}^{{R_n} \times {I_n} \times {R_{n+1}}}},\ n=1,2,\cdots,N.$ Here, $R_1 = R_{N+1}$ and $R = [{R_1, \cdots, R_N}]$ is called the TR rank of tensor $\mathcal{X}$.
This decomposition has a stronger ability to mine the intrinsic structure of data than other decomposition approaches.
Therefore, they have found extensive utilization in tasks related to third-order image processing.
Here the TR decomposition of the  tensor $\mathcal{X} \in {\mathbb{R}^{{I_1} \times {I_2} \times  \cdots  \times {I_N}}}$ can be expressed as
\begin{equation}
\mathcal{X}( {{i_1},{i_2}, \cdots ,{i_N}} ) = \mathrm{Tr}( {\mathbf{G}^{(1)}({i_1}){\mathbf{G}^{(2)}}({i_2}) \cdots {\mathbf{G}^{(N)}}({i_N})} ),
\end{equation}
where $\mathrm{Tr}(\cdot)$ signifies the matrix trace operation, and $\mathbf{G}^{(n)}({i_n})\in {\mathbb{R}^{{R_n} \times {R_{n+1}}}}$ means the $i_n$th lateral slicing matrix of the core tensor $\mathcal{G}^{(n)}$.
To simplify the expression, we denote the TR decomposition of a $N$-order tensor by $$\mathcal{X} = \mathbf{\Phi} (\mathcal{G}^{(1)}, \mathcal{G}^{(2)},\dots, \mathcal{G}^{(N)}).$$

The multilinear product of ${\mathcal{G}^{(n)}}$ and ${\mathcal{G}^{(n+1)}}$ is represented by ${\mathcal{G}^{(n,n+1)}}\in {\mathbb{R}^{{R_n} \times {I_n}{I_{n+1}} \times {R_{n+2}}}}$. Specifically,
\begin{equation}
\mathcal{G}^{(n,n + 1)}(({j_n} - 1){I_n} + {i_n}) = {\mathcal{G}^{(n)}}({i_n}) \times {\mathcal{G}^{(n+1)}}({j_n}),
\end{equation}
for ${i_n} = 1, \cdots ,{I_n},\ {j_n} = 1, \cdots ,{I_{n+1}}$.
In addition,
$\overleftarrow{\mathcal{X}}_n \in\mathbb{R}{^{{I_n} \times {I_{n+1}} \times  \cdots  \times {I_N} \times {I_1} \times  \cdots  \times {I_{n-1}}}}$
is a tensor derived from $\mathcal{X}$ by cyclically shifting its dimensions by $n$ steps.
Thus, the following equation can be obtained:
\begin{equation}
\overleftarrow{\mathcal{X}}_n= \mathbf{\Phi} ( \mathcal{G}^{(n)},\mathcal{G}^{(n+1)}, \cdots ,\mathcal{G}^{(N)},\mathcal{G}^{(1)}, \cdots ,\mathcal{G}^{(n-1)}).
\end{equation}
Based on this property, the second mode-n expansion matrix of the tensor $\mathcal{X}$ can be written as
\begin{equation}\label{XN}
\mathbf{X}_{<n>}= \mathbf{G}_{(2)}^{(n)}{\left({\mathbf{G}_{<2>}^{(\ne n)}}\right)^T},
\end{equation}
where $\mathcal{G}^{( \ne n)}\in {\mathbb{R}^{{R_{n+1}} \times \prod^N_{i=1,i\ne n}{I_i} \times {R_n}}}$ is a subchain tensor acquired by fusing all $N-1$ tensors excluding the $n$th core tensor.

Next, the other crucial property is introduced. Namely, given that the TR decomposition of $\mathcal{X}$ is $\mathcal{X} = \mathbf{\Phi} \left( {\mathcal{G}^{(1)}},{\mathcal{G}^{(2)}},\cdots,{\mathcal{G}^{(N)}} \right)$, the operation of mode-$n$ multiplication on $\mathcal{X}$ is expressed as follows:
\begin{equation}
\mathcal{X}\times_n\mathbf{D} = \mathbf{\Phi} ( {\mathcal{G}^{(1)}},{\mathcal{G}^{(2)}}, \cdots ,{\mathcal{G}^{(n)}}{ \times_2}{\mathbf{D}}, \cdots ,{\mathcal{G}^{(N)}}).
\end{equation}

\subsection{Problem formulation}
In actual scenarios, HSI and MSI constitute third-dimensional tensor data, where the initial two modes represent the spatial dimensions of the image, while the third mode corresponds to the spectral dimension.
Within the context of HSI fusion, a pair of images is provided, namely an LR-HSI denoted as $\mathcal{Y}\in\mathbb{R}{^{w\times h\times B}}$ and an HR-MSI termed $\mathcal{Z}\in\mathbb{R}{^{W\times H\times b}}$, both captured within the same scene.  HSI-MSI fusion focuses on  restoring an HR-HSI
$\mathcal{X}\in\mathbb{R}{^{W\times H\times B}}$ from $\mathcal{Y}$ and $\mathcal{Z}$.
Here, $W > w$ and $H > h$ are the spatial sizes and $B > b$ stands for the number of spectral bands.
Specifically, the procedures for obtaining LR-HSI and HR-MSI can be viewed as the downsampling spatially and spectrally, respectively.

The separability applies to
both the point spread function (PSF) of the hyperspectral sensor and the downsampling matrices for the width and height modes.
As a result, we have
\begin{equation}\label{YZ}
\mathcal{Y} = \mathcal{X}{\times _1}{\mathbf{U}_1}{ \times _2}{\mathbf{U}_2},\
\mathcal{Z} = \mathcal{X}{\times _3}{\mathbf{U}_3},
\end{equation}
where ${{\mathbf{U}}_1} \in \mathbb{R}^{w \times W}$ and ${\mathbf{U}_2} \in \mathbb{R}^{h \times H}$ are the separable degradation operators along the width and height modes, respectively.
Separable operators are shown to offer more advantages in computing and optimization \cite{PSF}.
And ${\mathbf{U}_3} \in \mathbb{R}^{b \times B}$ represents the spectral downsampling matrix.

By utilizing the simplified expression of
the TR decomposition, it becomes possible to depict the HR-HSI as
\begin{equation}\label{TRX}
\mathcal{X} =\mathbf{\Phi} ( \mathcal{G}^{(1)},\mathcal{G}^{(2)},\mathcal{G}^{(3)} ),
\end{equation}
where
$\mathcal{G}^{(1)} \in {\mathbb{R}^{{R_1} \times W \times {R_2}}}$,
$\mathcal{G}^{(2)} \in {\mathbb{R}^{{R_2} \times H \times {R_3}}}$ and $\mathcal{G}^{(3)} \in {\mathbb{R}^{{R_3} \times B \times {R_1}}}$ denote TR factors associated with two spatial dimensions and one spectral dimension, respectively.
$R=[R_1,R_2,R_3]$ signifies the TR rank of the HR-HSI.
According to equations (\ref{YZ}) and (\ref{TRX}), it is easy to get the following expressions for the LR-HSI and HR-MSI:
\begin{equation}\label{TRY}
\mathcal{Y} =\mathbf{\Phi}( \mathcal{G}^{(1)}\times_2{\mathbf{U}_1},\mathcal{G}^{(2)}\times_2{\mathbf{U}_2},\mathcal{G}^{(3)} )
\end{equation}
and
\begin{equation}\label{TRZ}
\mathcal{Z} = \mathbf{\Phi} (\mathcal{G}^{(1)}, \mathcal{G}^{(2)},\mathcal{G}^{(3)}\times_2{\mathbf{U}_3}).
\end{equation}
In this way, the fusion problem is transformed into the estimation of three TR factors by making use of the LR-HSI $\mathcal{Y}$ and the HR-MSI $\mathcal{Z}$.

%%%%%%%%%%%%%%%%%%%%%%%%%%%%%%%%%%%%%%%%%%%%%%%%%%%%%%%%%%%%%%%%%%

\subsection{Related models }

As previously mentioned, the aim of HSI-MSI fusion is to calculate the three TR factors via
(\ref{TRY}) and (\ref{TRZ}).
With the downsampling matrices $\mathbf{U}_1$, ${{\mathbf{U}}_2}$ and $\mathbf{U}_3$ known, the three TR factors can be solved by \cite{CTRF}
\begin{equation}\label{TRKJ}
\min\limits_{\mathcal{G}^{(1)}, \mathcal{G}^{(2)}, \mathcal{G}^{(3)}}
\frac{1}{2}\left\| {\mathcal{Y} - \mathbf{\Phi} ( \mathcal{G}^{(1)}\times_2{\mathbf{U}_1},{\mathcal{G}^{(2)}}\times_2{\mathbf{U}_2},\mathcal{G}^{(3)} )} \right\|_F^2
+ \frac{\lambda}{2}\left\| {\mathcal{Z} - \mathbf{\Phi} ( \mathcal{G}^{(1)},\mathcal{G}^{(2)},\mathcal{G}^{(3)}\times_2\mathbf{U}_3 )} \right\|_F^2,
\end{equation}
where the constant ${\lambda>0}$ is employed to equilibrium the two terms concerning data fidelity.
Compared with the traditional matrix/tensor decomposition-based approaches, TR decomposition generates a more accurate representation for high-dimensional tensor data such as HR-HSI. Nevertheless,
the traditional TR decomposition model (\ref{TRKJ}) only captures the spatial-spectral high-correlation of HR-HSI.
Furthermore,
deriving the TR factors directly from the model (\ref{TRKJ}) poses an issue of instability. Therefore,
additional prior information should be integrated to acquire a stable solution \cite{FSTRD}.

% In particular, the spatial-spectral piecewise smooth structure serves as a significant prior
% when reconstructing the HR-HSI \cite{TVLR}.
% % However, HR-HSI is decomposed into a series of third-order tensor factors $\{\mathcal{G}^{(n)}\}_{n=1}^3$ by TR decomposition.
% % Therefore, this prior should act on the tensor factors, not directly on HR-HSI.
% Specifically,
In order to design an effective regularization of the tensor factors, it is necessary to explore the intrinsic relationship between tensor factors and HR-HSI \cite{TVLR}.
% According to (\ref{XN}), the expression of the TR decomposition of the HR-HSI is
% \begin{equation}\label{TRC}
% \mathbf{X}_{<n>} = \mathbf{G}_{(2)}^{(n)}{\left( \mathbf{G}_{<2>}^{(\ne n)} \right)^T},\ (n = 1,2,3).
% \end{equation}
% As the HR-HSI exhibits a piecewise smooth structure across two spatial dimensions and one spectral dimension, each column of matrix $\mathbf{X}_{<n>}$ represents continuous data.
In terms of matrix theory, one can derive from (\ref{XN}) that each column of
$\mathbf{X}_{<n>}$ can be formulated as a linear combination of all columns of the factor $\mathbf{G}_{(2)}^{(n)}$.
Therefore,
drawing from the principle that continuous bases have the capacity to depict continuous data, all columns of three factors $\{\mathbf{G}_{(2)}^{(n)}\}_{n=1}^3$  should be continuous.
For this purpose, the work presented in \cite{FSTRD} introduced an approach within the TR decomposition framework.
The weighted total variation (TV) with updating the weights was presented to constrain the TR factors so as to promote their continuity.
Specifically, the resulting model can be expressed as
\begin{equation}
\begin{split}
\min\limits_{\mathcal{G}^{(1)}, \mathcal{G}^{(2)}, \mathcal{G}^{(3)}}
&\frac{1}{2}\left\| {\mathcal{Y} - \mathbf{\Phi} ( \mathcal{G}^{(1)}\times_2\mathbf{U}_1,\mathcal{G}^{(2)}\times_2\mathbf{U}_2,\mathcal{G}^{(3)} )} \right\|_F^2
+ \frac{\lambda}{2}\left\| \mathcal{Z} - \mathbf{\Phi} ( \mathcal{G}^{(1)},\mathcal{G}^{(2)},\mathcal{G}^{(3)}\times _2{\mathbf{U}_3} ) \right\|_F^2 \\
&+\alpha\sum_{n=1}^3 \left\| \mathcal{W}^{(n)} \odot(\mathcal{G}^{(n)}\times_2\mathbf{D}) \right\|_1,
\end{split}
\end{equation}
where $\mathcal{W}^{(n)} \in {\mathbb{R}^{{R_n} \times {I_n} \times {R_{n+1}}}}$ stands for a non-negative weighted tensor, and $\mathbf{D}\in \mathbb{R}^{{I_n} \times {I_n}}$ is a first-order difference square matrix. Here, ${I_n}$ refers to the second dimension of $\mathcal{G}^{(n)}$.

%%%%%%%%%%%%%%%%%%%%%%%%%%%%%%%%%%%%%%%%%%%%%%%%%%%%
\setcounter{equation}{0}
\section{The proposed method}\label{method}
In this section, we first propose a novel HSI fusion model by combining logarithmic nuclear norm with the weighted TV. Then an efficient numerical algorithm is designed.

\subsection{Proposed model }
To delve deeper into the latent information held within the TR factors $\mathcal{G}^{(n)}$ ($n=1,2,3$), we introduce a low-rank constraint to each of these factors. 
In \cite{TRLRF}, Yuan et al. applied the matrix nuclear norm to the rank-mode unfoldings of all TR factors for tensor completion.
In \cite{CTRF}, the nuclear norm regularization along model-2 of only the third TR factor was proposed for HSI-MSI fusion.
Due to the three-dimensional structure of each $\mathcal{G}^{(n)}$, utilizing a matrix nuclear norm on its unfolded representation could potentially disrupt its inherent high-order structural correlation, consequently impacting the ultimate reconstruction outcomes. To address this, we instead apply a tensor nuclear norm (TNN) constraint to every individual TR factor within the framework of TR decomposition.

Nowadays, the TNN-based HSI-MSI fusion has attracted increasing attention owing to its advantages in preserving high-dimensional low-rank structures \cite{TNN,LRTRTNN,TRPCA}.
Nevertheless, the general TNN method merely represents a skewed approximation of tensor low-rank and does not effectively enhance the solution's low-rank characteristics.
Therefore, we design a logarithmic tensor nuclear norm (LTNN) and apply it to each TR factor $\mathcal{G}^{(n)}$. The proposed LTNN provides a more accurate low-rank approximation, since the logarithmic function produces an approximation that is more closely aligned with the $\ell_0$-norm compared to the $\ell_1$-norm.
Particularly, we employ the model-2 tensor singular value thresholding (t-SVT) to
handle singular values differently.
To clarify, we apply the milder reduction to the significant singular values, retaining the primary data components, while implementing a more substantial reduction to the smaller ones to mitigate random errors.
%He et al. \cite{LTMR}, collected coefficients from the same cluster into 3D tensors and imposed non-convex tensor rank constraints on these collected tensors to obtain excellent reconstruction results.
%In view of this, we further propose to introduce non-convex tensor nuclear norm into the fusion framework of TR decomposition to better preserve important data contained in larger singular values and to improve the fusion quality.
Specifically, our proposed HSI-MSI fusion model can be generated as
\begin{equation}\label{LogLRTR}
\begin{split}
\min\limits_{\mathcal{G}^{(1)}, \mathcal{G}^{(2)}, \mathcal{G}^{(3)}}
&\frac{1}{2}\left\| {\mathcal{Y} - \mathbf{\Phi} ( \mathcal{G}^{(1)}\times_2\mathbf{U}_1,\mathcal{G}^{(2)}\times_2\mathbf{U}_2,\mathcal{G}^{(3)})} \right\|_F^2
+ \frac{\lambda}{2}\left\| {\mathcal{Z} - \mathbf{\Phi}( \mathcal{G}^{(1)},\mathcal{G}^{(2)},\mathcal{G}^{(3)}\times_2\mathbf{U}_3)} \right\|_F^2 \\
&+\sum_{n=1}^{3}\left(\alpha\left\|\mathcal{W}^{(n)} \odot(\mathcal{G}^{(n)}\times_2\mathbf{D}) \right\|_1+\beta\left\|\mathcal{G}^{(n)}\right\|_\mathrm{LTNN}\right),
\end{split}
\end{equation}
where $\beta$ is a regularization parameter and
\begin{equation}
\left\| \mathcal{G} \right\|_{\mathrm{LTNN}} = \frac{1}{S}\sum\limits_{i=1}^S {\sum\limits_j {\log ({\sigma_j}({\overline {\mathbf{G}}}^{[i]}) + \varepsilon)}}.
\end{equation}
Here, $S$ represents the size of the second mode of tensor $\mathcal{G}$, $\varepsilon > 0$ is an extremely small positive value, and ${\sigma_j}({\overline {\mathbf{G}} }^{[i]})$ signifies the $j$th singular value of the $i$th lateral slice of tensor $\overline{\mathcal{G}}=\mathrm{fft}\left( \mathcal{G},[\enspace],2 \right)$. Note that our LTNN is different from the one in \cite{LTMR} since the corresponding   Fourier transform is along the second dimension instead of the third one.

%\tr{to do list: comparison on the singular value on the different directions of fft. }

%%%%%%%%%%%%%%%%%%%%%%%%%%%%%%%%%%%%%%%%%%%%%%%%%%%%%%%%%%%%%%%%%%%%%

\subsection{Algorithm development}

In this subsection, we introduce an iterative scheme for solving the proposed model (\ref{LogLRTR}).
Even though this model exhibits joint non-convexity with respect to the three factors $\mathcal{G}^{(n)}$ (where $n=1,2,3$), it demonstrates convexity for each individual separable variable.
Thus, we utilize the proximal alternating minimization (PAM) framework \cite{PAM1}, ensuring the convergence of the solution to a stationary point of the objective function.

For the sake of convenience,
we consider the objective function of (\ref{LogLRTR}) as $f(\mathcal{G}^{(1)},\mathcal{G}^{(2)},\mathcal{G}^{(3)})$. Additionally, the proximal term is introduced to update each factor within the PAM framework. In this way, we can solve the proposed model through alternately iterating  three subproblems as below:

\begin{equation}
\left\{
\begin{aligned}
&\mathcal{G}^{(1),k+1} = \arg\min\limits_{\mathcal{G}^{(1)}} f( \mathcal{G}^{(1)},\mathcal{G}^{(2),k},\mathcal{G}^{(3),k} ) + \frac{\eta}{2}\left\| \mathcal{G}^{(1)} - \mathcal{G}^{(1),k} \right\|_F^2,\\
&\mathcal{G}^{(2),k+1} = \arg\min\limits_{\mathcal{G}^{(2)}} f( \mathcal{G}^{(1),k+1},\mathcal{G}^{(2)},\mathcal{G}^{(3),k} ) + \frac{\eta}{2}\left\| \mathcal{G}^{(2)} - \mathcal{G}^{(2),k} \right\|_F^2,\\
&\mathcal{G}^{(3),k+1} = \arg\min\limits_{\mathcal{G}^{(3)}} f( \mathcal{G}^{(1),k+1},\mathcal{G}^{(2),k+1},\mathcal{G}^{(3)} ) + \frac{\eta}{2}\left\|\mathcal{G}^{(3)} - \mathcal{G}^{(3),k} \right\|_F^2,
\end{aligned}
\right.
\end{equation}
where $\eta>0$ is a parameter.
And $\frac{\eta}{2}\left\| { \cdot } \right\|_F^2$ signifies the proximal term.
In the following, we describe the solution of each subproblem in detail.

First, we discuss the solution of the $\mathcal{G}^{(1)}$-subproblem. This subproblem can be expressed as
\begin{equation}\label{G11}
\begin{split}
\min\limits_{\mathcal{G}^{(1)}}
&\frac{1}{2}\left\| {\mathcal{Y} - \mathbf{\Phi}( \mathcal{G}^{(1)}\times_2{\mathbf{U}_1},\mathcal{G}^{(2),k}\times_2\mathbf{U}_2,\mathcal{G}^{(3),k} )} \right\|_F^2 + \frac{\lambda}{2}\left\| {\mathcal{Z} - \mathbf{\Phi}( {\mathcal{G}^{(1)},\mathcal{G}^{(2),k},\mathcal{G}^{(3),k}\times_2{\mathbf{U}}_3})} \right\|_F^2\\
&+ \alpha {\left\| {\mathcal{W}^{(1)} \odot( \mathcal{G}^{(1)}\times_2{\mathbf{D}})} \right\|_1} + \beta{\left\| \mathcal{G}^{(1)} \right\|_\mathrm{LTNN}} + \frac{\eta }{2}\left\| {\mathcal{G}^{(1)} - \mathcal{G}^{(1),k}} \right\|_F^2.
\end{split}
\end{equation}
To efficiently solve (\ref{G11}), we employ the alternating direction method of multipliers (ADMM) \cite{ADMM1,ADMM2}.
The ADMM is a fast approach for handling the minimization of non-differentiable and high-order functionals.
To begin, the unconstrained problem (\ref{G11}) can be transformed into the following constrained optimization problem by introducing two auxiliary variables $\mathcal{R}_{1}$ and $\mathcal{V}_{1}$:
\begin{equation}
\begin{split}
\min \limits_{\mathcal{G}^{(1)}, \mathcal{R}_1,\mathcal{V}_1}
&\frac{1}{2}\left\|\mathcal{Y}-\mathbf{\Phi}(\mathcal{G}^{(1)} \times_{2} \mathbf{U}_1, \mathcal{G}^{(2),k} \times_2 \mathbf{U}_2, \mathcal{G}^{(3), k})\right\|_F^2+\frac{\lambda}{2}\left\|\mathcal{Z}-\mathbf{\Phi}(\mathcal{G}^{(1)}, \mathcal{G}^{(2),k}, \mathcal{G}^{(3),k} \times_2 \mathbf{U}_3)\right\|_F^2\\
&+\alpha\left\|\mathcal{W}^{(1)}\odot\mathcal{R}_1\right\|_1+\beta\left\|\mathcal{V}_{1}\right\|_{\mathrm{LTNN}}+\frac{\eta}{2}
\left\|\mathcal{G}^{(1)}-\mathcal{G}^{(1),k}\right\|_F^2,\\
\hbox{s.t.}\quad &\mathcal{R}_1=\mathcal{G}^{(1)} \times_2 \mathbf{D},\ \mathcal{V}_1=\mathcal{G}^{(1)}.
\end{split}
\end{equation}
Here we can obtain its augmented Lagrangian function:
\begin{equation}
\begin{split}
L_{\mu}&(\mathcal{G}^{(1)}, \mathcal{R}_1, \mathcal{V}_1; \mathcal{M}_1,  \mathcal{N}_1)\\
&=\frac{1}{2}\left\|\mathcal{Y}-\mathbf{\Phi}(\mathcal{G}^{(1)} \times_2\mathbf{U}_1, \mathcal{G}^{(2),k} \times_2 \mathbf{U}_2, \mathcal{G}^{(3),k})\right\|_F^2
+\frac{\lambda}{2}\left\|\mathcal{Z}-\mathbf{\Phi}(\mathcal{G}^{(1)}, \mathcal{G}^{(2),k}, \mathcal{G}^{(3), k} \times_2 \mathbf{U}_3)\right\|_F^2 \\
&+\frac{\eta}{2}\left\|\mathcal{G}^{(1)}-\mathcal{G}^{(1), k}\right\|_F^2
+\alpha\left\|\mathcal{W}^{(1)} \odot \mathcal{R}_1\right\|_1+\frac{\mu}{2}\left\|\mathcal{R}_1-\mathcal{G}^{(1)} \times_2 \mathbf{D}+\frac{\mathcal{M}_1}{\mu}\right\|_F^2\\
&+\beta\left\|\mathcal{V}_1\right\|_{\mathrm{LTNN}}+\frac{\mu}{2}\left\|\mathcal{V}_1-\mathcal{G}^{(1)}
+\frac{\mathcal{N}_1}{\mu}\right\|_F^2,
\end{split}
\end{equation}
where $\mathcal{M}_1$ , $\mathcal{N}_1$ are the Lagrangian multipliers and $\mu>0$ is a constant.
Then the update of $\mathcal{G}^{(1)}$ can be acquired by iteratively calculating the following three subproblems and updating the two Lagrangian multipliers:

a) The $\mathcal{R}_1$-subproblem is written as
\begin{equation}
\min\limits_{\mathcal{R}_1}\alpha\left\|\mathcal{W}^{(1)}\odot\mathcal{R}_1\right\|_1+\frac{\mu}{2}\left\|\mathcal{R}_1
-\mathcal{G}^{(1)} \times_2 \mathbf{D}+\frac{\mathcal{M}_1}{\mu}\right\|_F^2.
\end{equation}
This is a weighted ${\ell_1}$-norm minimization, and its closed-form solution is achieved by the soft-shrinkage operator, that is
\begin{equation}\label{R1}
\mathcal{R}_1=\mathrm{soft}(\mathcal{J}_1,\frac{\alpha}{\mu} \mathcal{W}^{(1)}).
\end{equation}
Here, $\mathcal{J}_1=\mathcal{G}^{(1)} \times_2 \mathbf{D}-\left(\mathcal{M}_1 / \mu\right)$, soft($a$, $b$)=sign($a$) $\ast$ $\max$($|a|$$-$$b$, 0).
Following the reweighting approach \cite{W1}, the weight coefficient is assigned as
\begin{equation}
\mathcal{W}^{(1)}=1 /\left(\left|\mathcal{J}_1\right|+\varsigma\right),
\end{equation}
where $\varsigma>0$ is a small constant used to eliminate singularity.

b) The $\mathcal{G}^{(1)}$-subproblem is formulated as
\begin{equation}\label{G1sub}
\begin{aligned}
\min_{\mathcal{G}^{(1)}} & \frac{1}{2}\left\|\mathcal{Y}-\mathbf{\Phi}(\mathcal{G}^{(1)} \times_2 \mathbf{U}_1, \mathcal{G}^{(2),k} \times_2 \mathbf{U}_2, \mathcal{G}^{(3), k})\right\|_F^2
+\frac{\lambda}{2}\left\|\mathcal{Z}-\mathbf{\Phi}(\mathcal{G}^{(1)}, \mathcal{G}^{(2),k}, \mathcal{G}^{(3),k} \times_2 \mathbf{U}_3)\right\|_F^2\\
\quad+& \frac{\eta}{2}\left\|\mathcal{G}^{(1)}-\mathcal{G}^{(1), k}\right\|_F^2+\frac{\mu}{2}\left\|\mathcal{R}_1-\mathcal{G}^{(1)} \times_2 \mathbf{D}+\frac{\mathcal{M}_1}{\mu}\right\|_F^2+\frac{\mu}{2}\left\|\mathcal{V}_1-\mathcal{G}^{(1)}
+\frac{\mathcal{N}_1}{\mu}\right\|_F^2.
\end{aligned}
\end{equation}
For convenience, let $\mathbf{P}_1=\left(\left(\mathcal{G}^{(2),k} \times_2\mathbf{U}_2\right) \mathcal{G}^{(3), k}\right)_{<2>}^T$ and
$ \mathbf{P}_2= \left(\mathcal{G}^{(2), k}\left(\mathcal{G}^{(3),k} \times_2 \mathbf{U}_3\right)\right)_{<2>}^T$.
Utilizing the matrix representation of TR decomposition (\ref{XN}), the above problem (\ref{G1sub}) can be rewritten as
\begin{equation}
\begin{aligned}
\min_{\mathbf{G}_{(2)}^{(1)}} & \frac{1}{2}\left\|\mathbf{Y}_{<1>}-\mathbf{U}_1 \mathbf{G}_{(2)}^{(1)} \mathbf{P}_1\right\|_F^2+\frac{\lambda}{2}\left\|\mathbf{Z}_{<1>}
-\mathbf{G}_{(2)}^{(1)} \mathbf{P}_2\right\|_F^2+\frac{\eta}{2}\left\|\mathbf{G}_{(2)}^{(1)}-\mathbf{G}_{(2)}^{(1),k}\right\|_F^2\\
&+\frac{\mu}{2}\left\|\mathbf{R}_\mathbf{1(2)}-\mathbf{D} \mathbf{G}_{(2)}^{(1)}
+\frac{\mathbf{M}_{1(2)}}{\mu}\right\|_F^2+\frac{\mu}{2}\left\|\mathbf{V}_{1(2)}
-\mathbf{G}_{(2)}^{(1)}+\frac{\mathbf{N}_{1(2)}}{\mu}\right\|_F^2.
\end{aligned}
\end{equation}
Its Euler-Lagrange equation is
\begin{equation}\label{G1}
\begin{aligned}
\mathbf{U}_1^T \mathbf{U}_{1} \mathbf{G}_{(2)}^{(1)} \mathbf{P}_1\mathbf{P}_1^T+\lambda \mathbf{G}_{(2)}^{(1)} \mathbf{P}_2 \mathbf{P}_2^T+\eta\mathbf{G}_{(2)}^{(1)}+\mu \mathbf{D}^T \mathbf{D} \mathbf{G}_{(2)}^{(1)}+\mu \mathbf{G}_{(2)}^{(1)}
=\mathbf{U}_1^T \mathbf{Y}_{<1>} \mathbf{P}_1^T\\
+\lambda \mathbf{Z}_{<1>} \mathbf{P}_{2}^T+\eta\mathbf{G}_{(2)}^{(1),k}+\mu\mathbf{D}^T\left(\mathbf{R}_{1(2)}+\frac{\mathbf{M}_{1(2)}}{\mu}\right)
+\mu\left(\mathbf{V}_{1(2)} +\frac{\mathbf{N}_{1(2)}}{\mu}\right).
\end{aligned}
\end{equation}
As a general Sylvester equation, this linear system is solved using the conjugate gradient (CG) method.
Subsequently, the TR factor $\mathcal{G}^{(1)}$ is acquired by folding the solution ${\mathbf{G}}_{(2)}^{(1)}$: $\mathcal{G}^{(1)}=\operatorname{Fold}_2(\mathbf{G}_{(2)}^{(1)})$.

c) The $\mathcal{V}_1$-subproblem can be expressed as
\begin{equation}
\min\limits_{\mathcal{V}_1}\beta\left\|\mathcal{V}_1\right\|_\mathrm{LTNN}+\frac{\mu}{2}\left\|\mathcal{V}_1-\mathcal{G}^{(1)}+\frac{\mathcal{N}_1}{\mu}\right\|_F^2.
\end{equation}
According to the work in \cite{V1}, its solution can be given by
\begin{equation}\label{V1}
\begin{aligned}
\mathcal{V}_1=\mathcal{D}_{\mathrm{LTNN}}^{\beta/\mu}\left(\mathcal{G}^{(1)}-\frac{\mathcal{N}_1}{\mu}\right)=\mathcal{U} \mathcal{S}_{\mathrm{LTNN}}^{\beta/ \mu, \varepsilon} \mathcal{V}^T,
\end{aligned}
\end{equation}
where $\mathcal{G}^{(1)}-\frac{\mathcal{N}_1}{\mu}=\mathcal{U} \mathcal{S} \mathcal{V}^T, \mathcal{S}_\mathrm{LTNN}^{\beta/\mu, \varepsilon}=\mathrm{ifft}(\overline{\mathcal{S}}_\mathrm{LTNN}^{\beta/\mu, \varepsilon},[\enspace],2)$.
The thresholding operator $\overline{\mathcal{S}}_\mathrm{LTNN}^{\beta/\mu ,\varepsilon}$ is defined as
\begin{equation}
\overline{\mathcal{S}}_\mathrm{LTNN}^{\beta / \mu, \varepsilon}(i_1,i_2,i_3)=
\left\{
\begin{array}{ll}
0, & \mathrm{if}\ c_2 \leq 0 ,\\
\mathrm{sign}(\overline{\mathcal{S}}\ (i_1,i_2,i_3))\left(\frac{c_1+\sqrt{c_2}}{2}\right),
& \mathrm{if}\ c_2>0.
\end{array}
\right.
\end{equation}
Among them, $\overline{\mathcal{S}}=\mathrm{fft}(\mathcal{S},[\enspace],2)$,    $c_{1}=|\overline{\mathcal{S}}(i_1,i_2,i_3)|-\varepsilon$,
$c_{2}=c_1^2-4(\beta/\mu-\varepsilon|\overline{\mathcal{S}}(i_1,i_2,i_3)|)$.
In this way,
the thresholding operator shrinks less for larger singular values and more for smaller ones \cite{V2}.
%Algorithm 1 summarizes the algorithm for estimating $\mathcal{V}_{1}$.

d) The update of Lagrangian multipliers $\mathcal{M}_1$, $\mathcal{N}_1$ unfolds in the subsequent manner:
\begin{equation}\label{MN1}
\left\{
\begin{aligned}
&{\mathcal{M}_1} \leftarrow {\mathcal{M}_1} + \mu ( {\mathcal{R}_1} - {\mathcal{G}^{(1)}}\times_2{\mathbf{D}}) ,\\
&{\mathcal{N}_1} \leftarrow {\mathcal{N}_1} + \mu ( {\mathcal{V}_1} - {\mathcal{G}^{(1)}}).
\end{aligned}
\right.
\end{equation}
Through iteratively updating  $\mathcal{R}_1$, $\mathcal{G}^{(1)}$, $\mathcal{V}_1$, $\mathcal{M}_1$, and $\mathcal{N}_1$, the first factor $\mathcal{G}^{(1),k+1}$ could be acquired.

Second, we focus on solving the $\mathcal{G}^{(2)}$-subproblem. This subproblem can be rewritten as
\begin{equation}
\begin{split}
\min\limits_{\mathcal{G}^{(2)}}
&\frac{1}{2}\left\| {\mathcal{Y} - \mathbf{\Phi}( {\mathcal{G}^{(1),k+1}\times_2{\mathbf{U}_1},\mathcal{G}^{(2)}\times_2{\mathbf{U}_2},\mathcal{ G}^{(3),k}})} \right\|_F^2 + \frac{\lambda}{2}\left\| {\mathcal{Z} - \mathbf{\Phi}( {\mathcal{G}^{(1),k+1},\mathcal{G}^{(2)},\mathcal{G}^{(3),k}\times_2\mathbf{U}_3} )} \right\|_F^2\\
&+ \alpha {\left\| \mathcal{W}^{(2)} \odot( {\mathcal{G}^{(2)}\times_2\mathbf{D}}) \right\|_1} + \beta {\left\| \mathcal{G}^{(2)} \right\|_\mathrm{LTNN}} + \frac{\eta}{2}\left\| \mathcal{G}^{(2)} - \mathcal{G}^{(2),k} \right\|_F^2.
\end{split}
\end{equation}
Similarly, we can obtain the process of optimizing $\mathcal{G}^{(2)}$.
By employing the ADMM, the update for each subproblem is outlined as
\begin{equation}\label{R2}
\mathcal{R}_2=\mathrm{soft}(\mathcal{J}_2,\frac{\alpha}{\mu} \mathcal{W}^{(2)}),
\end{equation}
\begin{equation}\label{G2}
\begin{aligned}
\mathbf{U}_2^T \mathbf{U}_2 \mathbf{G}_{(2)}^{(2)} \mathbf{Q}_1 \mathbf{Q}_1^T+\lambda \mathbf{G}_{(2)}^{(2)} \mathbf{Q}_2 \mathbf{Q}_2^T+\eta\mathbf{G}_{(2)}^{(2)}+\mu \mathbf{D}^T \mathbf{D} \mathbf{G}_{(2)}^{(2)}+\mu \mathbf{G}_{(2)}^{(2)}=\mathbf{U}_2^T \mathbf{Y}_{<2>} \mathbf{Q}_1^T\\
+\lambda\mathbf{Z}_{<2>}\mathbf{Q}_2^T+\eta\mathbf{G}_{(2)}^{(2),k}+\mu\mathbf{D}^T\left(\mathbf{R}_{2(2)}
+\frac{\mathbf{M}_{2(2)}}{\mu}\right)+\mu\left(\mathbf{V}_{2(2)}+\frac{\mathbf{N}_{2(2)}}{\mu}\right),
\end{aligned}
\end{equation}
\begin{equation}\label{V2}
\mathcal{V}_2=\mathcal{D}_\mathrm{LTNN}^{\beta/\mu}\left(\mathcal{G}^{(2)}-\frac{\mathcal{N}_2}{\mu}\right),
\end{equation}
\begin{equation}\label{MN2}
\left\{
\begin{aligned}
&\mathcal{M}_2 \leftarrow \mathcal{M}_2 + \mu ( \mathcal{R}_2 - \mathcal{G}^{(2)}\times_2\mathbf{D}) ,\\
&\mathcal{N}_2 \leftarrow {\mathcal{N}_2} + \mu ( \mathcal{V}_2 - \mathcal{G}^{(2)}),
\end{aligned}
\right.
\end{equation}
where $\mathcal{R}_2$, $\mathcal{V}_2$ and $\mathcal{M}_2$, $\mathcal{N}_2$ are auxiliary variables and Lagrangian multipliers, respectively. In addition, some variables in the aforementioned equations are defined as
\begin{equation}
\left\{
\begin{aligned}
&\mathcal{J}_2 = \mathcal{G}^{(2)}\times_2\mathbf{D} - (\mathcal{M}_2/\mu ),\ \mathcal{W}^{(2)} = 1/( \left| \mathcal{J}_2 \right| + \varsigma),\\
&\mathbf{Q}_1 = ( \mathcal{G}^{(3),k}( \mathcal{G}^{(1),k+1}\times_2\mathbf{U}_1 ))^T_{<2>},\ \mathbf{Q}_2 = ( {(\mathcal{G}^{(3),k}\times_2\mathbf{U}_3)\mathcal{G}^{(1),k+1}} )^T_{<2>},\\
&\mathcal{G}^{(2)} - \frac{\mathcal{N}_2}{\mu} = \mathcal{U}\mathcal{S}\mathcal{V}^T.
%\ \mathcal{S}_\mathrm{LTNN}^{\beta/\mu,\varepsilon } = \mathrm{ifft}(\overline{\mathcal{S}} _\mathrm{LTNN}^{ \beta/\mu ,\varepsilon },[\enspace],2).
\end{aligned}
\right.
\end{equation}

Finally, we consider the $\mathcal{G}^{(3)}$-subproblem, which can be written as
\begin{equation}
\begin{aligned}
\min\limits_{\mathcal{G}^{(3)}}
&\frac{1}{2}\left\| {\mathcal{Y} - \mathbf{\Phi}( \mathcal{G}^{(1),k+1}\times_2{\mathbf{U}_1},\mathcal{G}^{(2),k+1}\times_2{\mathbf{U}_2},\mathcal{G}^{(3)})} \right\|_F^2 + \alpha {\left\| {\mathcal{W}^{(3)} \odot ({\mathcal{G}^{(3)}}\times_2{\mathbf{D}})} \right\|_1} + \beta {\left\| \mathcal{G}^{(3)} \right\|_\mathrm{LTNN}}\\
&+ \frac{\lambda}{2}\left\| {\mathcal{Z} - \mathbf{\Phi}( \mathcal{G}^{(1),k+1},\mathcal{G}^{(2),k+1},\mathcal{G}^{(3)}\times_2{\mathbf{U}_3} )} \right\|_F^2  + \frac{\eta}{2}\left\| {\mathcal{G}^{(3)} - \mathcal{G}^{(3),k}} \right\|_F^2.
\end{aligned}
\end{equation}
Similarly, the update of each subproblem is as follows:
\begin{equation}\label{R3}
\mathcal{R}_3=\mathrm{soft}(\mathcal{J}_3,\frac{\alpha}{\mu} \mathcal{W}^{(3)}),
\end{equation}
\begin{equation}\label{G3}
\begin{aligned}
\lambda \mathbf{U}_3^T\mathbf{U}_{3} \mathbf{G}_{(2)}^{(3)} \mathbf{T}_2 \mathbf{T}_2^T+\mathbf{G}_{(2)}^{(3)} \mathbf{T}_1 \mathbf{T}_1^T+\eta\mathbf{G}_{(2)}^{(3)}+\mu\mathbf{D}^T \mathbf{D} \mathbf{G}_{(2)}^{(3)}+\mu \mathbf{G}_{(2)}^{(3)}= \mathbf{Y}_{<3>} \mathbf{T}_1^T\\
+\lambda\mathbf{U}_3^T\mathbf{Z}_{<3>}\mathbf{T}_2^T+\eta\mathbf{G}_{(2)}^{(3),k}+\mu\mathbf{D}^T\left(\mathbf{R}_{3(2)}+\frac{\mathbf{M}_{3(2)}}{\mu}\right)+\mu\left(\mathbf{V}_{3(2)}+\frac{\mathbf{N}_{3(2)}}{\mu}\right),
\end{aligned}
\end{equation}
\begin{equation}\label{V3}
\mathcal{V}_3=\mathcal{D}_\mathrm{LTNN}^{\beta/\mu}\left(\mathcal{G}^{(3)}-\frac{\mathcal{N}_3}{\mu}\right),
\end{equation}
\begin{equation}\label{MN3}
\left\{
\begin{aligned}
&{\mathcal{M}_3} \leftarrow {\mathcal{M}_3} + \mu ( \mathcal{R}_3 -\mathcal{G}^{(3)}\times_2\mathbf{D} ) ,\\
&{\mathcal{N}_3} \leftarrow {\mathcal{N}_3} + \mu ( \mathcal{V}_3 - \mathcal{G}^{(3)}).
\end{aligned}
\right.
\end{equation}
Here, $\mathcal{R}_3$, $\mathcal{V}_3$ and $\mathcal{M}_3$, $\mathcal{N}_3$ serve as auxiliary variables and Lagrangian multipliers, respectively.
Additionally, some variables in the aforementioned equations are defined as
\begin{equation}
\left\{
\begin{aligned}
&\mathcal{J}_3 = \mathcal{G}^{(3)}\times_2{\mathbf{D}} - \left( {\mathcal{M}_3}/\mu \right),\ \mathcal{W}^{(3)} = 1/\left({\left| {\mathcal{J}_3} \right| + \varsigma} \right),\\
&\mathbf{T}_1 = ( {( \mathcal{G}^{(1),k+1}\times_2{\mathbf{U}_1})( \mathcal{G}^{(2),k+1}\times_2{\mathbf{U}_2} )} )_{<2>}^T,\ \mathbf{T}_2 = ( \mathcal{G}^{(1),k+1}{\mathcal{G}^{(2),k+1}} )^T_{<2>},\\
&\mathcal{G}^{(3)} - \frac{\mathcal{N}_3}{\mu} = \mathcal{U}\mathcal{S}\mathcal{V}^T.
%\mathcal{S}_\mathrm{LTNN}^{\beta/\mu ,\varepsilon } = \mathrm{ifft} (\overline{\mathcal{S}}_{\mathrm{LTNN}}^{ \beta/\mu ,\varepsilon },[\enspace],2).
\end{aligned}
\right.
\end{equation}

We summarize the algorithm for solving the proposed HSI-MSI fusion model (\ref{LogLRTR}) in Algorithm \ref{algorithm1}.
Due to the non-convex nature of this optimization problem, a well-chosen initialization will result in a favorable solution.
In experiments, we decompose LR-HSI $\mathcal{Y}$ and HR-MSI $\mathcal{Z}$ using TR-SVD proposed in the literature \cite{TRSVD} to initialize the variables
$\left\{\mathcal{G}^{(1),0},\mathcal{G}^{(2),0},\mathcal{G}^{(3),0} \right\}$.
\begin{algorithm}[!ht]\caption{Optimization of the proposed model (\ref{LogLRTR})} \label{algorithm1}
    \renewcommand{\algorithmicrequire}{\textbf{Input:}}
	\renewcommand{\algorithmicensure}{\textbf{Output:}}
    \begin{algorithmic}[1]
        \REQUIRE LR-HSI $\mathcal{Y}$, HR-MSI $\mathcal{Z}$, $\mathbf{U}_1,\mathbf{U}_2,\mathbf{U}_3$

        \STATE Parameters: $\mathrm{R}= [\mathrm{R}_1, \mathrm{R}_2, \mathrm{R}_3]$, $\lambda, \alpha, \beta, \eta, \mu$, and kMax

        \STATE Initialization: $\mathcal{G}^{(1),0},\mathcal{G}^{(2),0},\mathcal{G}^{(3),0}$, and $k=0$

        \STATE \textbf{while} $k<$kMax \textbf{do}
        \STATE \quad Initialization: $\mathcal{G}^{(1)}=\mathcal{G}^{(1),k},\ \mathcal{V}_1=\mathcal{M}_1=\mathcal{N}_1=0$
        \STATE \quad\textbf{while} not coverged \textbf{do}
        \STATE  \qquad Update $\mathcal{R}_1$ via (\ref{R1})
        \STATE  \qquad Update $\mathcal{G}^{(1)}$ via (\ref{G1})
        \STATE  \qquad Update $\mathcal{V}_1$ via (\ref{V1})
        \STATE  \qquad Update $\mathcal{M}_1,\ \mathcal{N}_1$ via (\ref{MN1})
        \STATE \quad\textbf{end while}
        \STATE \quad\textbf{Output:} $\mathcal{G}^{(1),k+1}$
         \STATE \quad Initialization: $\mathcal{G}^{(2)}=\mathcal{G}^{(2),k},\ \mathcal{V}_2=\mathcal{M}_2=\mathcal{N}_2=0$
        \STATE \quad\textbf{while} not coverged \textbf{do}
        \STATE  \qquad Update $\mathcal{R}_2$ via (\ref{R2})
        \STATE  \qquad Update $\mathcal{G}^{(2)}$ via (\ref{G2})
        \STATE  \qquad Update $\mathcal{V}_2$ via (\ref{V2})
        \STATE  \qquad Update $\mathcal{M}_2,\ \mathcal{N}_2$ via (\ref{MN2})
        \STATE \quad\textbf{end while}
        \STATE \quad\textbf{Output:} $\mathcal{G}^{(2),k+1}$
        \STATE \quad Initialization: $\mathcal{G}^{(3)}=\mathcal{G}^{(3),k},\ \mathcal{V}_3=\mathcal{M}_3=\mathcal{N}_3=0$
        \STATE \quad\textbf{while} not coverged \textbf{do}
        \STATE  \qquad Update $\mathcal{R}_3$ via (\ref{R3})
        \STATE  \qquad Update $\mathcal{G}^{(3)}$ via (\ref{G3})
        \STATE  \qquad Update $\mathcal{V}_3$ via (\ref{V3})
        \STATE  \qquad Update $\mathcal{M}_3,\ \mathcal{N}_3$ via (\ref{MN3})
        \STATE \quad\textbf{end while}
        \STATE \quad\textbf{Output:} $\mathcal{G}^{(3),k+1}$
        \STATE \quad {Assign $k$  by $k+1$ }
		\STATE \quad {If $\| \mathbf{\Phi}(\mathcal{G})^{k}- \mathbf{\Phi}(\mathcal{G})^{k-1}\|_F/\| \mathbf{\Phi}(\mathcal{G})^{k}\|_F \leq \epsilon$, stop the iteration}
        \STATE \textbf{end while}
        \ENSURE HR-HSI $\mathcal{X}=\mathbf{\Phi}(\mathcal{G}^{(1),k}, \mathcal{G}^{(2),k}, \mathcal{G}^{(3),k})$
    \end{algorithmic}
\end{algorithm}

%%%%%%%%%%%%%%%%%%%%%%%%%%%%%%%%%%%%%%%%%%%%%%%%%%%%%%%%%%%%%%%

\section{Experiments}\label{Experiments}

In this section, we commence by providing an overview of the datasets utilized in the conducted experiments. Following that, we introduce five prominent quantitative metrics that have been employed to evaluate the quality of the fusion outcomes.
Furthermore, we delve into an analysis of the impact of various parameters on the experimental results, subsequently presenting a list of the identified optimal parameters.
Finally, we show the experimental results obtained by the proposed method (named LogLRTR) and provide a convergence analysis.
Specifically, to establish the superiority of our approach, we compare it with three advanced methods, including coupled sparse tensor factorization (CSTF) \cite{CSTF}, coupled TR factorization (CTRF) \cite{CTRF}, and factor smoothed tensor ring decomposition (FSTRD) \cite{FSTRD}.
All experiments were performed on a desktop computer equipped with a 2.70 GHz Intel(R) Core(TM) i5-11400U CPU and 16 GB RAM using MATLAB (R2018a).

\subsection{Datasets}
The University of Pavia dataset \cite{Pavia} was acquired using the ROSIS sensor during an aerial campaign conducted over the University of Pavia, Italy.
The HSI encompasses $115$ spectral bands, encompassing an image dimension of $610\times340$ pixels.
After excluding bands with low signal-to-noise ratio (SNR) and delineating the specific subregion, we select the upper-left 256×256 spatial pixels containing 93 spectral bands as our HR-HSI.
Employing a Gaussian blur (with a size of $7\times7$ and a standard deviation of 2), we simulated the low-resolution HSI. Subsequently, the image underwent a downsampling process with a reduction factor of 4.
The HR-MSI consisting of four bands, is generated using the IKONOS-like reflectance spectral response filter \cite{MSI}.

The CAVE dataset \cite{CAVE}, which comprises 32 indoor HSIs captured by a generalized combined pixel camera within real scenes, has become a prevalent choice for HSI-MSI fusion \cite{LRSR,LTMR}. Each HSI within this dataset has dimensions of $512\times512\times31$, with $512\times512$ indicating spatial pixels and 31 signifying spectral bands. These 31 spectral bands cover the wavelength range of $400\sim700$ nm in 10 nm intervals.
For our testing requirements, we have selected three distinctive HSI scenes (Superballs, Flowers, and Peppers) as the primary real-world dataset. To derive the corresponding LR-HSI, we apply a symmetric Gaussian blur with a standard deviation of 2, utilizing a $7\times7$ kernel to convolve the HR-HSI.
During the downsampling process, we reduce the spatial resolution by selecting every 8 pixels in both spatial dimensions for each HSI band.
To generate the HR-MSI, we employ spectral downsampling on the HR-HSI from the CAVE dataset, utilizing the spectral response matrix associated with the Nikon D700 camera.

In addition, we add Gaussian noise with SNRs of 25dB and 30dB to the LR-HSI and the HR-MSI, respectively.

%%%%%%%%%%%%%%%%%%%%%%%%%%%%%%%%%%%%%%%%%%%%%%%%%%%%%%%%%%%%%
\subsection{Quantitative indicators}
In order to fully assess the quality of HR-HSI obtained by fusing LR-HSI and HR-MSI, five quality metrics are used in our study, including the average peak signal-to-noise ratio (PSNR) \cite{CTRF}, structural similarity index (SSIM) \cite{SSIM}, relative dimensionless global error in synthesis (ERGAS) \cite{ERGAS}, spectral angle mapper (SAM) \cite{SAM}, and universal image quality index (UIQI) \cite{UIQI}.
Among these metrics, higher values of SSIM, PSNR, and UIQI indicate superior quality, while lower values of ERGAS and SAM correspond to enhanced quality measures.
Before initiating the simulation, the pixel values of the HR-HSI are normalized within the range [0, 1]. During the computation of quality metrics, all image elements are subsequently rescaled to span the range [0, 255].

%%%  The computational efficiency of the comparison method is evaluated in terms of running time (seconds).

\subsection{Parameters analysis}

Within our devised method, there are six regularization parameters $\lambda$, $\alpha$, $\beta$, $\mathrm{R}_1$, $\mathrm{R}_2$ and $\mathrm{R}_3$, as well as two algorithmic parameters $\eta$, $\mu$. Notably, $\mathrm{R}_1$, $\mathrm{R}_2$ and $\mathrm{R}_3$ make up the TR rank $\mathrm{R}$.
Initially, it is imperative to establish a sensible choice for the TR rank.
Subsequently, a stepwise adjustment is conducted for the remaining five parameters.
While fine-tuning these parameters, our approach began with informed estimations for their initial values, relying on our prior expertise.
% Then we gradually adjusted the parameters through a trial-and-error approach, ultimately identifying the optimal parameter set that produced the highest PSNR value.
We iteratively fine-tuned the parameters using a trial-and-error methodology, systematically searching for the optimal parameter configuration that yielded the highest peak signal-to-noise ratio (PSNR) value.

\begin{figure}[h!]
\centering
{\includegraphics[width=2.5in]{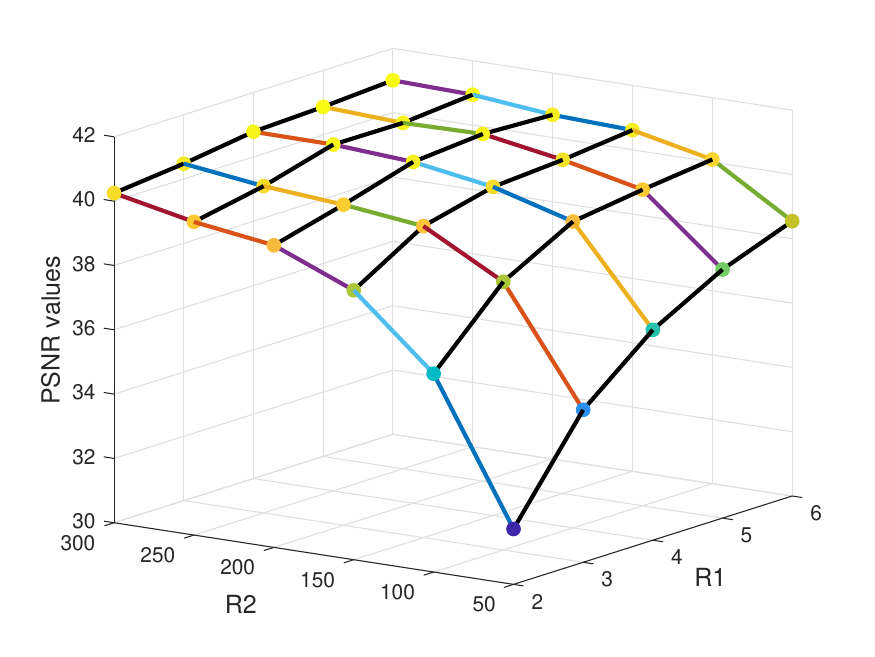}\vspace{-1em}}
\caption{Sensitivity analysis of TR rank.}\vspace{-0.5em}
\label{fig1}
\end{figure}

\textbf{Analysis of TR rank:}
In experiments, we find that the TR rank is usually needed into a pattern that sets the two sides of the rank smaller and the middle larger.
To simplify the complexity of the parameters analysis, we choose $\mathrm{R}_3=\mathrm{R}_1$ to maintain the consistency of the spatial TR core tensor.
$\mathrm{R}_1$ and $\mathrm{R}_3$ undergo adjustments within the range of 2 to 6, while $\mathrm{R}_2$ is varied within the interval of 50 to 300.
Fig. \ref{fig1} shows the PSNR curves versus TR rank $[\mathrm{R}_1,\mathrm{R}_2,\mathrm{R}_1]$ for the reconstructed Pavia University dataset.
From this figure, we find that a larger value should be chosen for both $\mathrm{R}_1$ and $\mathrm{R}_2$ to obtain satisfactory results.

Specifically, Table \ref{table1} lists the optimal parameter sets chosen for the LogLRTR.
\begin{table}[htbp]\vspace{-1em}
\caption{Selected parameter sets for the proposed method.}
\centering
\begin{tabular}{ccccccc}
\hline
{\rule[-1mm]{0mm}{6mm}}Image & $\mathrm{R}$ &$\lambda$  &$\alpha$ &$\beta$ &$\eta$&$\mu$     \\
\hline
{\rule[-1mm]{0mm}{6mm}}Pavia University &[6,300,6]&0.01 &$1\cdot10^{-5}$ &3&20 &0.2 \\
{\rule[-1mm]{0mm}{6mm}}Flowers     &[5,300,5]&1.3 &$7\cdot10^{-4}$  &1   &0.8 &0.04 \\
{\rule[-1mm]{0mm}{6mm}}Superballs  &[5,300,5]&1   &$3\cdot10^{-5}$  &0.8  &0.7  &0.04 \\
{\rule[-1mm]{0mm}{6mm}}Peppers     &[5,300,5]&1.2 &$3\cdot10^{-4}$  &0.3  &0.9&0.08  \\
\hline
\end{tabular}
\label{table1}
\end{table}

\subsection{Experimental results}
The numerical results of various fusion approaches for ``Pavia University", ``Flowers", ``Superballs", and ``Peppers" are displayed in Table \ref{table2}, with the optimal values highlighted in bold.
As shown in Table \ref{table2}, LogLRTR outperforms other comparative methods in all five evaluation indicators.
This suggests that our approach substantially improves both the spatial and spectral characteristics of the fused HR-HSI in the Pavia and CAVE datasets.
In particular, across the four images, LogLRTR exhibits a PSNR advantage over FSTRD by 0.76 dB, 0.69 dB, 0.76 dB, and 0.72 dB, respectively.

\begin{table}[htbp]\vspace{-1em}
\caption{Quantitative evaluation of four different fusion methods.}
\centering
\resizebox{\textwidth}{!}{
\newcommand{\rb}[1]{\raisebox{1.0ex}[0pt]{#1}}
\begin{tabular}{lllllllllllllll}
\hline
{\rule[-1mm]{0mm}{6mm}}\hbox{Image}&\multicolumn{5}{c}{Pavia University}        &\multicolumn{4}{c}{Flowers} \\
\cmidrule(r){2-6}\cmidrule(r){7-11}%\noalign{\smallskip}
{\rule[-1mm]{0mm}{6mm}}Method&PSNR&SSIM&ERGAS&SAM&UIQI&PSNR&SSIM&ERGAS&SAM&UIQI\\
\hline
{\rule[-1mm]{0mm}{6mm}}Best Values &$+\infty$ &1 &0 &0 &1  &$+\infty$ &1 &0 &0 &1  \\
{\rule[-1mm]{0mm}{6mm}}CSTF &39.462 &0.963 &1.605 &3.119  &0.983              &42.640 &0.963 &1.445 &14.350 &0.718   \\
{\rule[-1mm]{0mm}{6mm}}CTRF &39.797 &0.971 &1.555 &2.922  &0.986              &41.934 &0.952 &3.995 &18.832 &0.682   \\
{\rule[-1mm]{0mm}{6mm}}FSTRD &40.272 &0.973 &1.468 &2.717 &0.987              &43.867 &0.971 &1.210 &11.979 &0.728   \\
{\rule[-1mm]{0mm}{6mm}}LogLRTR     &\textbf{41.035}&\textbf{0.977}&\textbf{1.346}&\textbf{2.472}&\textbf{0.989}
&\textbf{44.557}&\textbf{0.976}&\textbf{1.127}&\textbf{10.839}&\textbf{0.746}\\
\hline
{\rule[-1mm]{0mm}{6mm}}\hbox{Image} &\multicolumn{5}{c}{Superballs}           &\multicolumn{4}{c}{Peppers}             \\
\cmidrule(r){2-6}\cmidrule(r){7-11}%\noalign{\smallskip}
{\rule[-1mm]{0mm}{6mm}}Method  &PSNR  &SSIM  &ERGAS  &SAM   &UIQI               &PSNR  &SSIM  &ERGAS  &SAM &UIQI\\
\hline
{\rule[-1mm]{0mm}{6mm}}Best Values &$+\infty$ &1     &0      &0      &1                  &$+\infty$ &1        &0        &0         &1      \\
{\rule[-1mm]{0mm}{6mm}}CSTF        &43.809    &0.970 &1.780  &9.424  &0.762              &43.754    &0.974    &1.466    &8.557     &0.799  \\
{\rule[-1mm]{0mm}{6mm}}CTRF        &43.088    &0.959 &1.995  &12.000 &0.753              &42.305    &0.959    &1.782    &11.876    &0.760  \\
{\rule[-1mm]{0mm}{6mm}}FSTRD       &44.690    &0.972 &1.619  &8.948  &0.765              &45.596    &0.984    &1.247    &6.687     &0.842  \\
{\rule[-1mm]{0mm}{6mm}}LogLRTR    &\textbf{45.458}&\textbf{0.976}&\textbf{1.474}&\textbf{8.546}&\textbf{0.791}
&\textbf{46.317}&\textbf{0.986}&\textbf{1.101}&\textbf{6.110}&\textbf{0.850}\\ \hline
\end{tabular}}\vspace{-0.5em}
\label{table2}
\end{table}

\begin{figure}
\centering
\subfigure[Pavia University]{
    \begin{minipage}[b]{0.225\linewidth}
    \includegraphics[width=4.25cm]{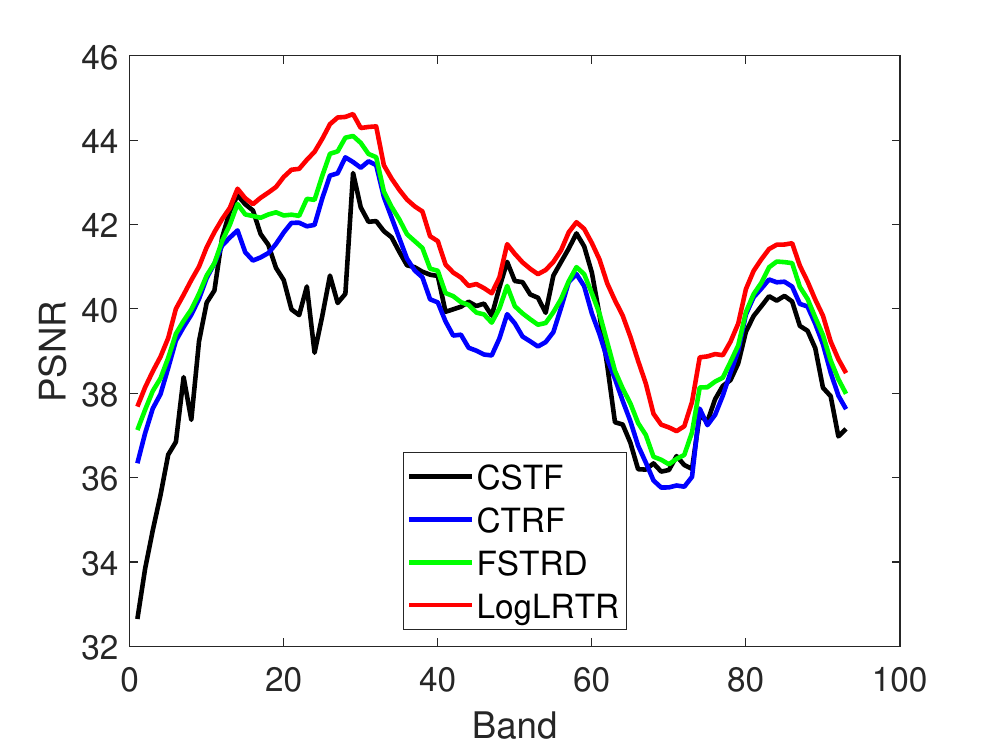}\vspace{1pt}
    \includegraphics[width=4.25cm]{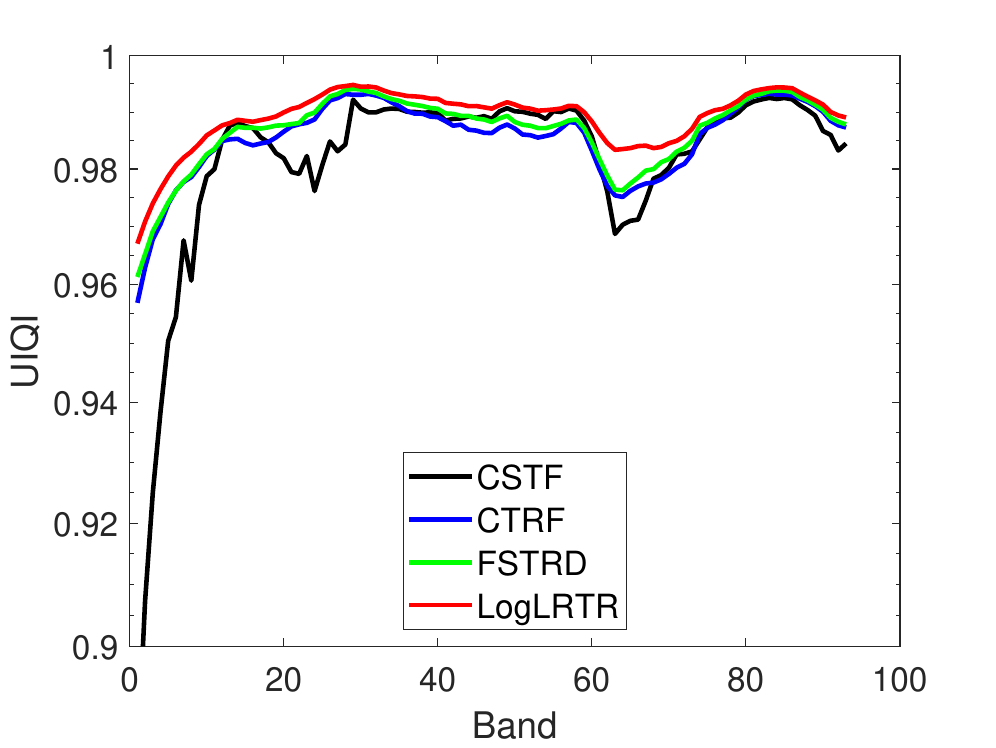}\vspace{1pt}
    \end{minipage}
}
\subfigure[Flowers]{
    \begin{minipage}[b]{0.225\linewidth}
    \includegraphics[width=4.25cm]{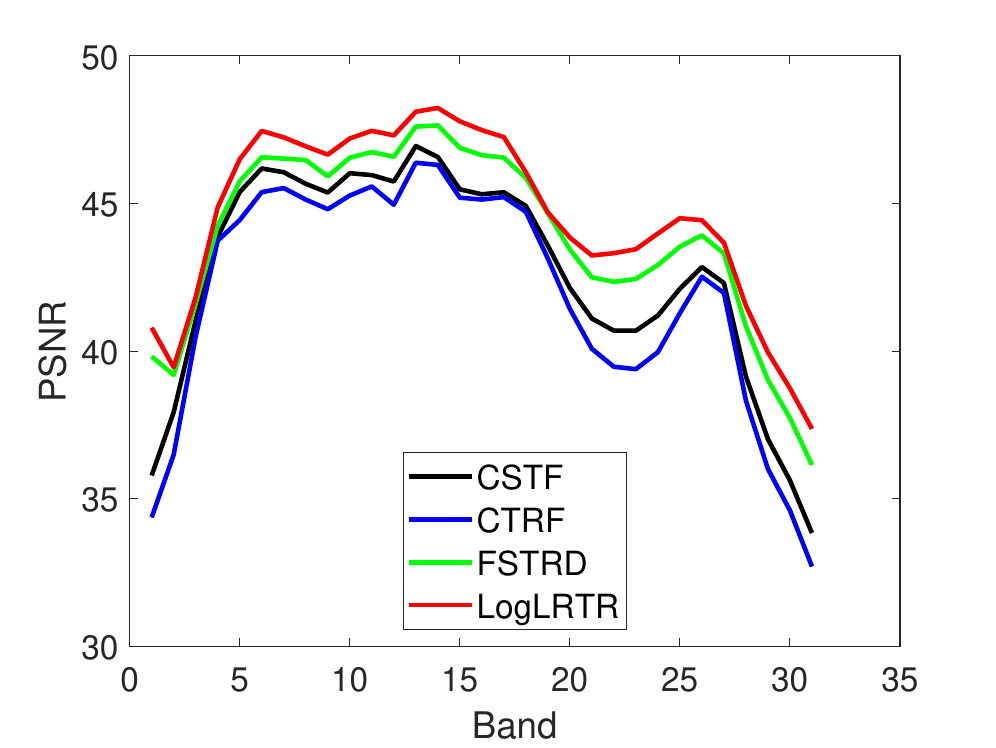}\vspace{1pt}
    \includegraphics[width=4.25cm]{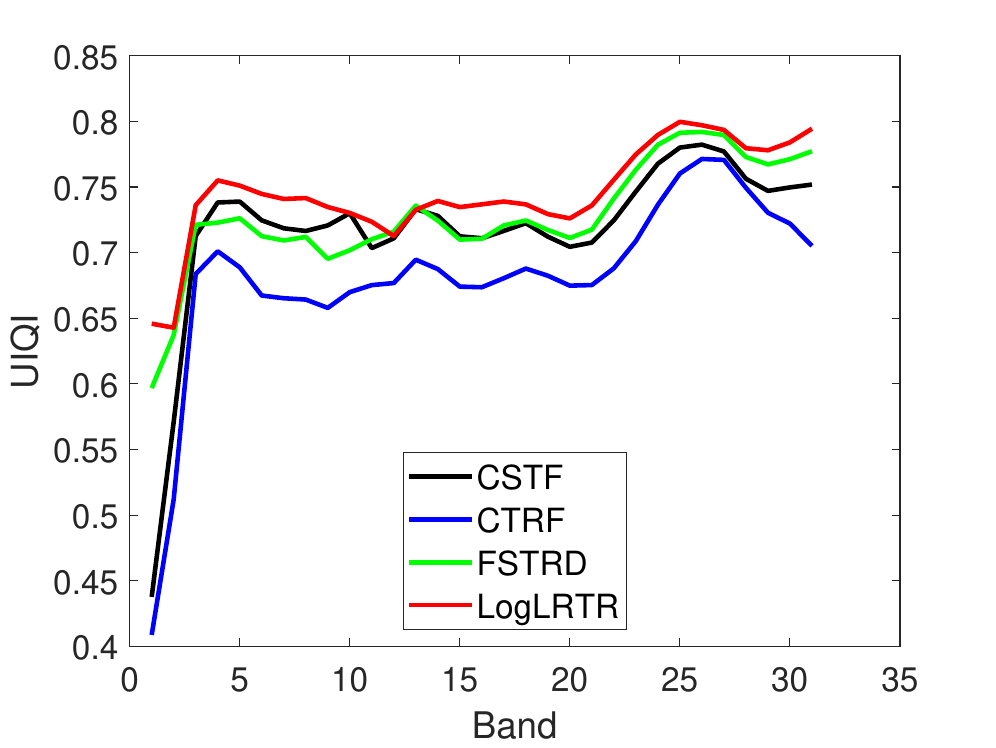}\vspace{1pt}
    \end{minipage}
}
\subfigure[Superballs]{
    \begin{minipage}[b]{0.225\linewidth}
    \includegraphics[width=4.25cm]{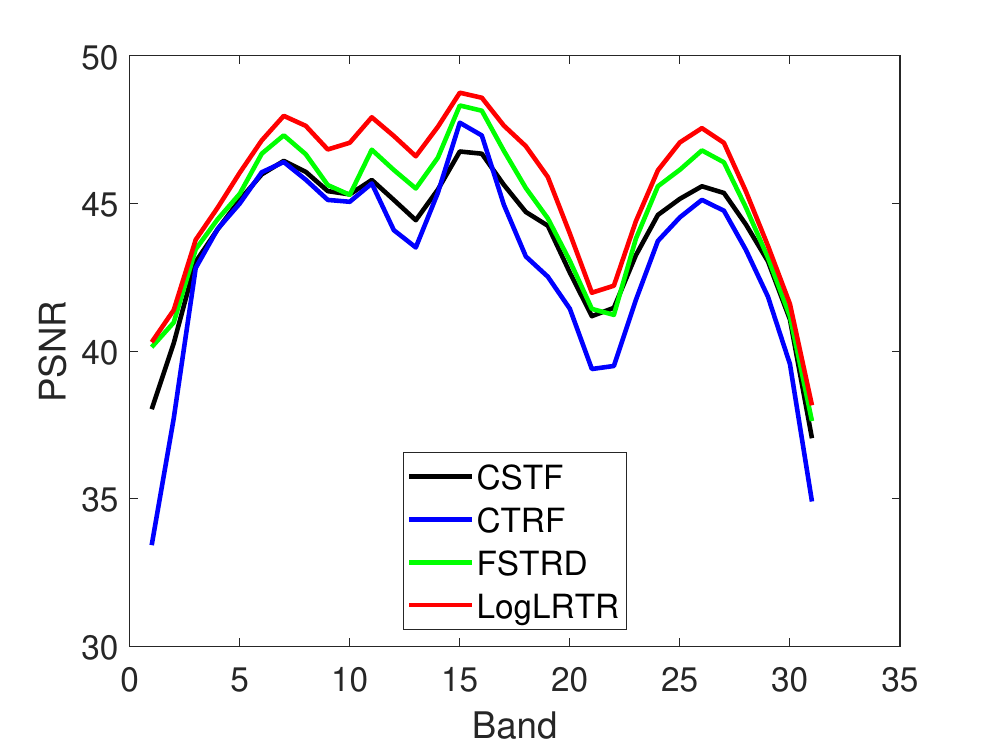}\vspace{1pt}
    \includegraphics[width=4.25cm]{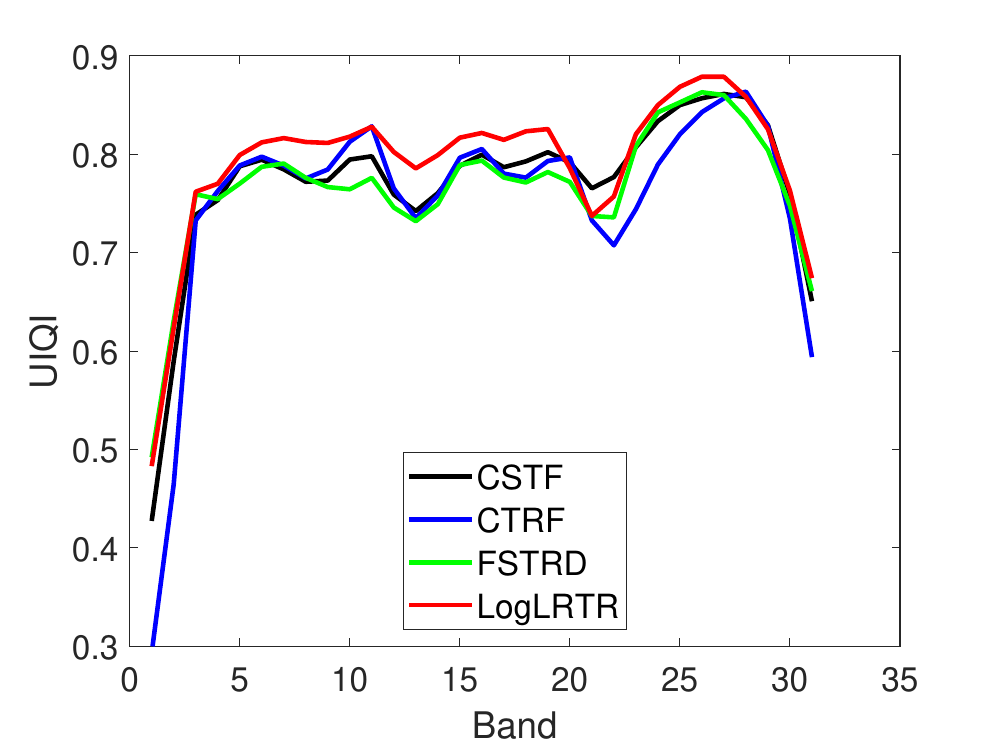}\vspace{1pt}
    \end{minipage}
}
\subfigure[Peppers]{
    \begin{minipage}[b]{0.225\linewidth}
    \includegraphics[width=4.25cm]{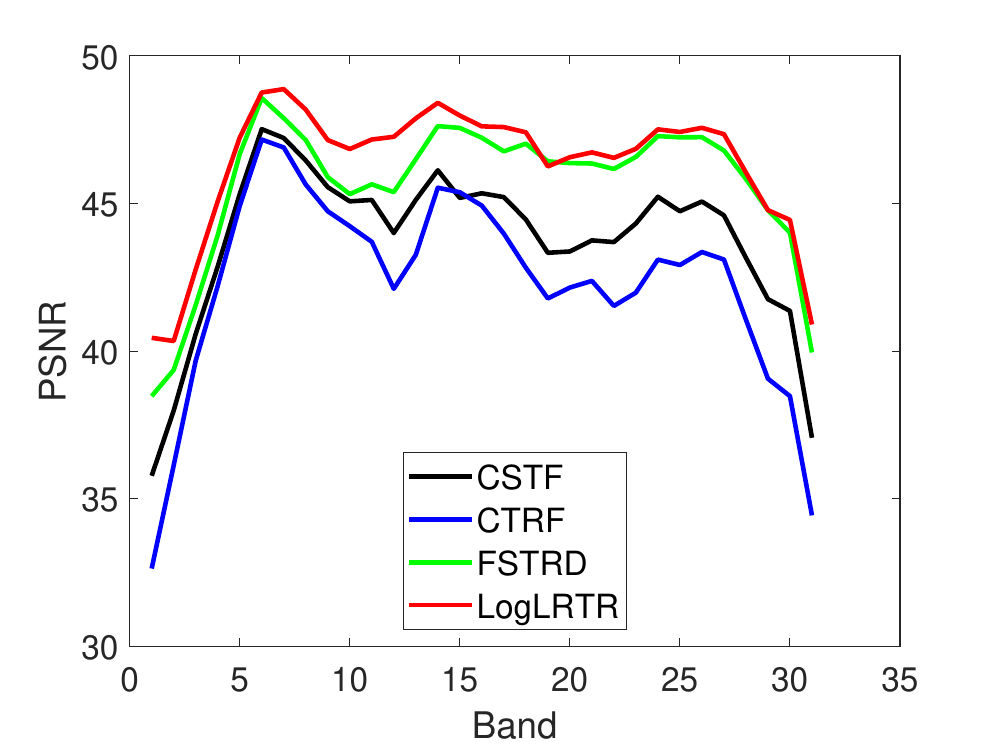}\vspace{1pt}
    \includegraphics[width=4.25cm]{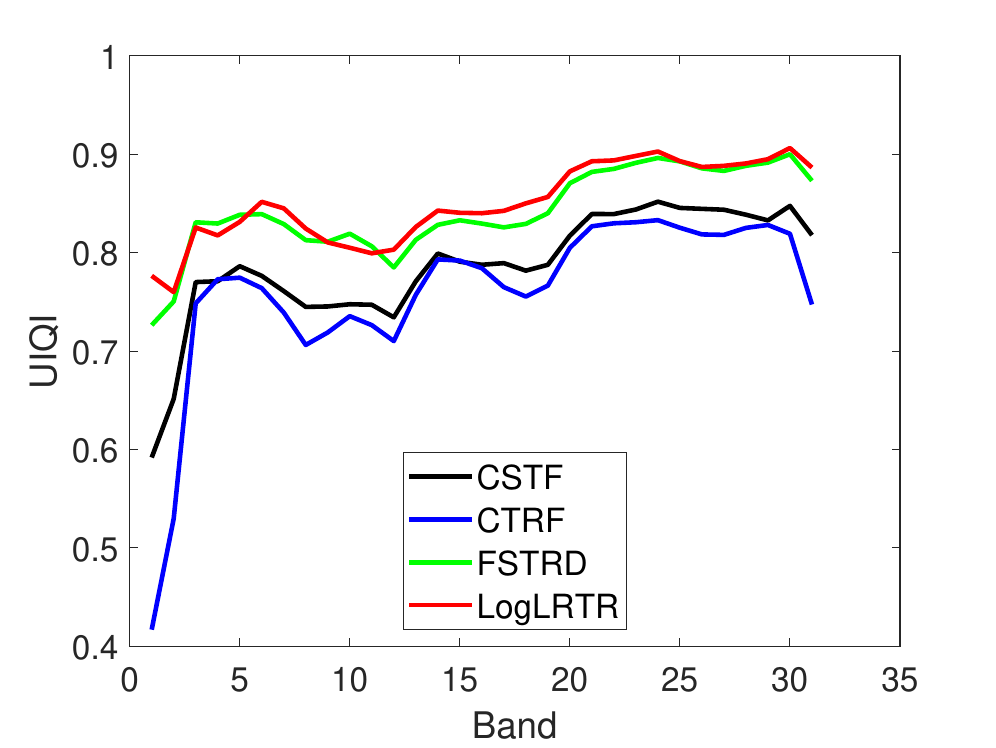}\vspace{1pt}
    \end{minipage}
}\vspace{-0.5em}
\caption{The plots of PSNR (top) and UIQI (bottom) values versus the spectral band for four test images.}\vspace{-1em}
\label{fig2}
\end{figure}

In addition to the numerical analysis, we depict the PSNR and UIQI versus band curves for the various methods in Fig. \ref{fig2}. This graphical representation offers compelling evidence that LogLRTR attains higher PSNR and UIQI values across almost all spectral bands when compared to other approaches. Particularly, the ``Pavia University" and ``Flowers" images display substantial enhancements. In summation, these findings prove the superior performance of our proposed method.

To further validate the ability of LogLRTR on maintaining spatial structures, we show the fused and error images acquired by different methods for each test image.
The error image effectively captures the disparities between the fused image and the ground truth reference image.
In addition, to facilitate visual comparison, we mark and enlarge meaningful regions from each fused the false color image.

\begin{figure}
\centering
\subfigure[CSTF]{
    \begin{minipage}[b]{0.17\linewidth}
    \includegraphics[width=3cm]{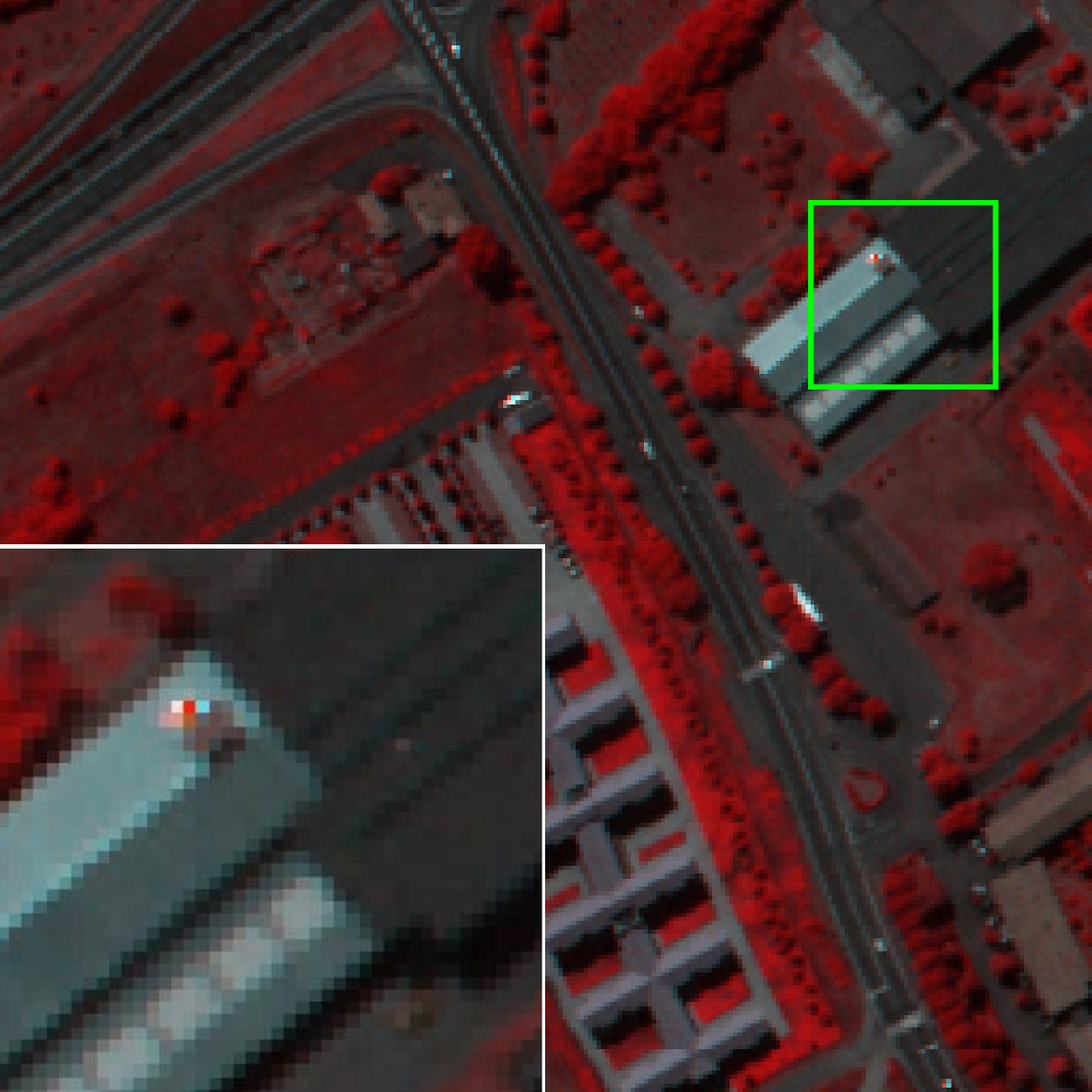}\vspace{1pt}
    \includegraphics[width=3cm]{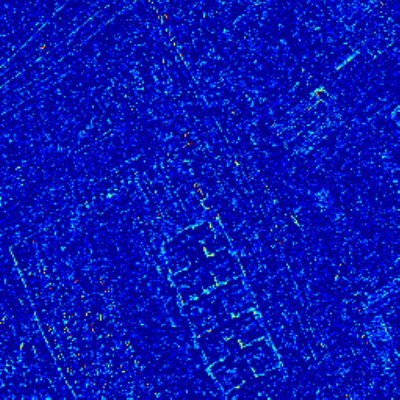}\vspace{1pt}
    \end{minipage}
}
\subfigure[CTRF]{
    \begin{minipage}[b]{0.17\linewidth}
    \includegraphics[width=3cm]{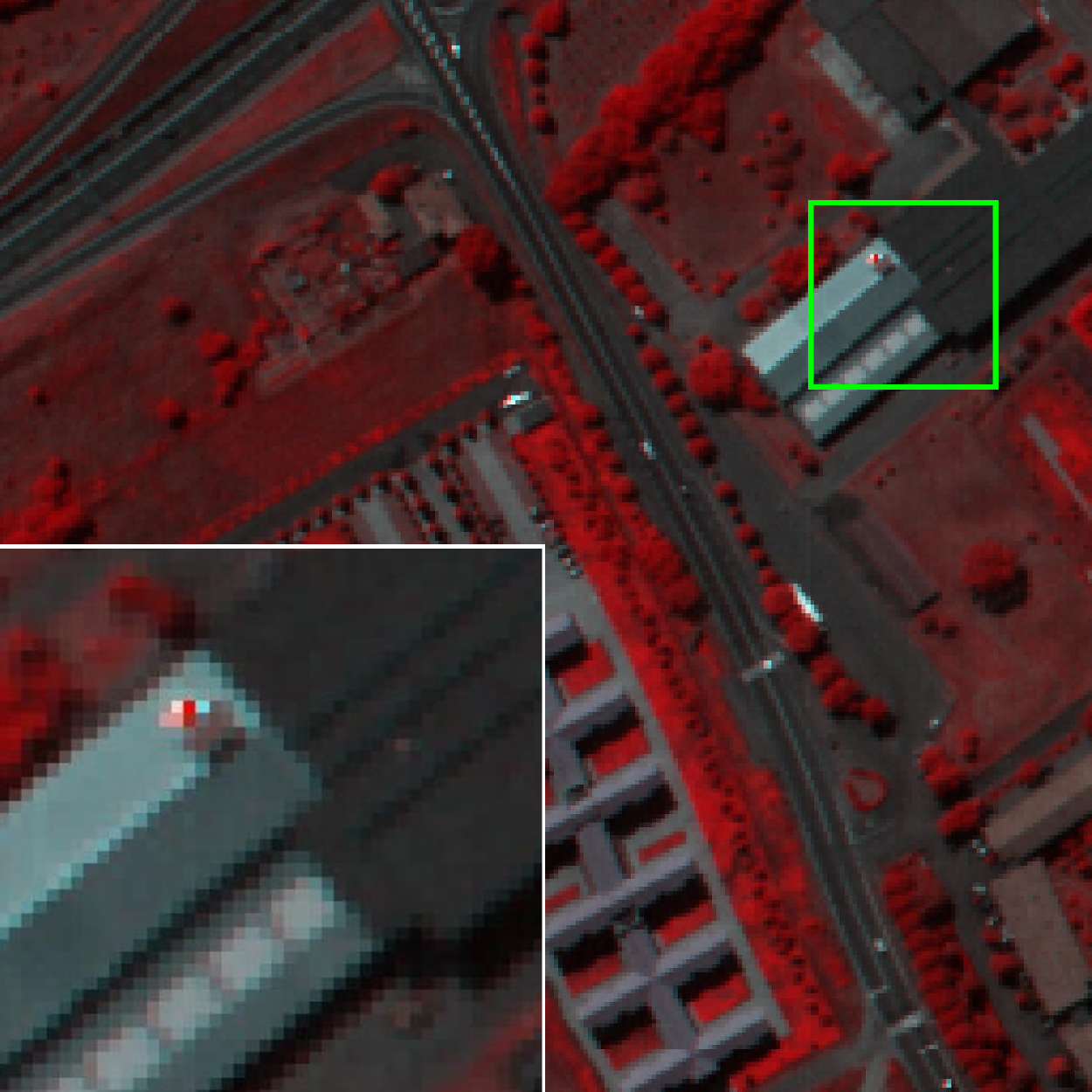}\vspace{1pt}
    \includegraphics[width=3cm]{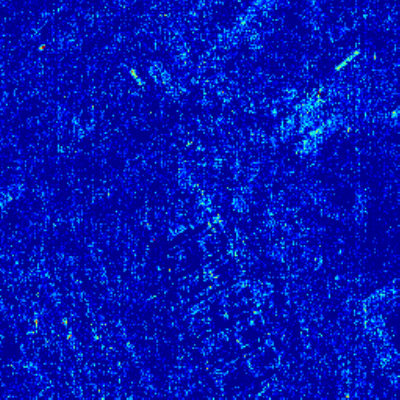}\vspace{1pt}
    \end{minipage}
}
\subfigure[FSTRD]{
    \begin{minipage}[b]{0.17\linewidth}
    \includegraphics[width=3cm]{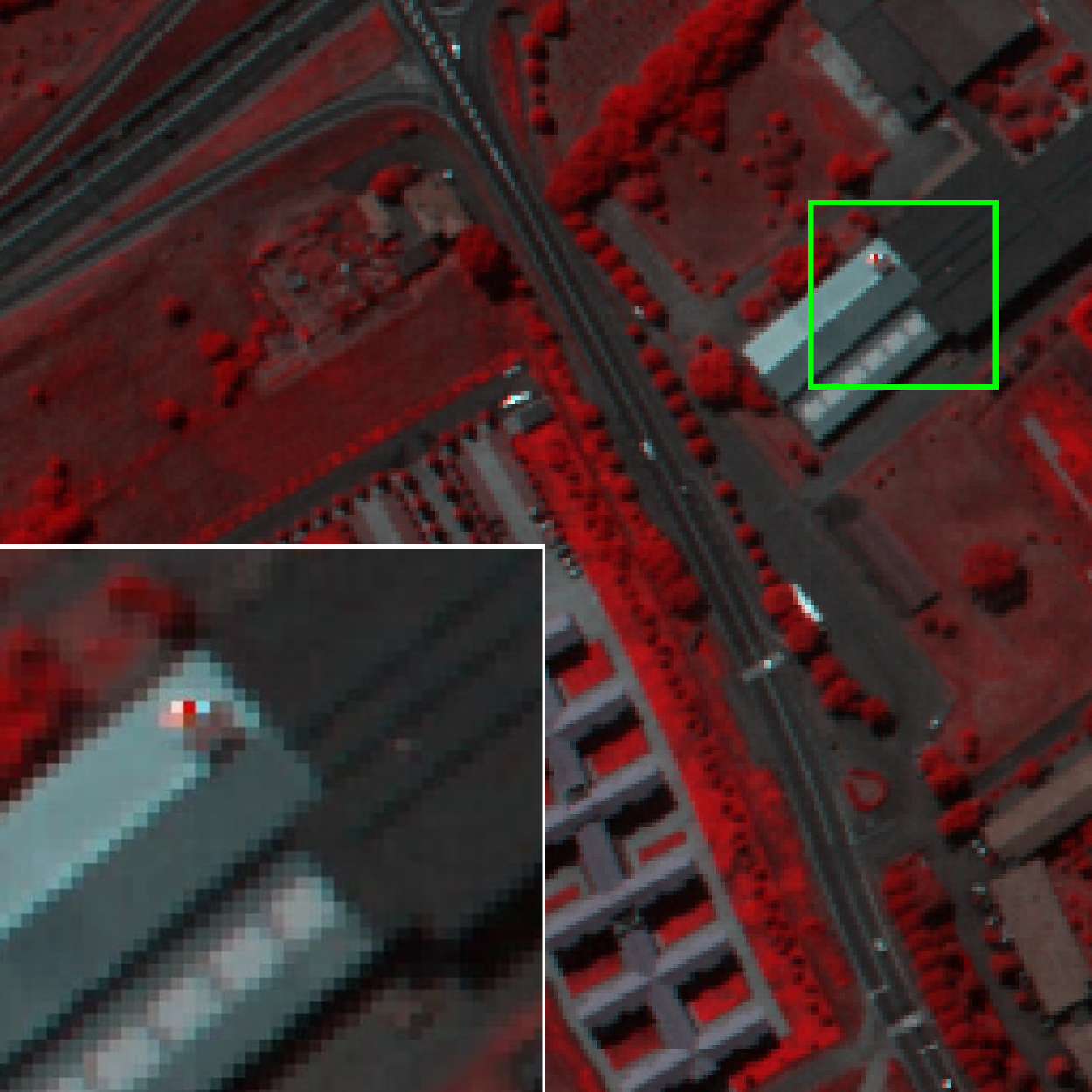}\vspace{1pt}
    \includegraphics[width=3cm]{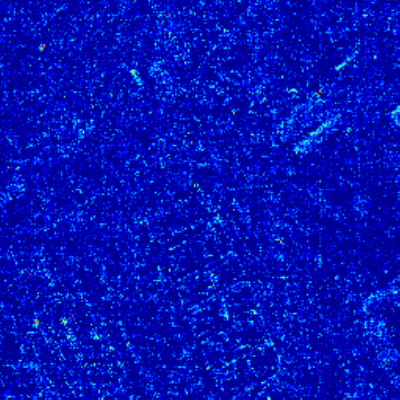}\vspace{1pt}
    \end{minipage}
}
\subfigure[LogLRTR]{
    \begin{minipage}[b]{0.17\linewidth}
    \includegraphics[width=3cm]{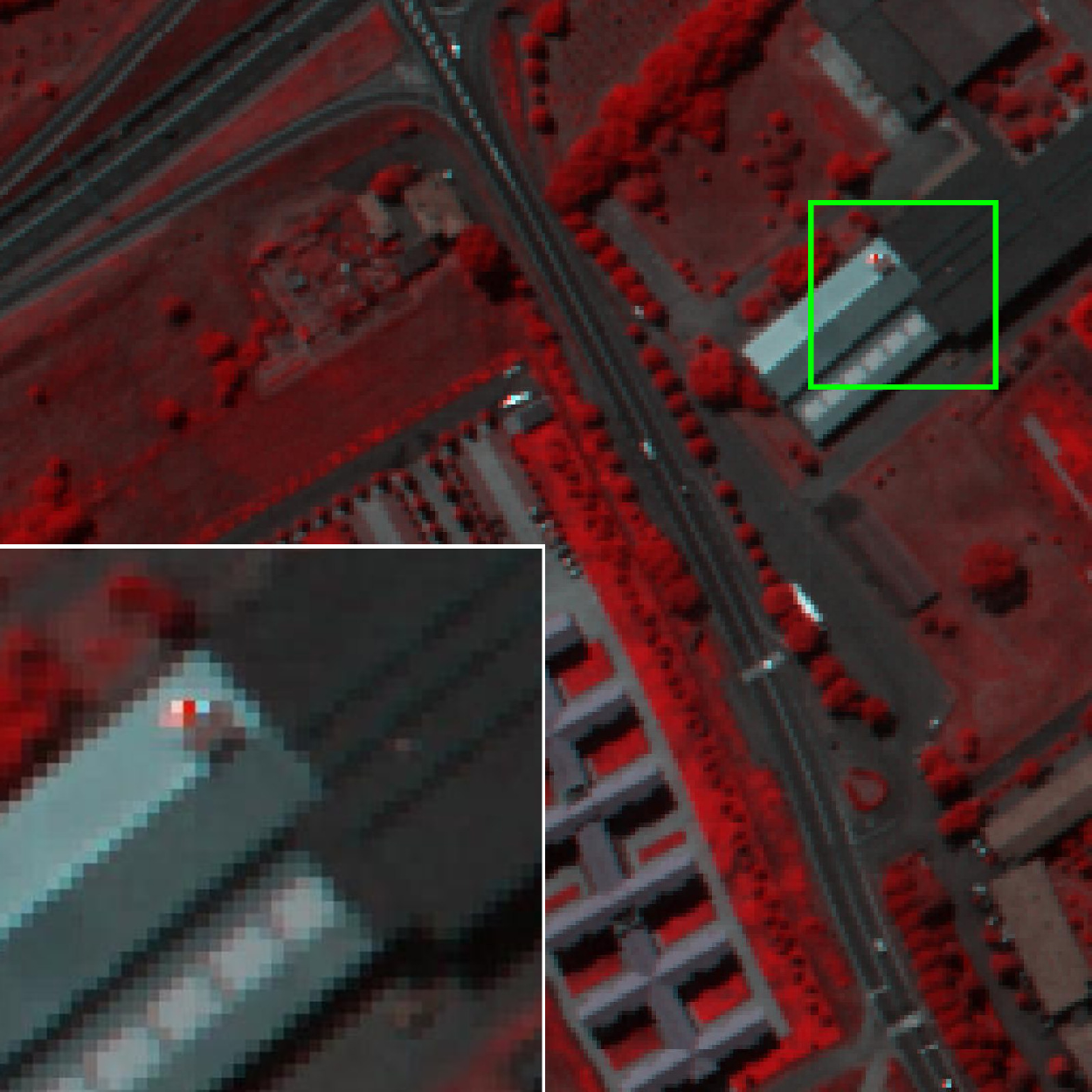}\vspace{1pt}
    \includegraphics[width=3cm]{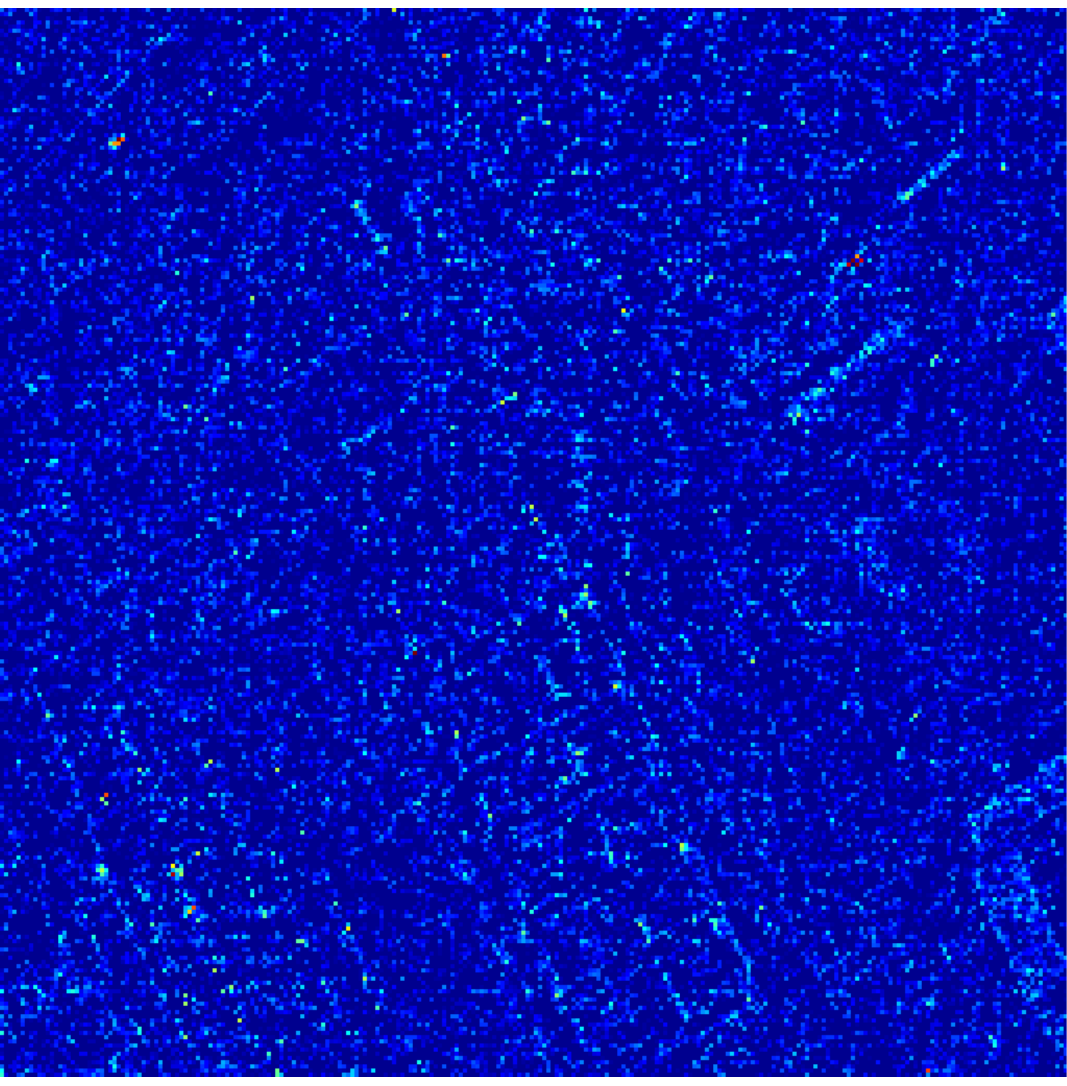}\vspace{1pt}
    \end{minipage}
}
\subfigure[Ground truth]{
    \begin{minipage}[b]{0.17\linewidth}
    \includegraphics[width=3cm]{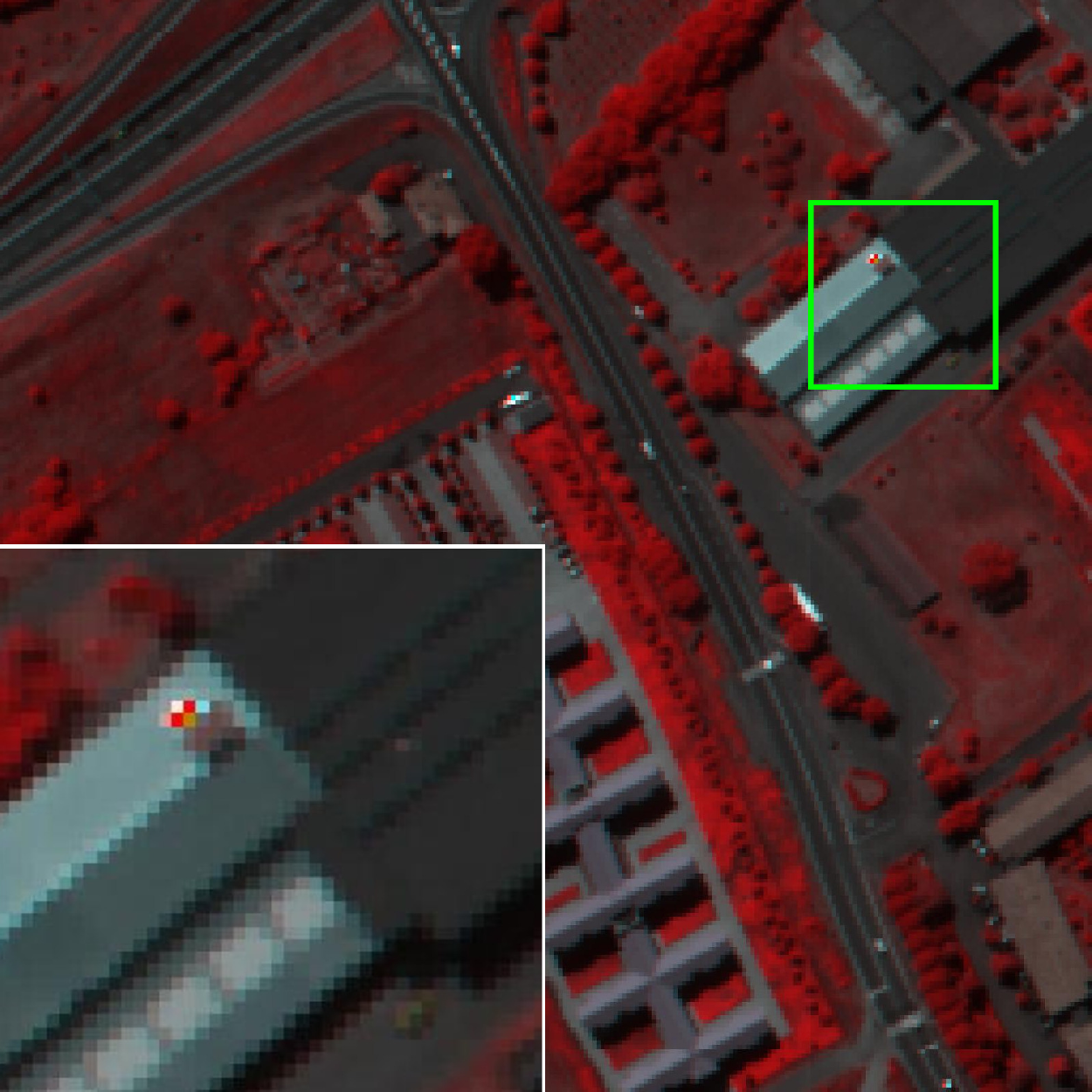}\vspace{1pt}
    \includegraphics[width=3cm]{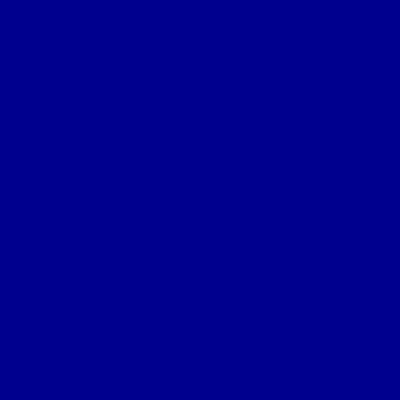}\vspace{1pt}
    \end{minipage}
}\\
    \includegraphics[width=12cm]{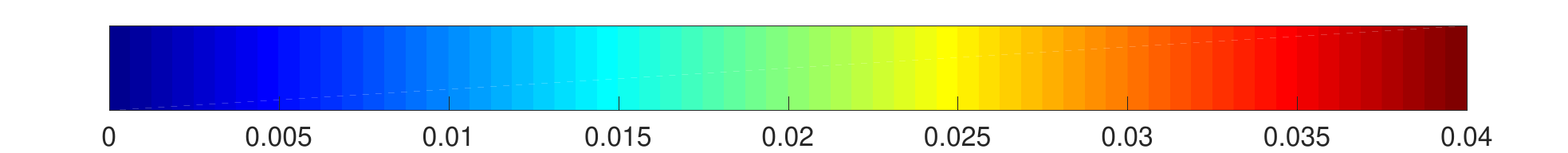}\vspace{1pt}
\caption{The first row  presents the false color images formed by bands 83, 38 and 34 for ``Pavia University", and the second  row  displays the corresponding error images.}
\label{fig3}
\end{figure}

In Fig. \ref{fig3}, we show the false color images of ``Pavia University" generated using bands 83, 38 and 34 obtained through different fusion methods, as well as the corresponding error images.
In the magnified sub-region of the false color image, the CSTF blurs the image details, while the roofs of the other methods are closer to the original image, proving once again the drawback of Tucker decomposition that does not sufficiently exploit the intrinsic information of the HSI.
From the second row of Fig. \ref{fig3}, it can be more clearly observed that LogLRTR achieves a smaller error, indicating that our proposed method exhibits an advantage in preserving most of the spatial details.

\begin{figure}
\centering
\subfigure[CSTF]{
    \begin{minipage}[b]{0.17\linewidth}
    \includegraphics[width=3cm]{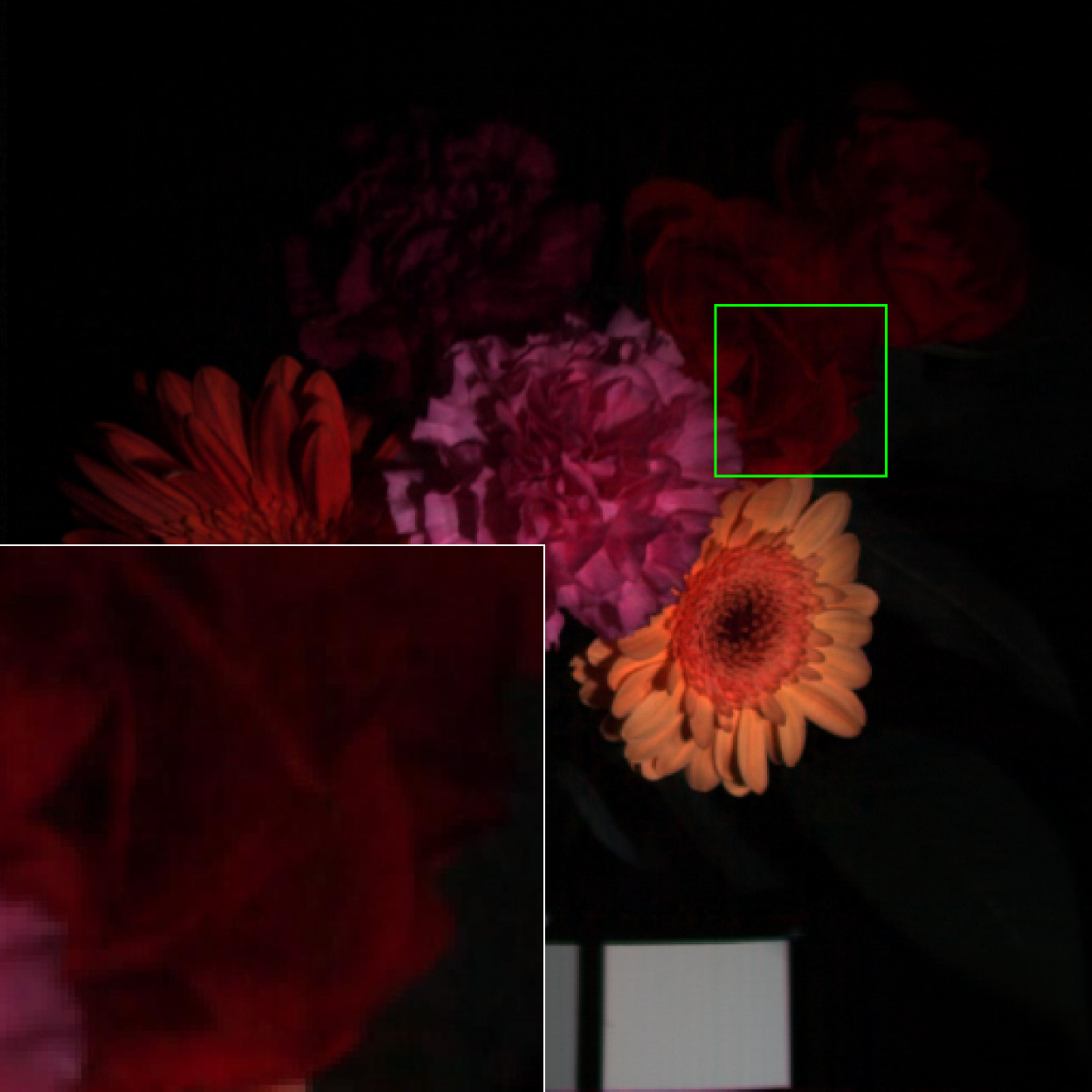}\vspace{1pt}
    \includegraphics[width=3cm]{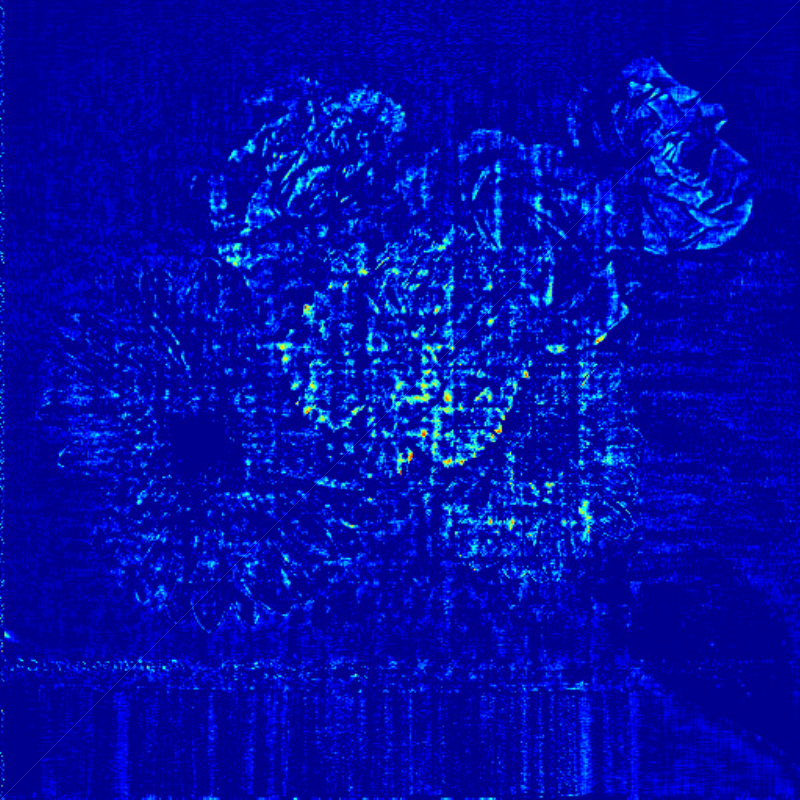}\vspace{1pt}
    \end{minipage}
}
\subfigure[CTRF]{
    \begin{minipage}[b]{0.17\linewidth}
    \includegraphics[width=3cm]{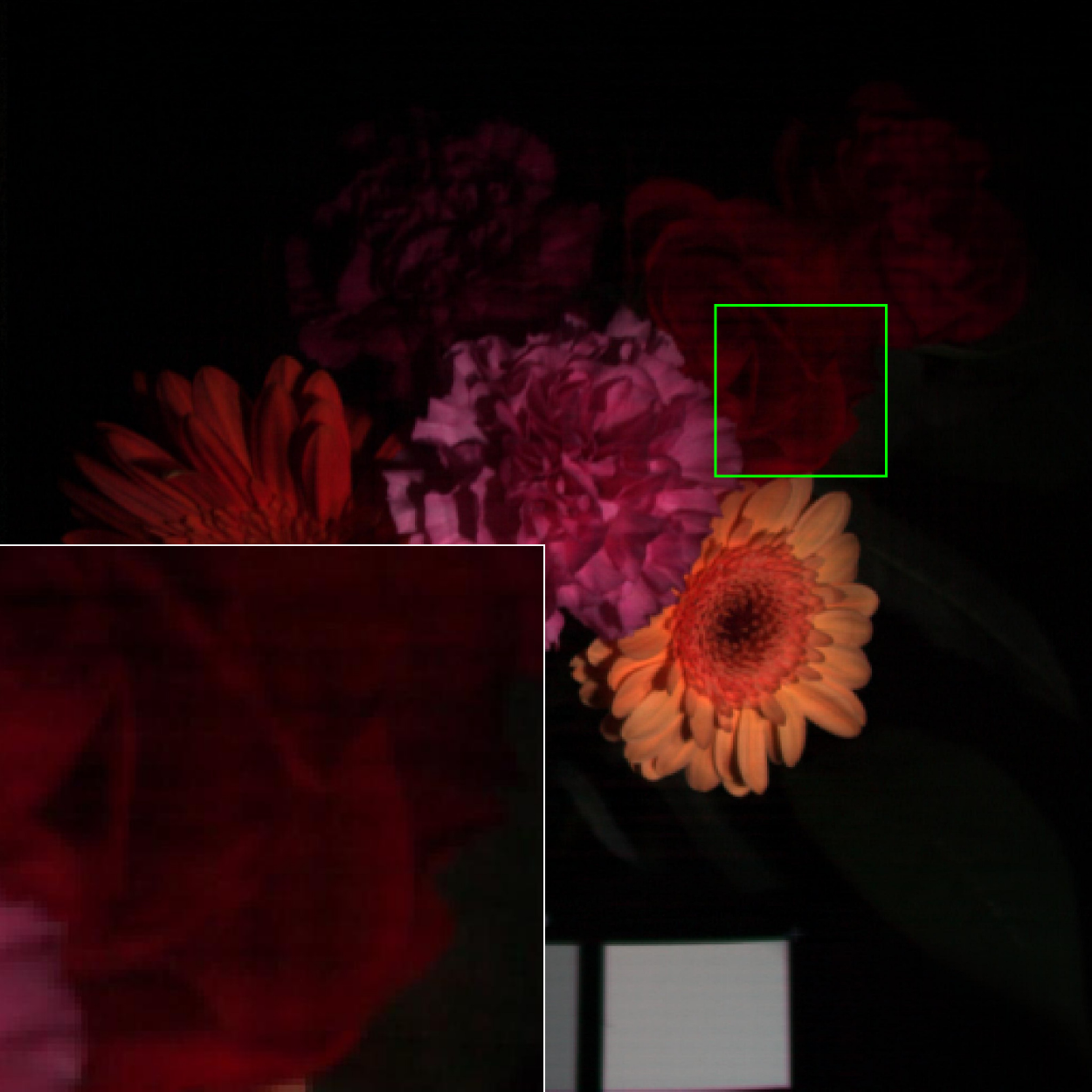}\vspace{1pt}
    \includegraphics[width=3cm]{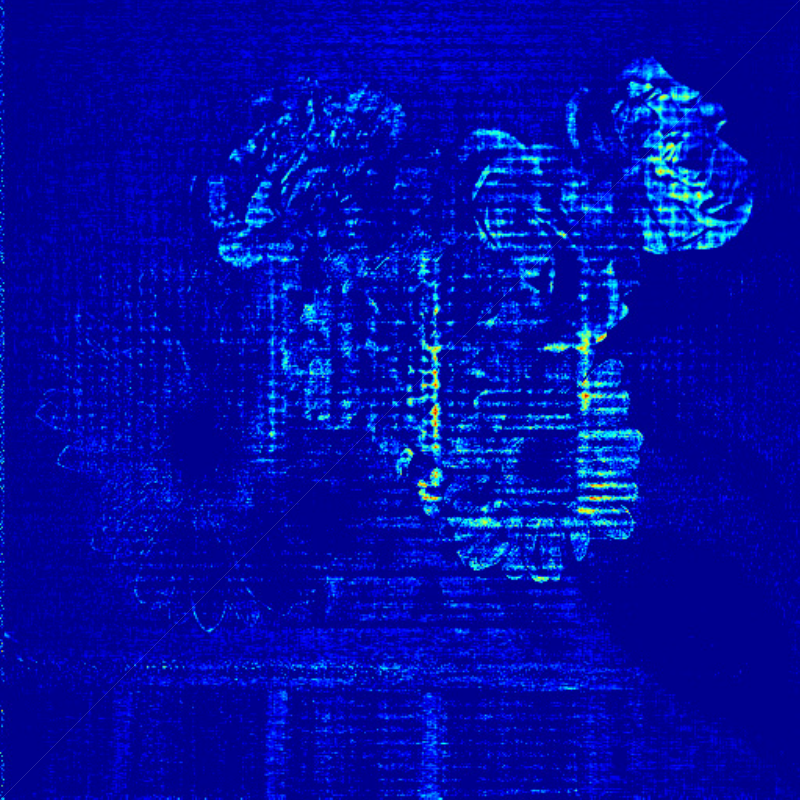}\vspace{1pt}
    \end{minipage}
}
\subfigure[FSTRD]{
    \begin{minipage}[b]{0.17\linewidth}
    \includegraphics[width=3cm]{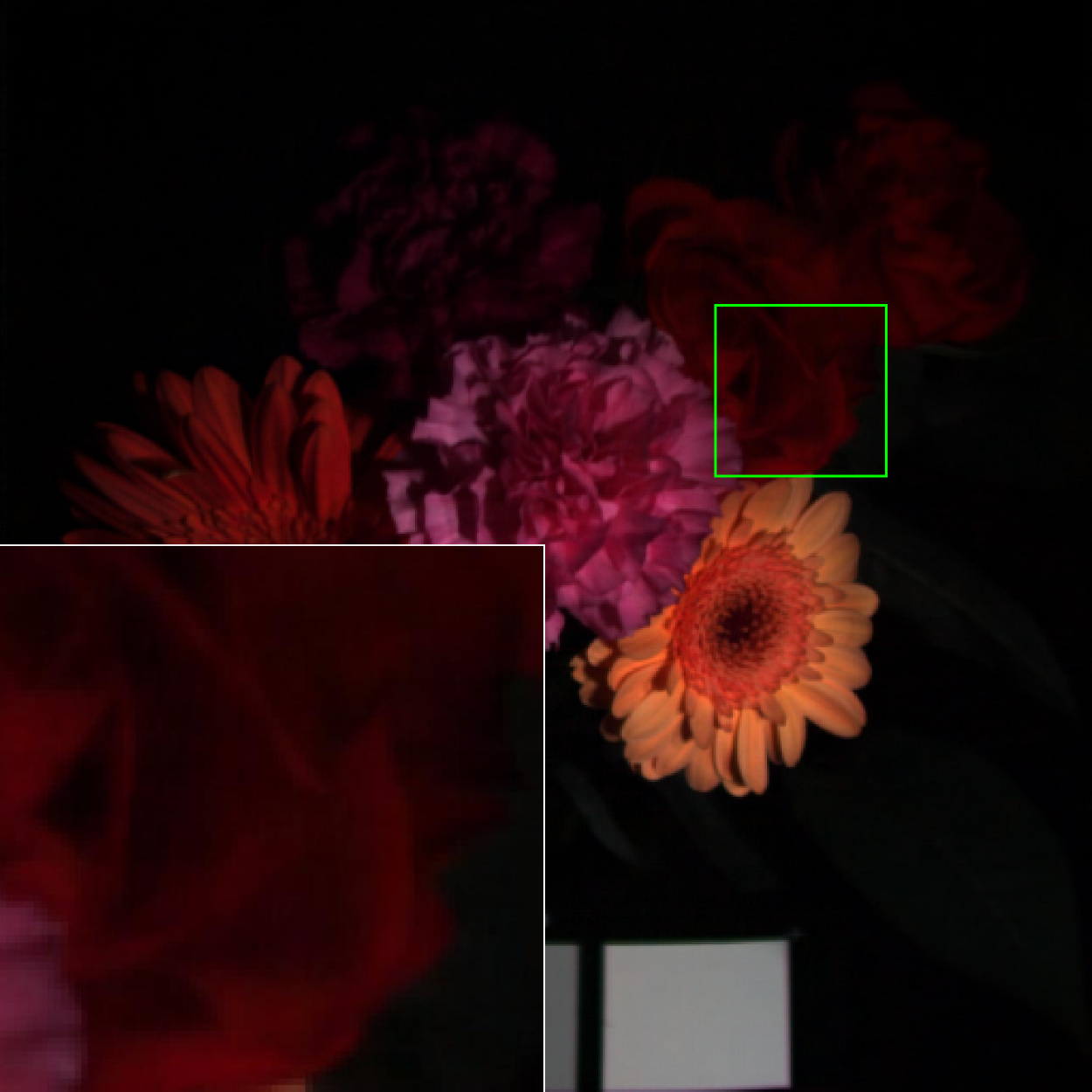}\vspace{1pt}
    \includegraphics[width=3cm]{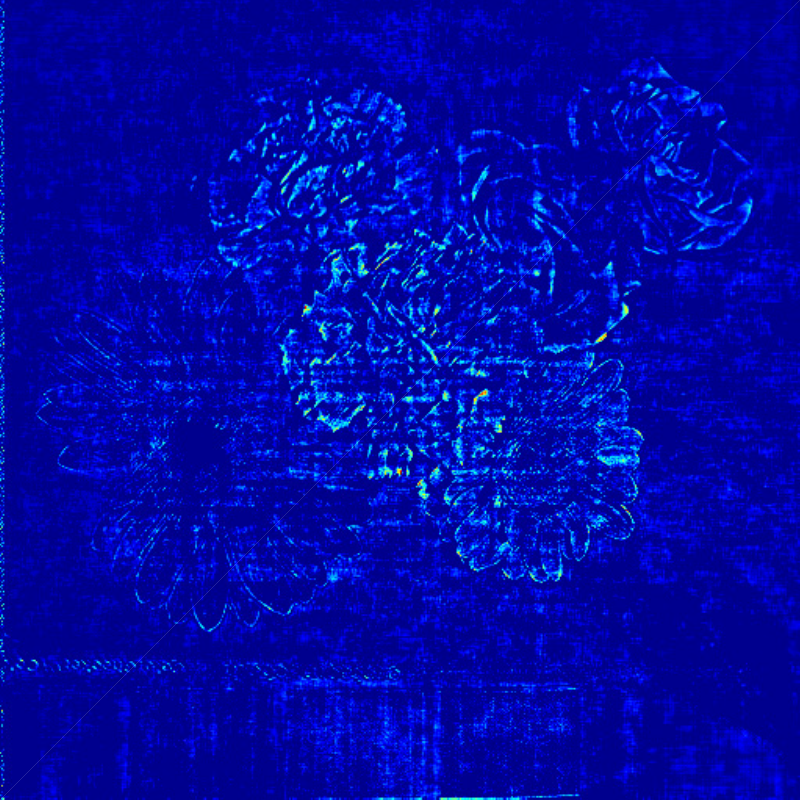}\vspace{1pt}
    \end{minipage}
}
\subfigure[LogLRTR]{
    \begin{minipage}[b]{0.17\linewidth}
    \includegraphics[width=3cm]{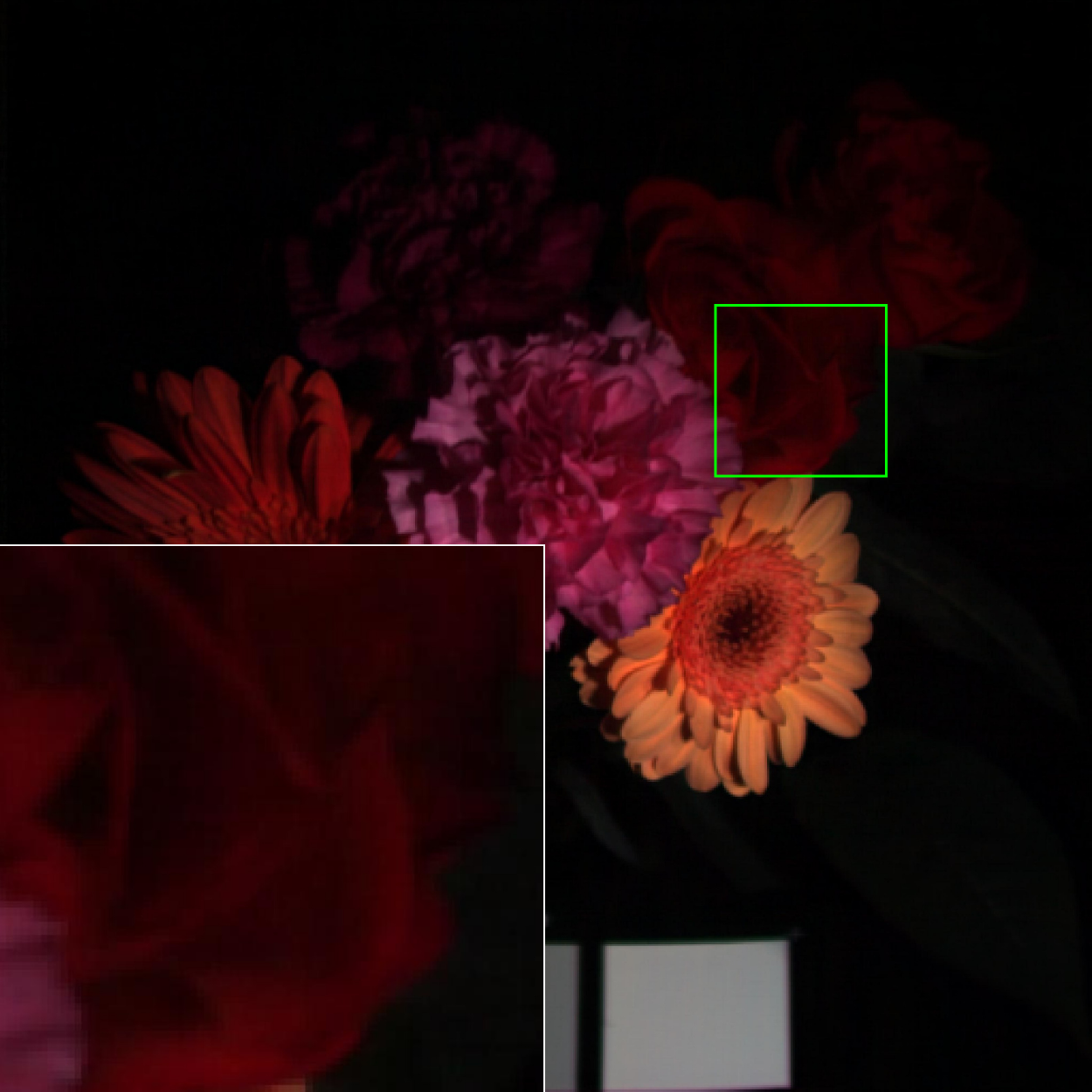}\vspace{1pt}
    \includegraphics[width=3cm]{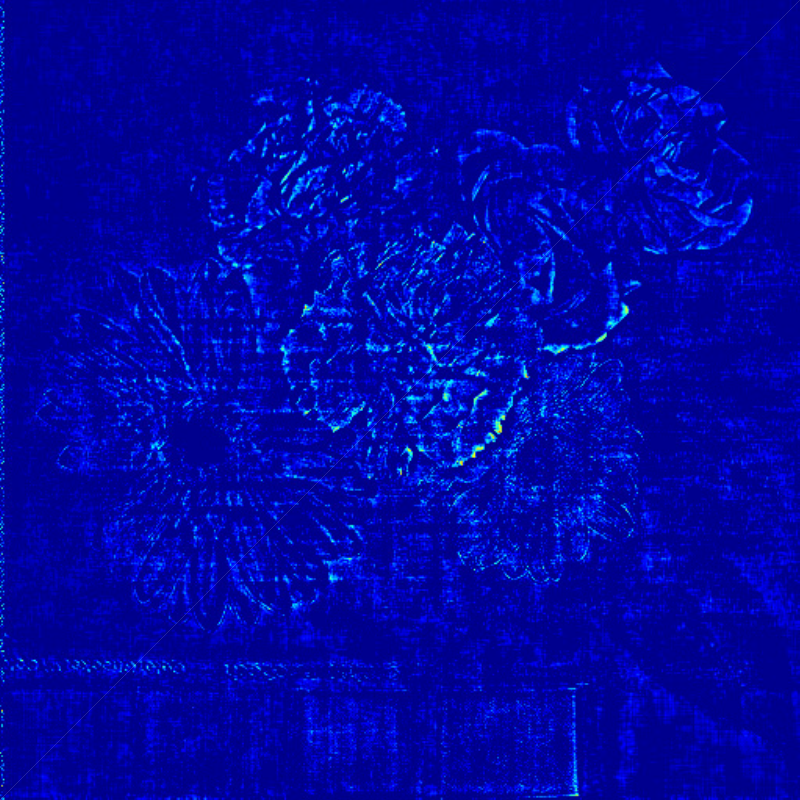}\vspace{1pt}
    \end{minipage}
}
\subfigure[Ground truth]{
    \begin{minipage}[b]{0.17\linewidth}
    \includegraphics[width=3cm]{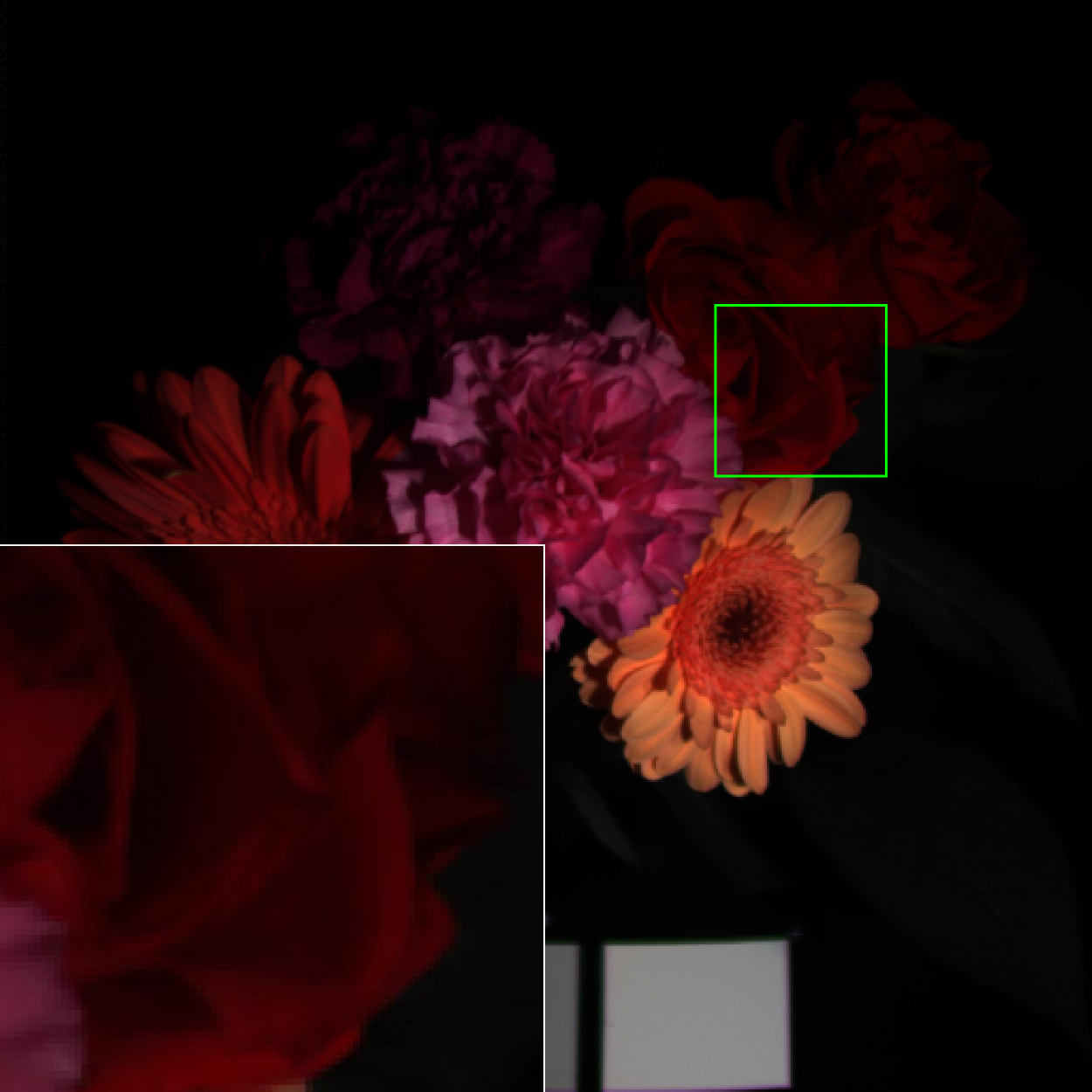}\vspace{1pt}
    \includegraphics[width=3cm]{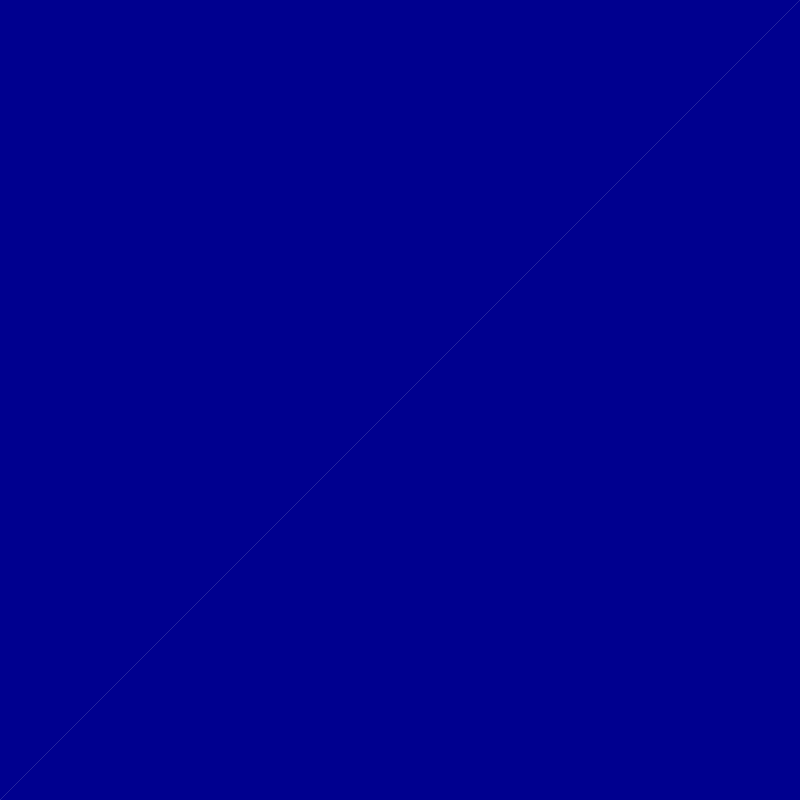}\vspace{1pt}
    \end{minipage}
}\\
    \includegraphics[width=12cm]{fig/colorbar0.04.pdf}\vspace{1pt}
\caption{The first row gives the false color images formed by bands 28, 12 and 5 for ``Flowers", and the second row shows the corresponding error images.}
\label{fig4}
\end{figure}

Additionally, we assess the reconstructed images of ``Flowers", ``Superballs" and ``Peppers" in the CAVE dataset and the visualisation of their error maps to verify the superiority of LogLRTR.
Fig. \ref{fig4} provides the false color images formed by bands 28, 12 and 5 in the ``Flowers" fusion images. Specifically, the enlarged regions show that some important details are lost in the results of  CSTF and CTRF.
The proposed method and FSTRD perform better in reconstructing high-resolution details. In other words, more detailed edge structures are preserved through these two methods.
In addition, we clearly observe from the corresponding error images that LogLRTR produces less error than FSTRD.

\begin{figure}
\centering
\subfigure[CSTF]{
    \begin{minipage}[b]{0.17\linewidth}
    \includegraphics[width=3cm]{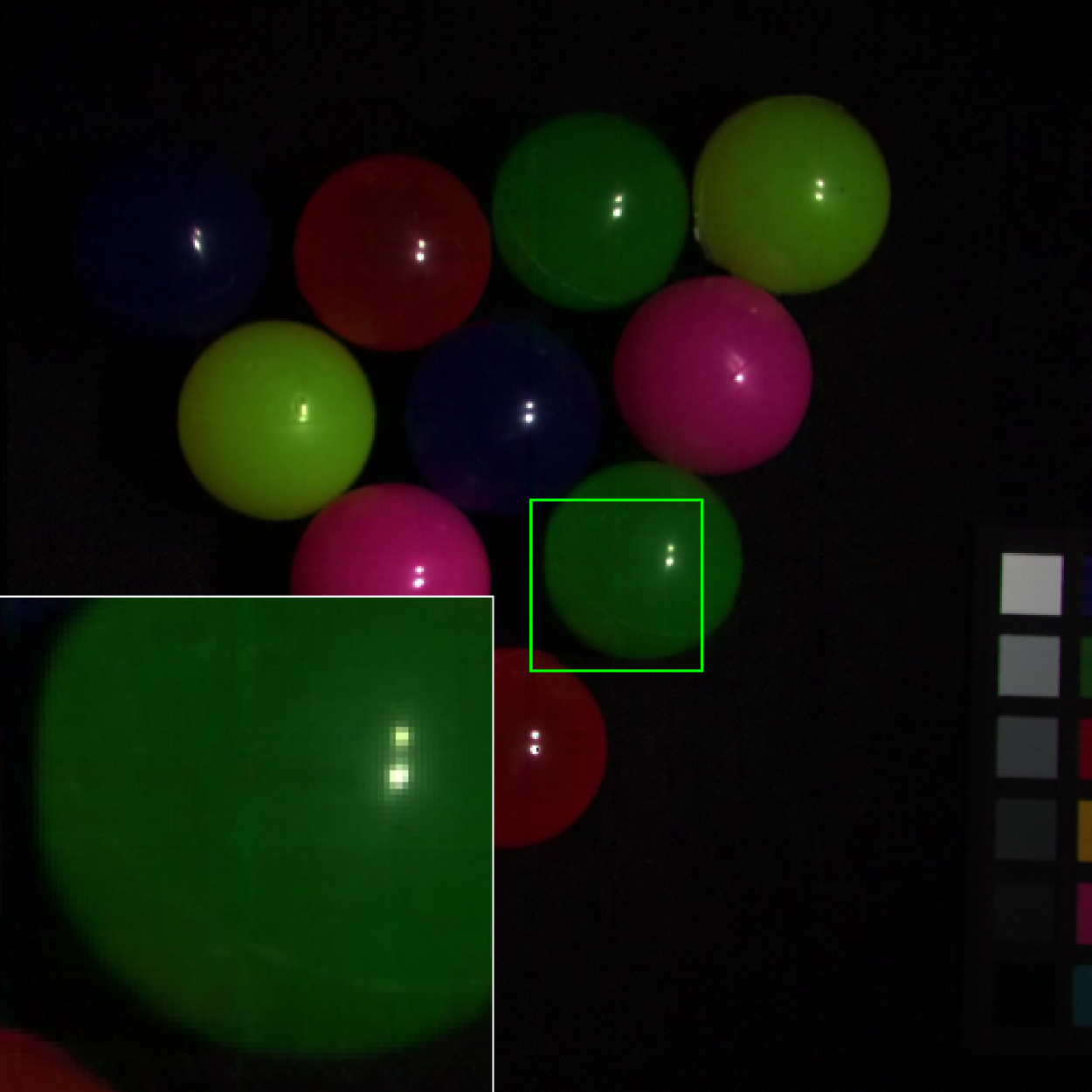}\vspace{1pt}
    \includegraphics[width=3cm]{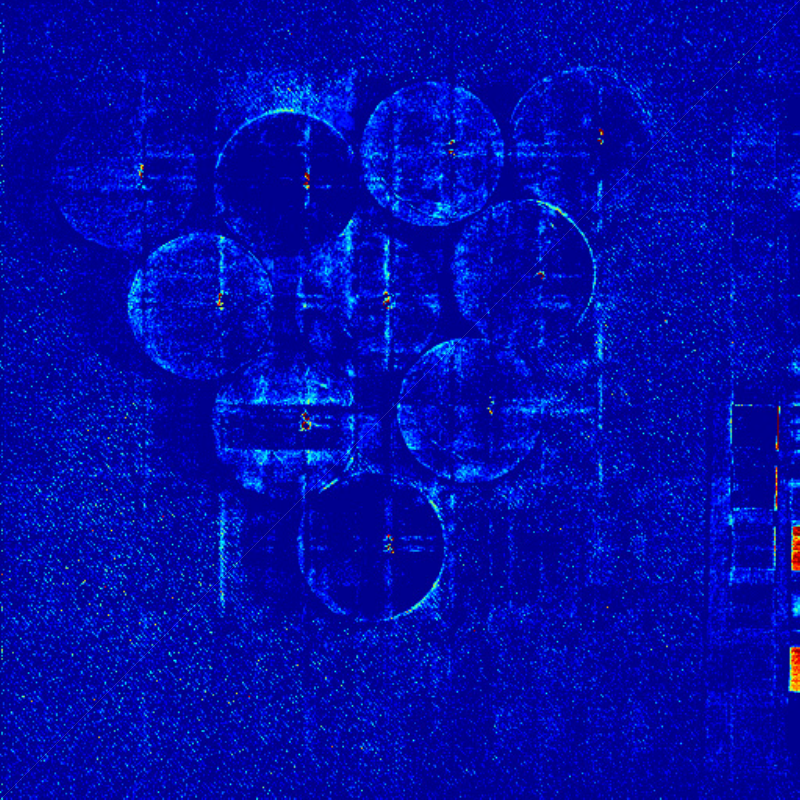}\vspace{1pt}
    \end{minipage}
}
\subfigure[CTRF]{
    \begin{minipage}[b]{0.17\linewidth}
    \includegraphics[width=3cm]{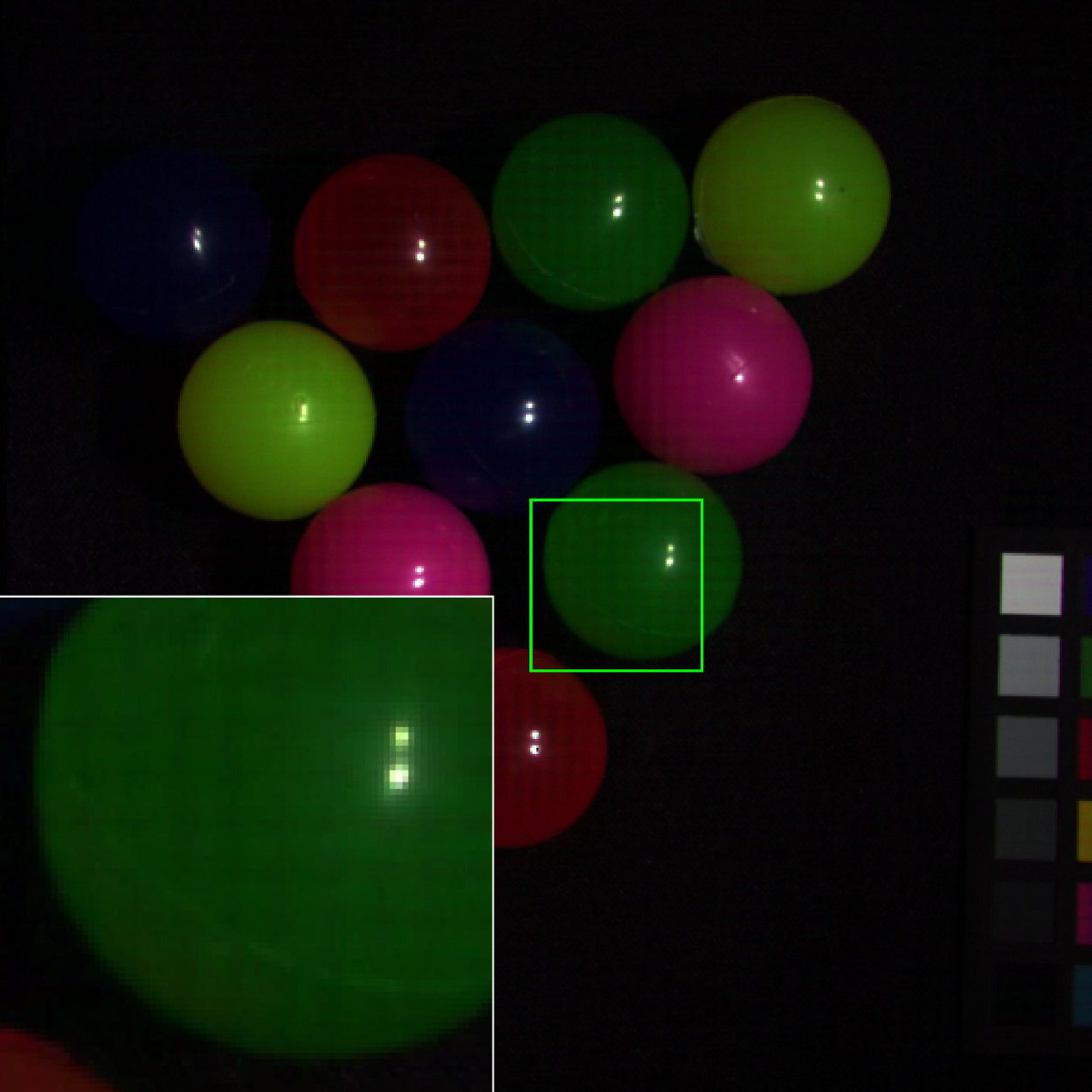}\vspace{1pt}
    \includegraphics[width=3cm]{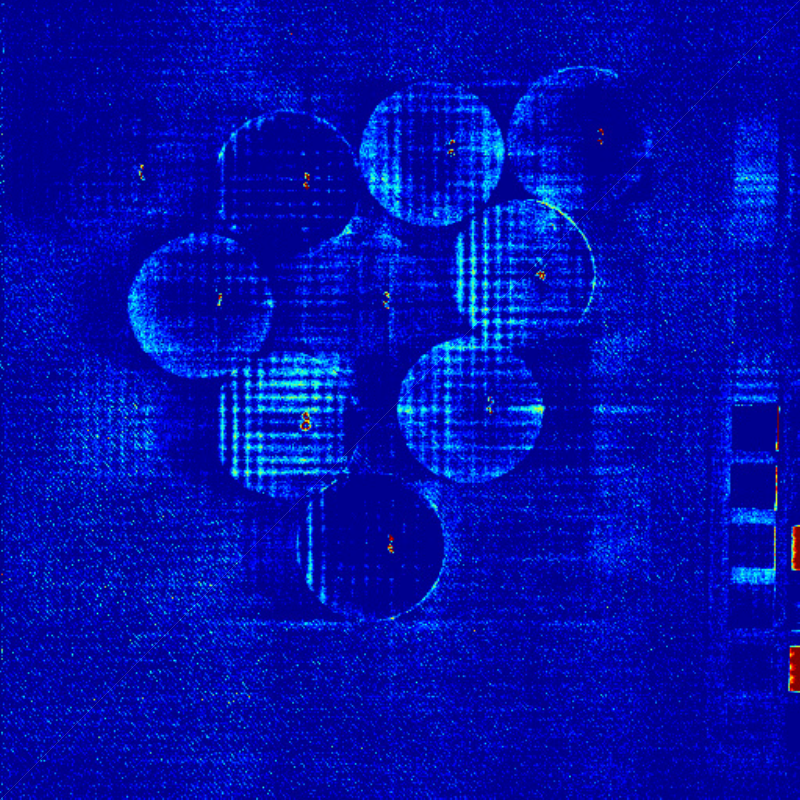}\vspace{1pt}
    \end{minipage}
}
\subfigure[FSTRD]{
    \begin{minipage}[b]{0.17\linewidth}
    \includegraphics[width=3cm]{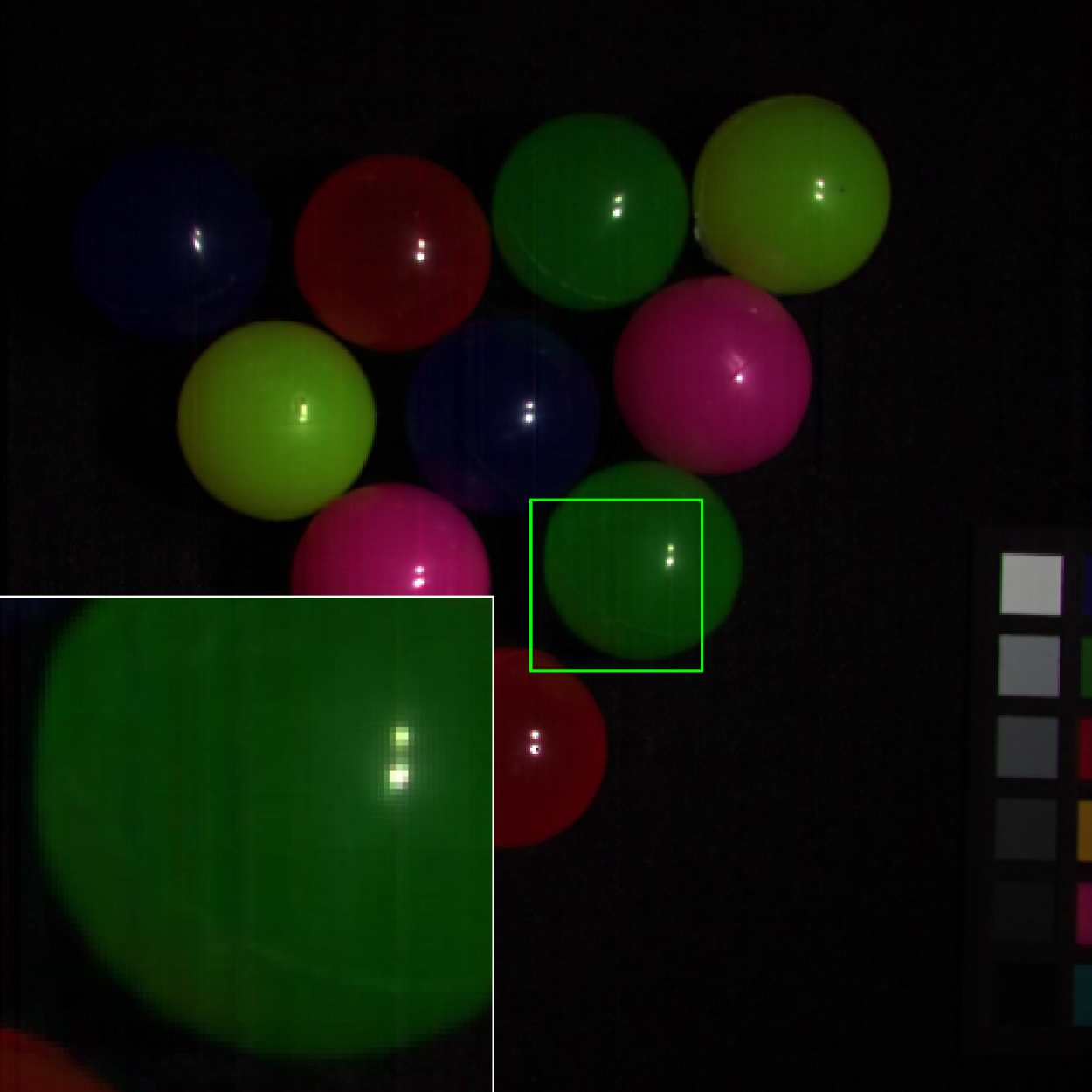}\vspace{1pt}
    \includegraphics[width=3cm]{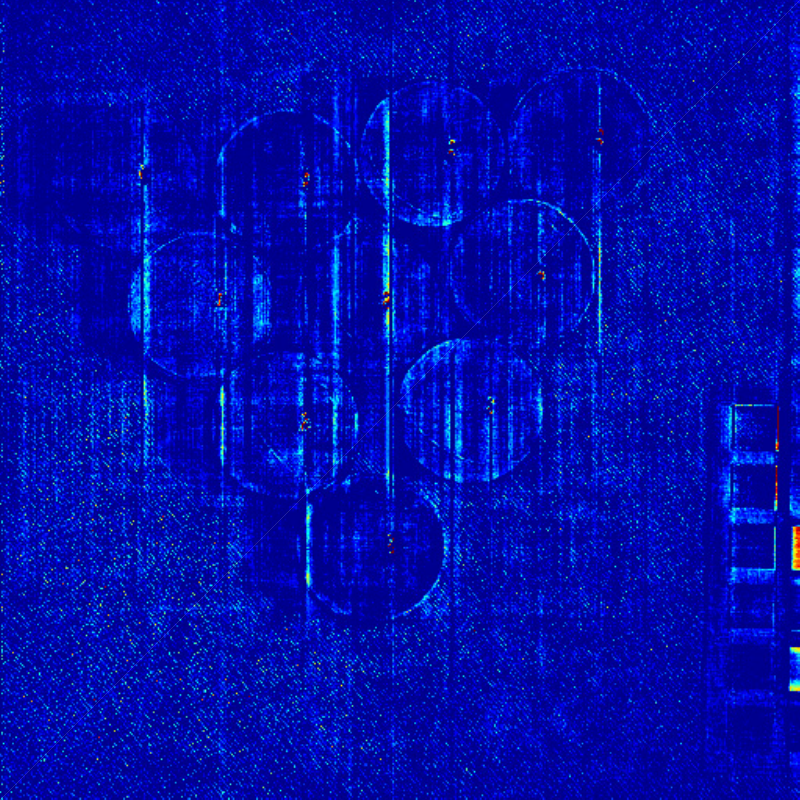}\vspace{1pt}
    \end{minipage}
}
\subfigure[LogLRTR]{
    \begin{minipage}[b]{0.17\linewidth}
    \includegraphics[width=3cm]{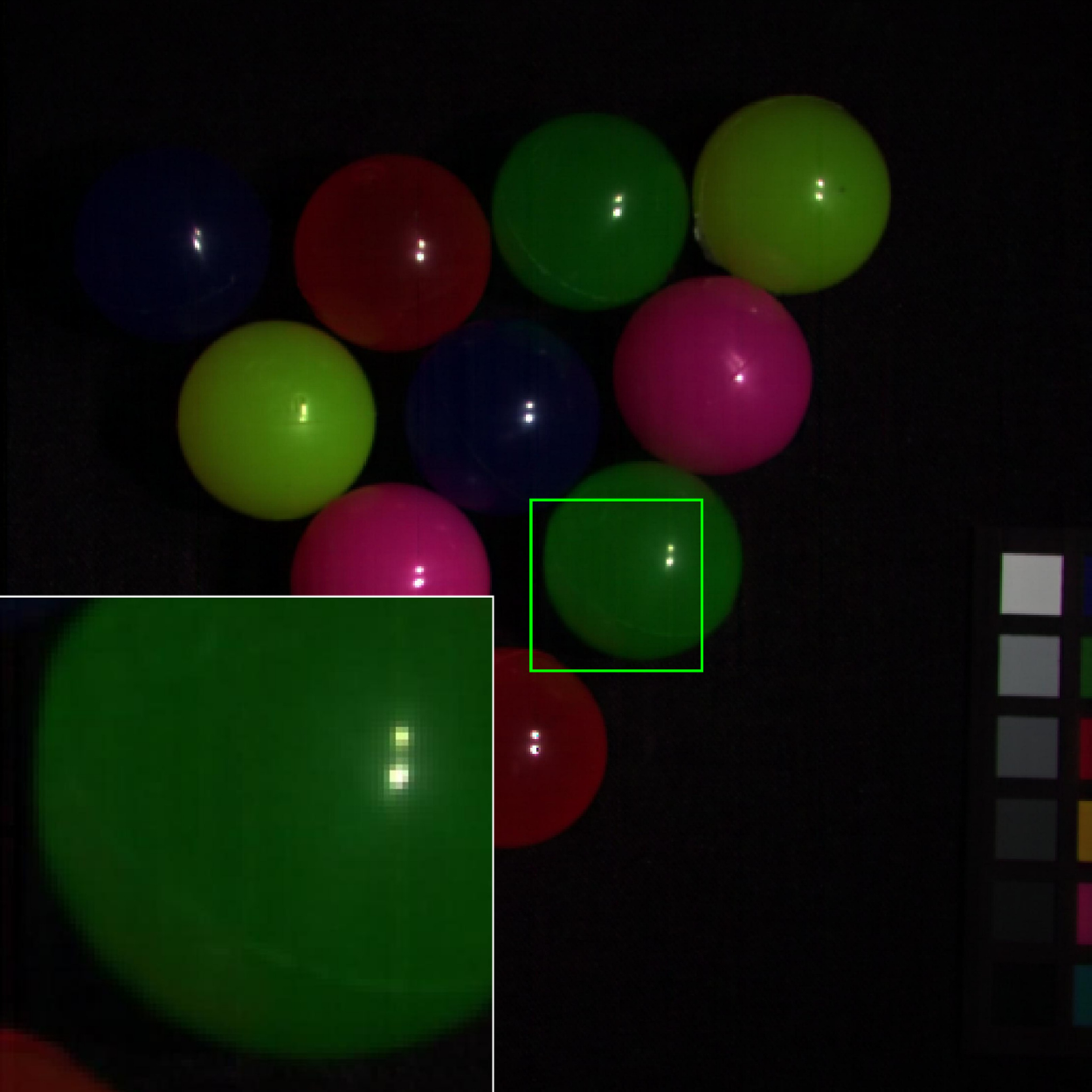}\vspace{1pt}
    \includegraphics[width=3cm]{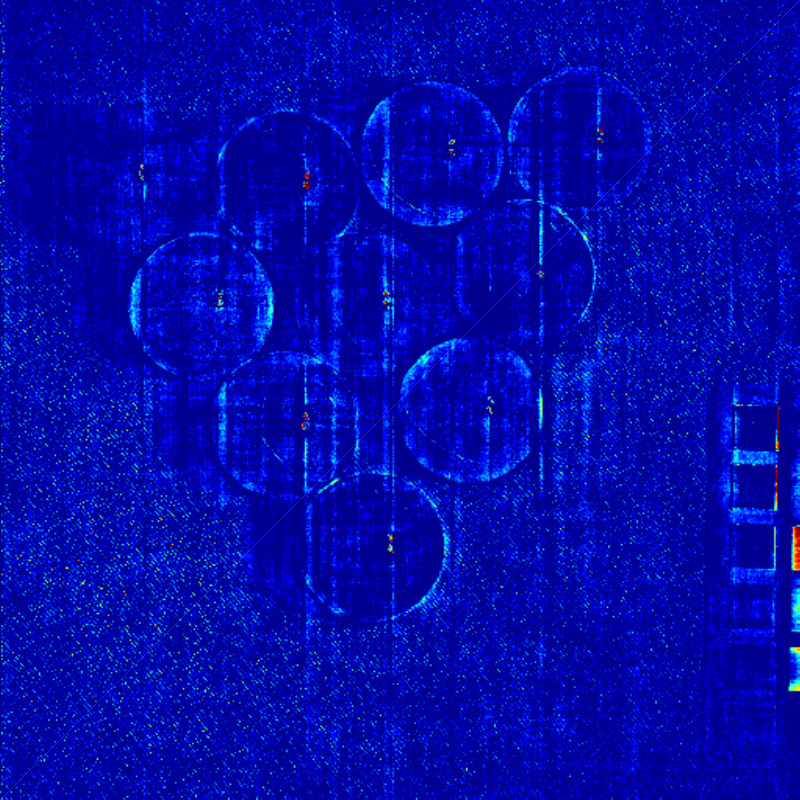}\vspace{1pt}
    \end{minipage}
}
\subfigure[Ground truth]{
    \begin{minipage}[b]{0.17\linewidth}
    \includegraphics[width=3cm]{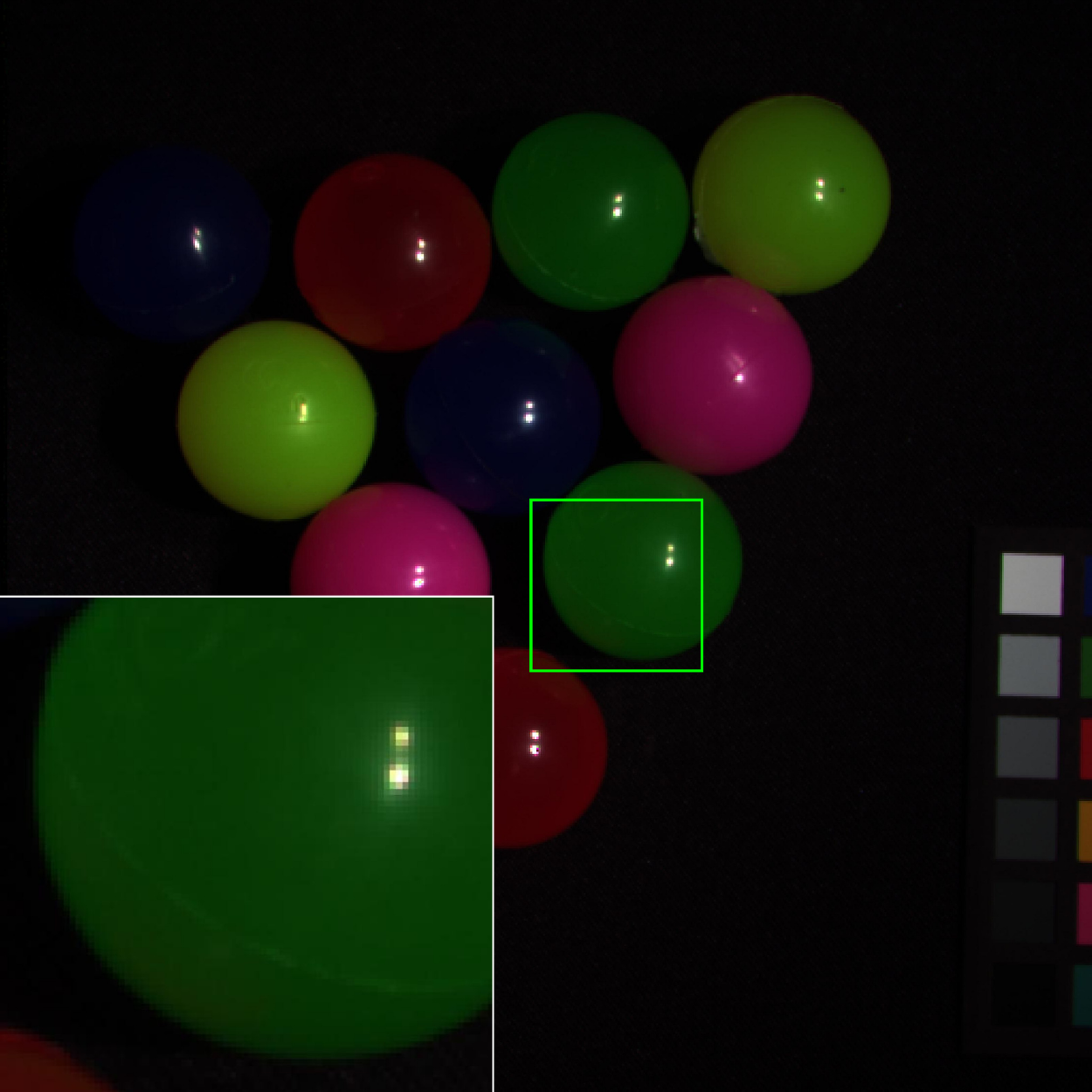}\vspace{1pt}
    \includegraphics[width=3cm]{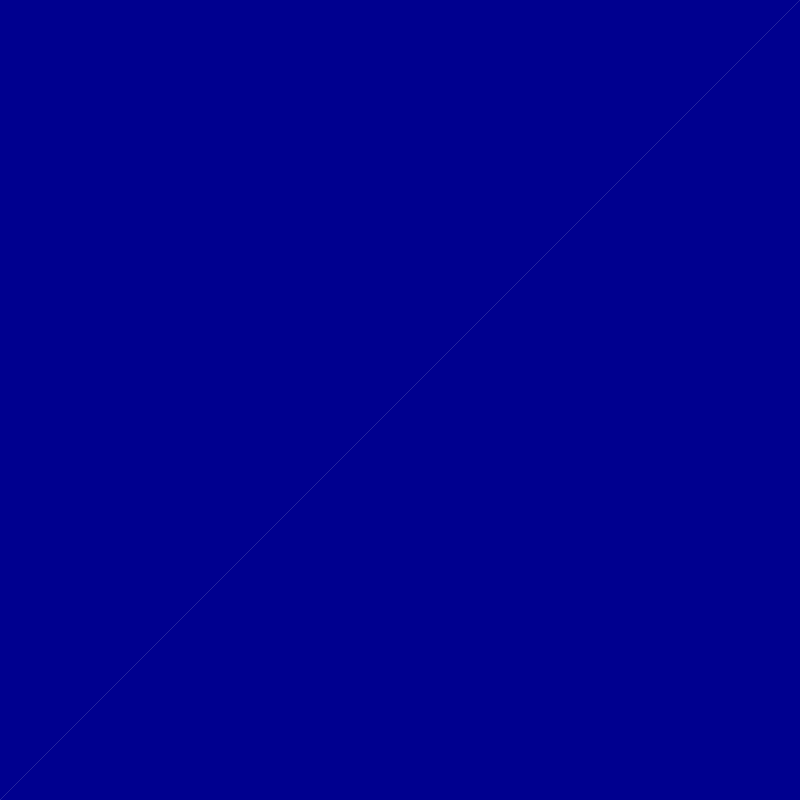}\vspace{1pt}
    \end{minipage}
}\\
    \includegraphics[width=12cm]{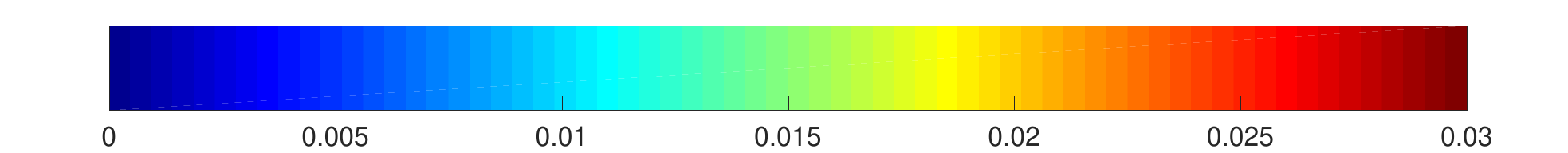}\vspace{1pt}
\caption{The first row shows the false color images formed by bands 29, 13 and 5 for ``Superballs", and the second row presents the corresponding error images.}
\label{fig5}
\end{figure}

The false color images consisting of bands 29, 13 and 5 of ``Superballs" and the corresponding residual images are displayed in Fig. \ref{fig5}.
The magnified regions of the fused images show that LogLRTR generates smoother result and less parallax compared to other approaches.
This conclusion is also supported by the fact that LogLRTR retains relatively more edge details in the error image.

\begin{figure}
\centering
\subfigure[CSTF]{
    \begin{minipage}[b]{0.17\linewidth}
    \includegraphics[width=3cm]{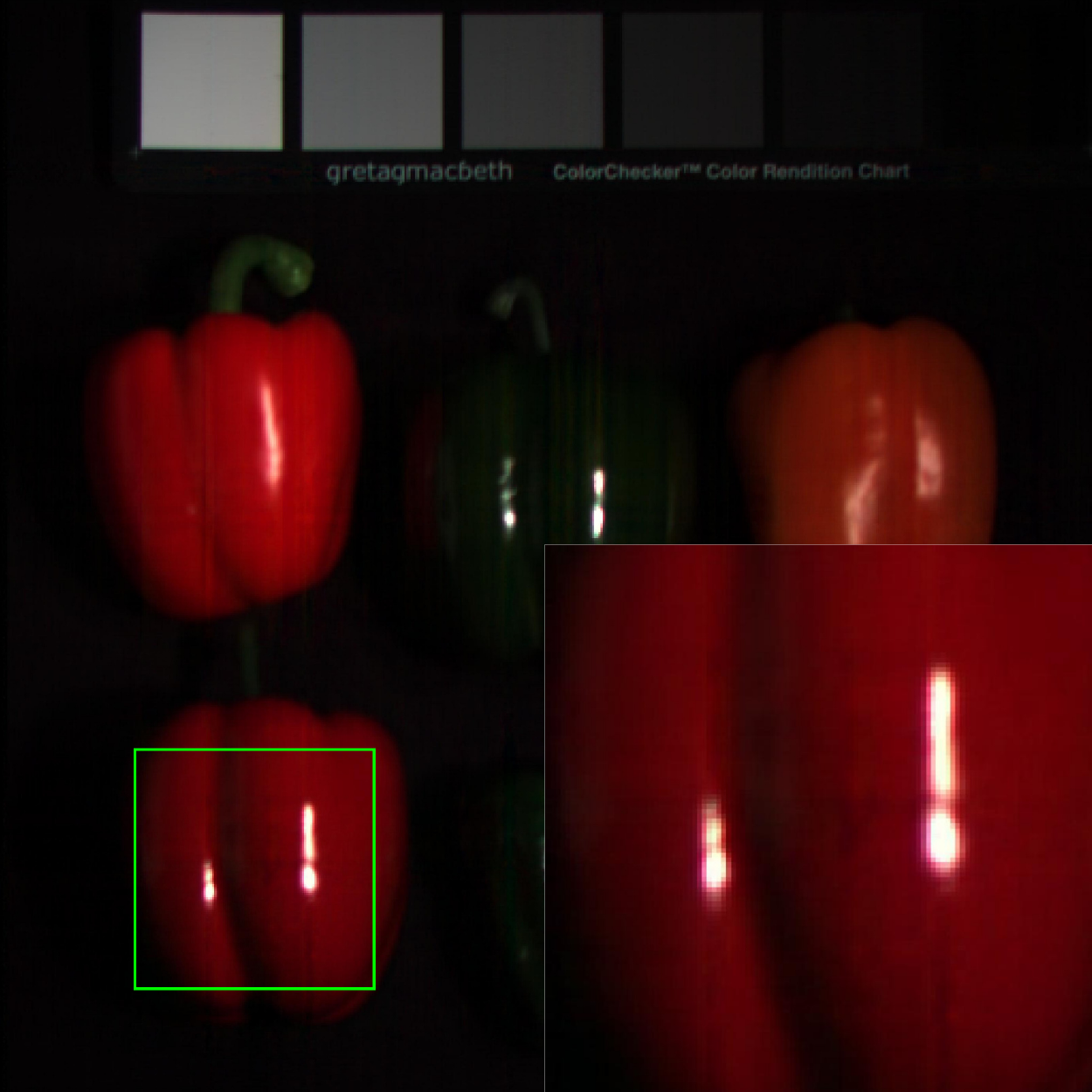}\vspace{1pt}
    \includegraphics[width=3cm]{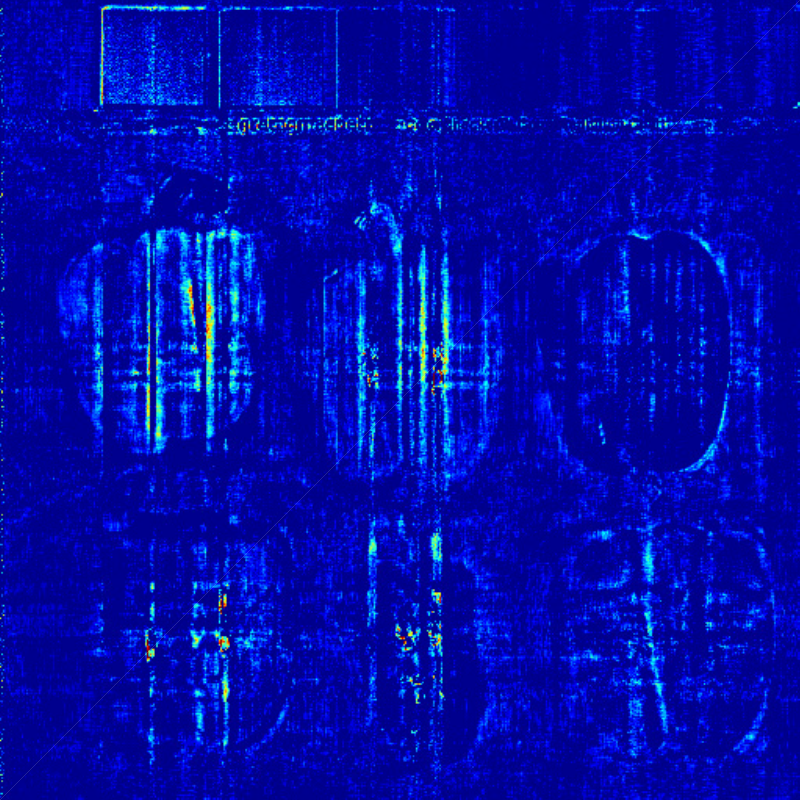}\vspace{1pt}
    \end{minipage}
}
\subfigure[CTRF]{
    \begin{minipage}[b]{0.17\linewidth}
    \includegraphics[width=3cm]{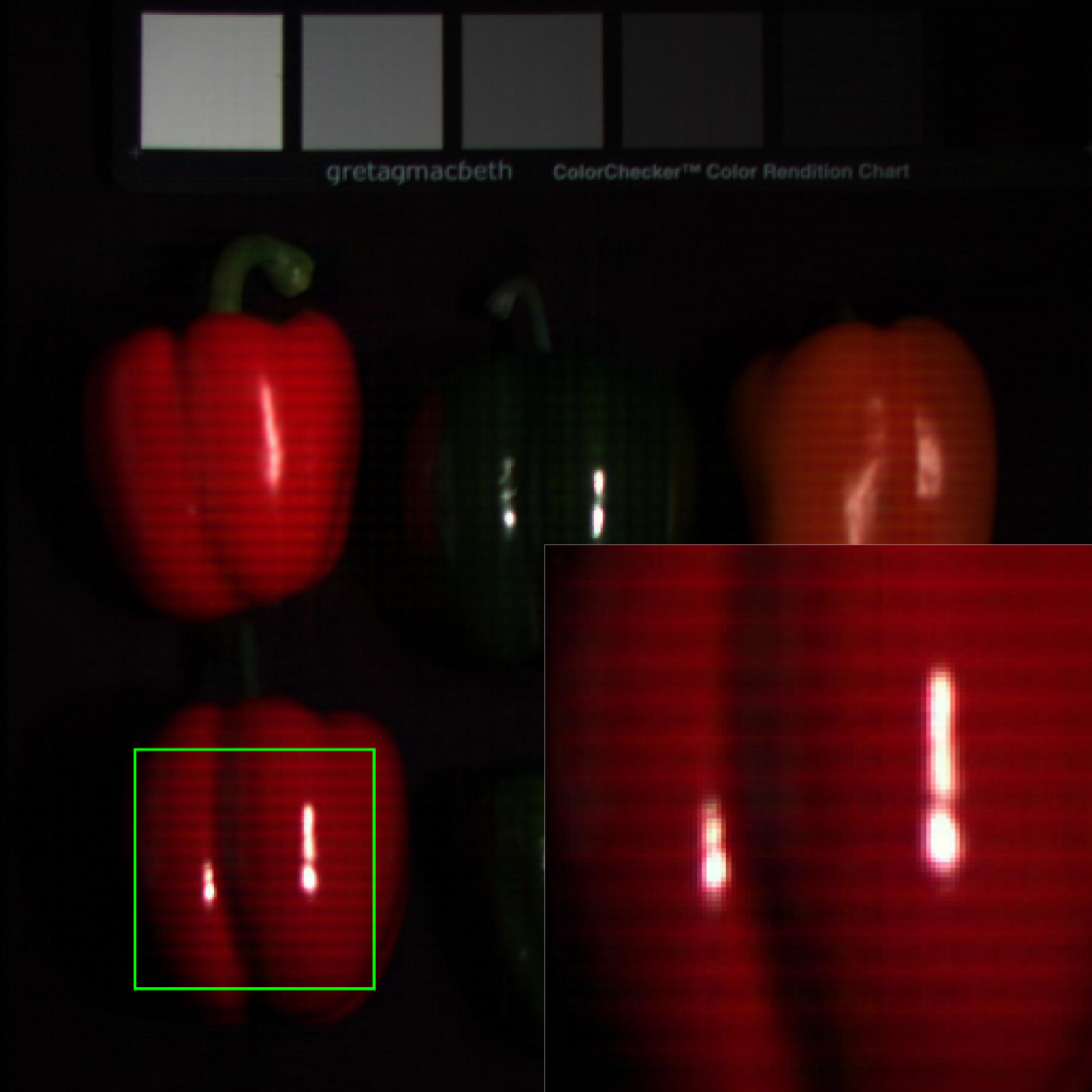}\vspace{1pt}
    \includegraphics[width=3cm]{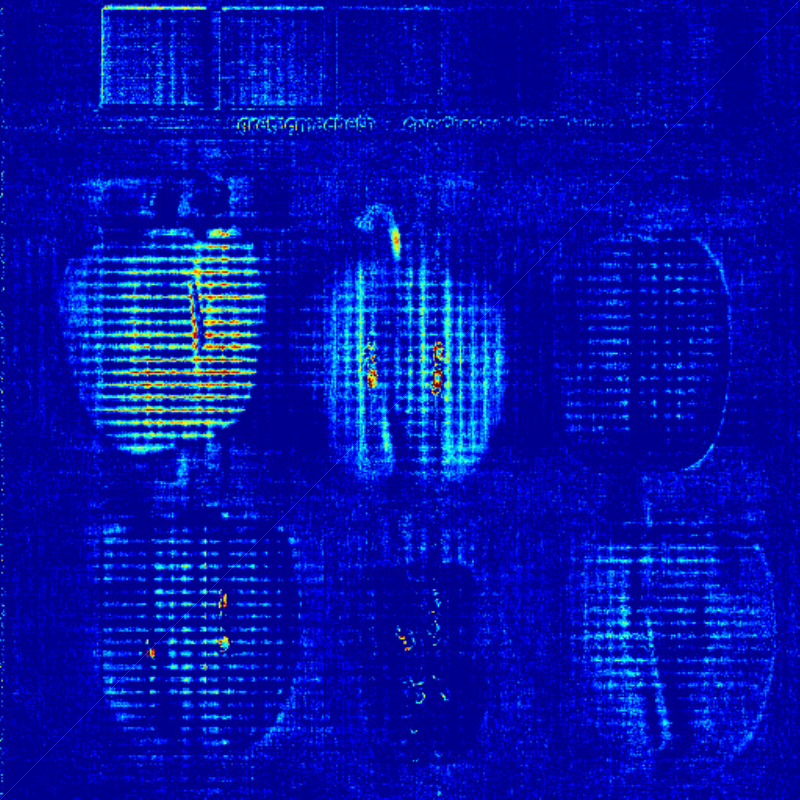}\vspace{1pt}
    \end{minipage}
}
\subfigure[FSTRD]{
    \begin{minipage}[b]{0.17\linewidth}
    \includegraphics[width=3cm]{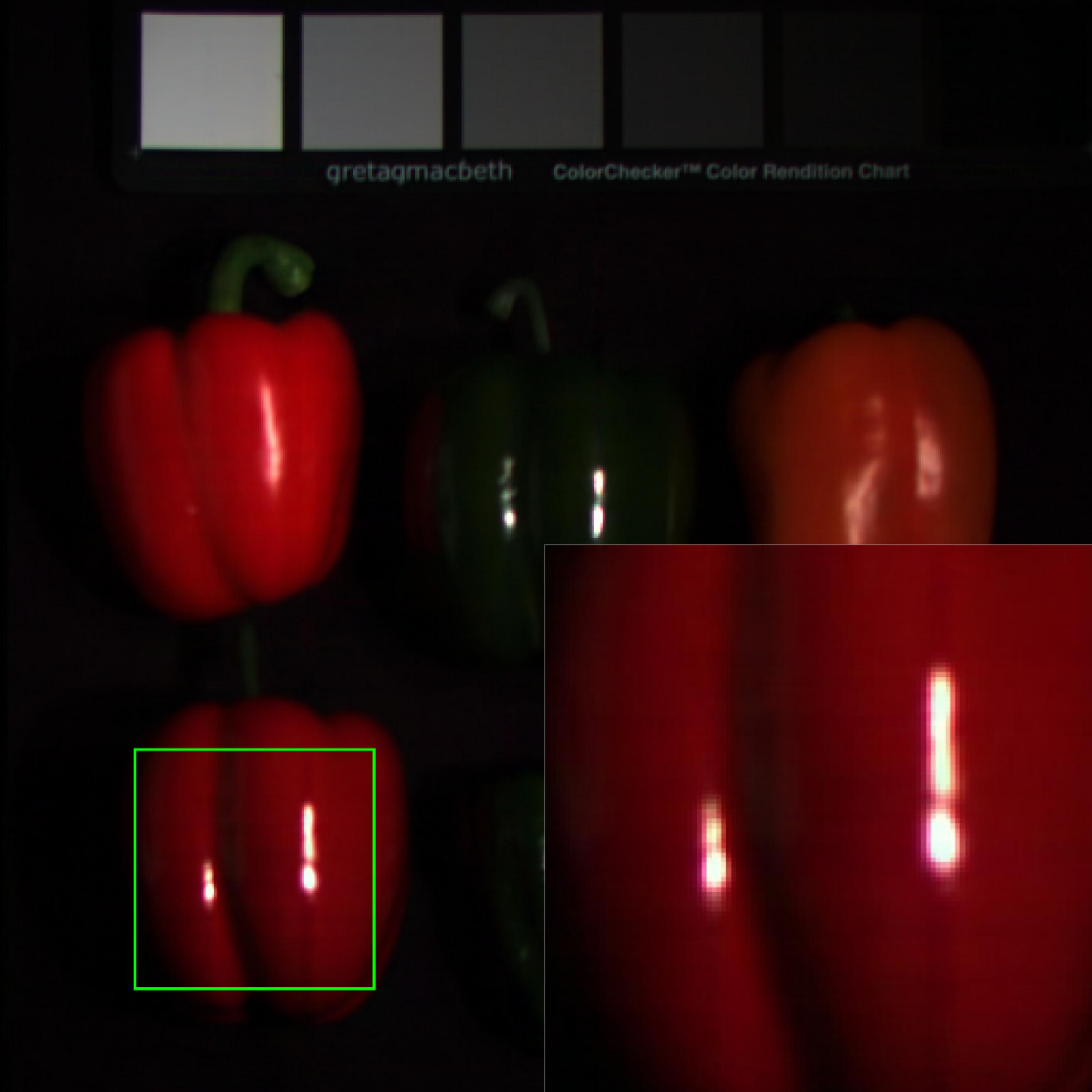}\vspace{1pt}
    \includegraphics[width=3cm]{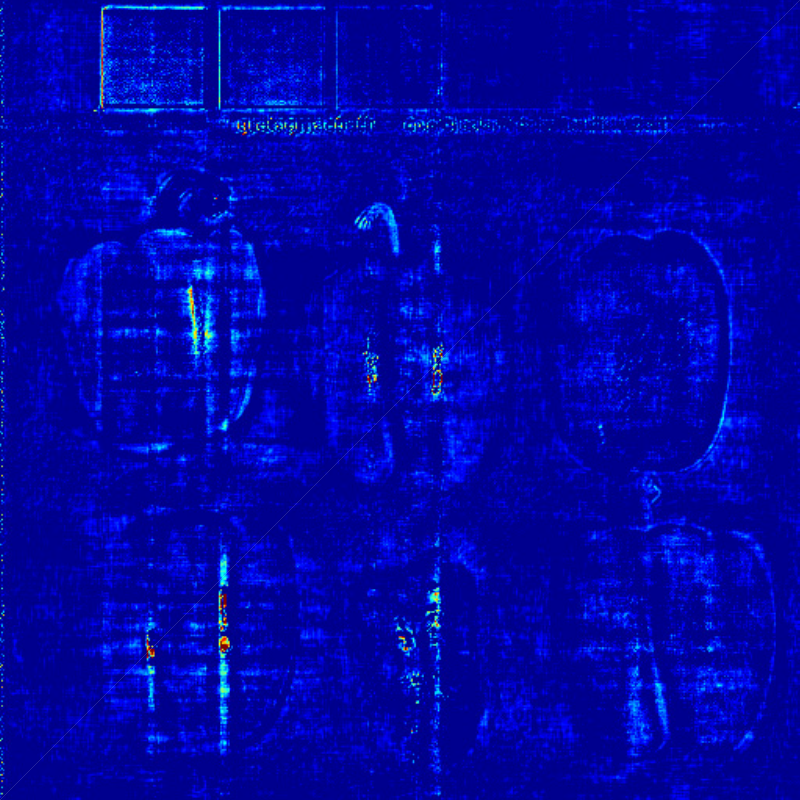}\vspace{1pt}
    \end{minipage}
}
\subfigure[LogLRTR]{
    \begin{minipage}[b]{0.17\linewidth}
    \includegraphics[width=3cm]{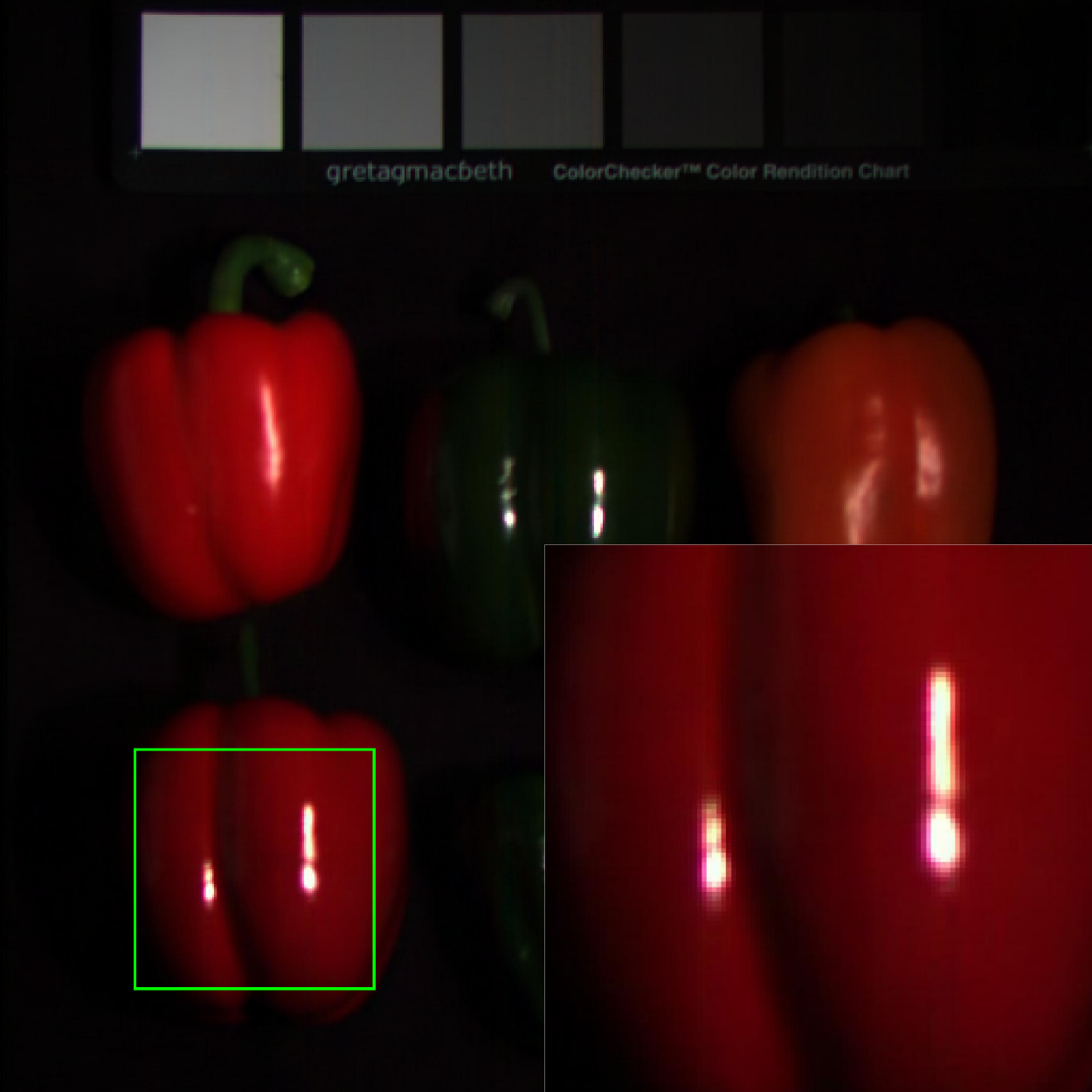}\vspace{1pt}
    \includegraphics[width=3cm]{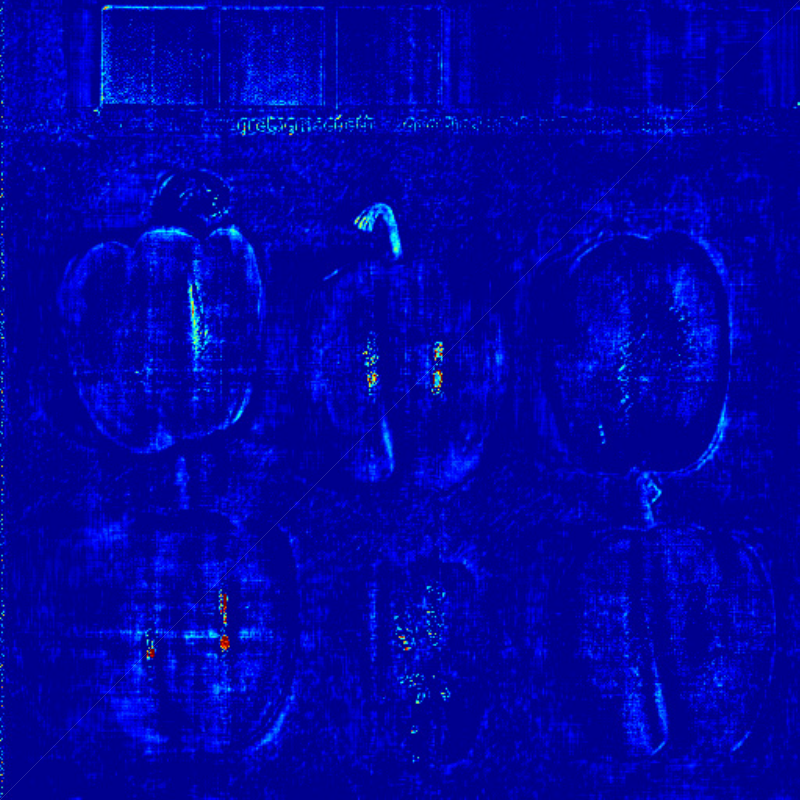}\vspace{1pt}
    \end{minipage}
}
\subfigure[Ground truth]{
    \begin{minipage}[b]{0.17\linewidth}
    \includegraphics[width=3cm]{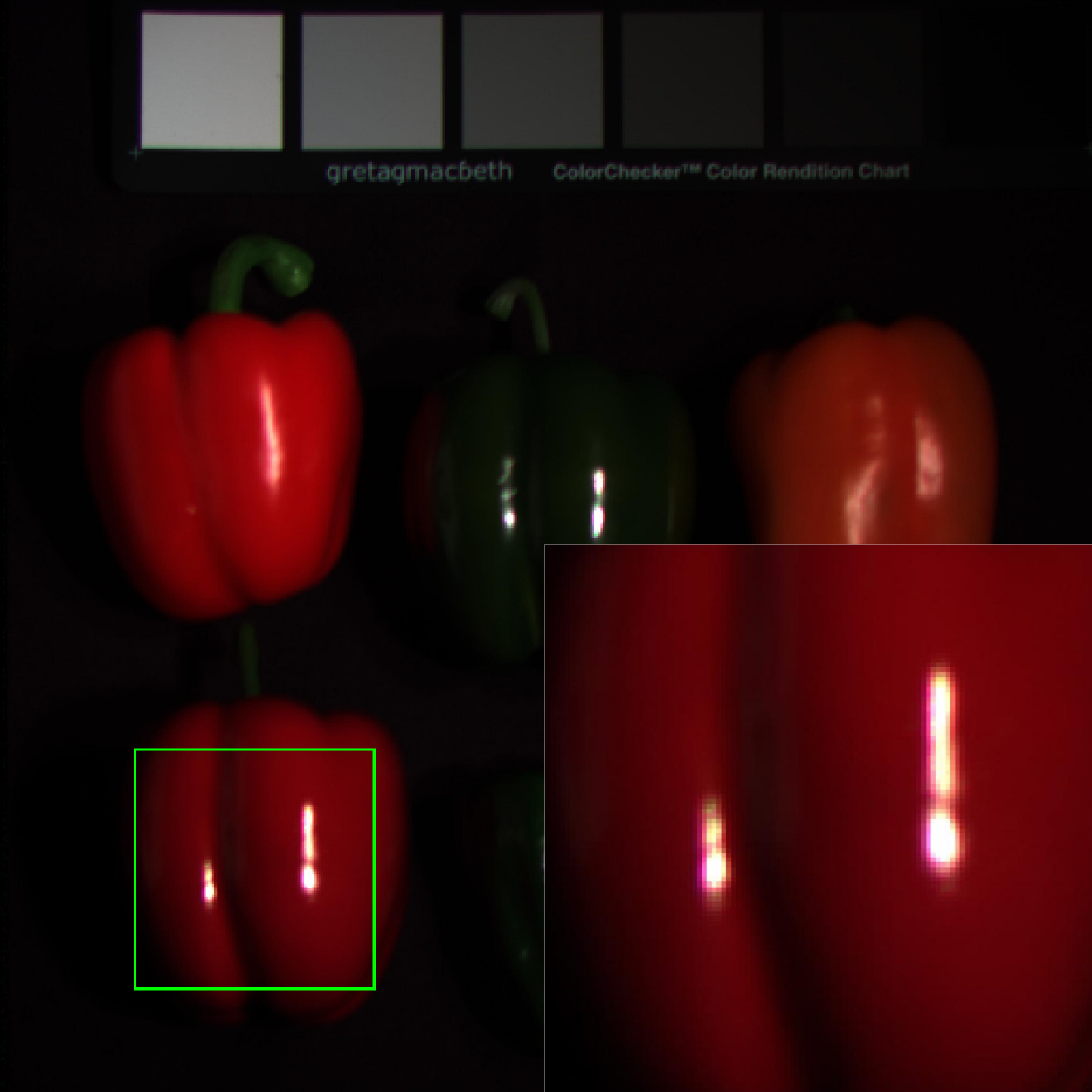}\vspace{1pt}
    \includegraphics[width=3cm]{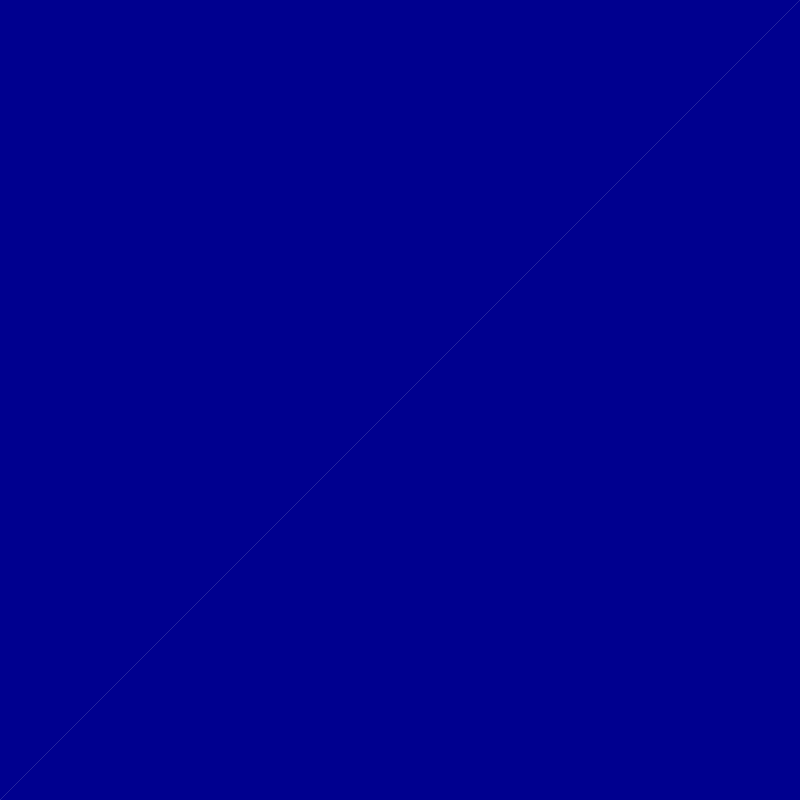}\vspace{1pt}
    \end{minipage}
}\\
    \includegraphics[width=12cm]{fig/colorbar0.03.pdf}\vspace{1pt}
\caption{The first row presents the false color images formed by bands 29, 13 and 5 for ``Peppers", and the second row displays their corresponding error images.}
\label{fig6}
\end{figure}

Fig. \ref{fig6} presents the false color images comprising of the 29th, 13th and 5th bands of ``Peppers" reconstructed by various fusion approaches.
The magnified regions show an evidence that the previous three methods generate artifacts on the surface of the pepper, while the fused image acquired by LogLRTR is closer to the ground truth image.
The error images clearly illustrate that CTRF produces less satisfactory fusion result, confirming its suboptimal performance on the CAVE dataset. Conversely, LogLRTR showcases a notable improvement, especially for the pepper in the upper left corner.

\subsection{Ablation experiments}

To highlight the efficacy of the three TR factor LTNN terms, we carry out ablation experiments. This involves disabling in turn the spectral and spatial TR factors in the proposed method, providing additional insights into their individual contributions.
More specifically, we ban the use of the one spectral TR factor LTNN term and the two spatial TR factor LTNN terms, which are respectively termed as LogLRTR$\cdot$ban$\cdot$spe and LogLRTR$\cdot$ban$\cdot$spa.
The numerical results of ablation experiments conducted on ``Flowers" and ``Superballs" by using FSTRD, the proposed LogLRTR and its variants are shown in Table \ref{table3}.
We easily observe that LogLRTR surpasses the other three approaches, and the performance of the proposed method benefits significantly from the inclusion of the three TR factor LTNN terms.

\begin{table}[htbp]\vspace{-1em}
\caption{Comparison of quantitative indices from ablation experiments.}
\centering
\begin{tabular}{ccccccc}
\hline
{\rule[-1mm]{0mm}{6mm}}Image & Method &PSNR  &SSIM &ERGAS &SAM &UIQI     \\
\hline
{\rule[-1mm]{0mm}{6mm}}Flowers  &FSTRD&43.867 &0.971 &1.210 &11.979 &0.728 \\
{\rule[-1mm]{0mm}{6mm}}     &LogLRTR$\cdot$ban$\cdot$spe&44.172 &0.973 &1.162 &11.586 &0.733 \\
{\rule[-1mm]{0mm}{6mm}}     &LogLRTR$\cdot$ban$\cdot$spa&44.148 &0.971 &1.210 &12.091 &0.729 \\
{\rule[-1mm]{0mm}{6mm}}     &LogLRTR&\textbf{44.557}&\textbf{0.976}&\textbf{1.127}&\textbf{10.839}&\textbf{0.746}\\
{\rule[-1mm]{0mm}{6mm}}Superballs &FSTRD   &44.690   &0.972 &1.619  &8.948  &0.765  \\
{\rule[-1mm]{0mm}{6mm}}     &LogLRTR$\cdot$ban$\cdot$ spe&45.196&0.975&1.497&\textbf{8.404}&0.781  \\
{\rule[-1mm]{0mm}{6mm}}     &LogLRTR$\cdot$ban$\cdot$ spa&45.269&0.975&1.504&8.571&0.770\\
{\rule[-1mm]{0mm}{6mm}}     &LogLRTR&\textbf{45.458}&\textbf{0.976}&\textbf{1.474}&8.546&\textbf{0.791}  \\
\hline
\end{tabular}
\label{table3}
\end{table}

\subsection{Convergence}

\begin{figure}[h!]
\centering\vspace{-1em}
\subfigure[]{
\includegraphics[width=2.25in]{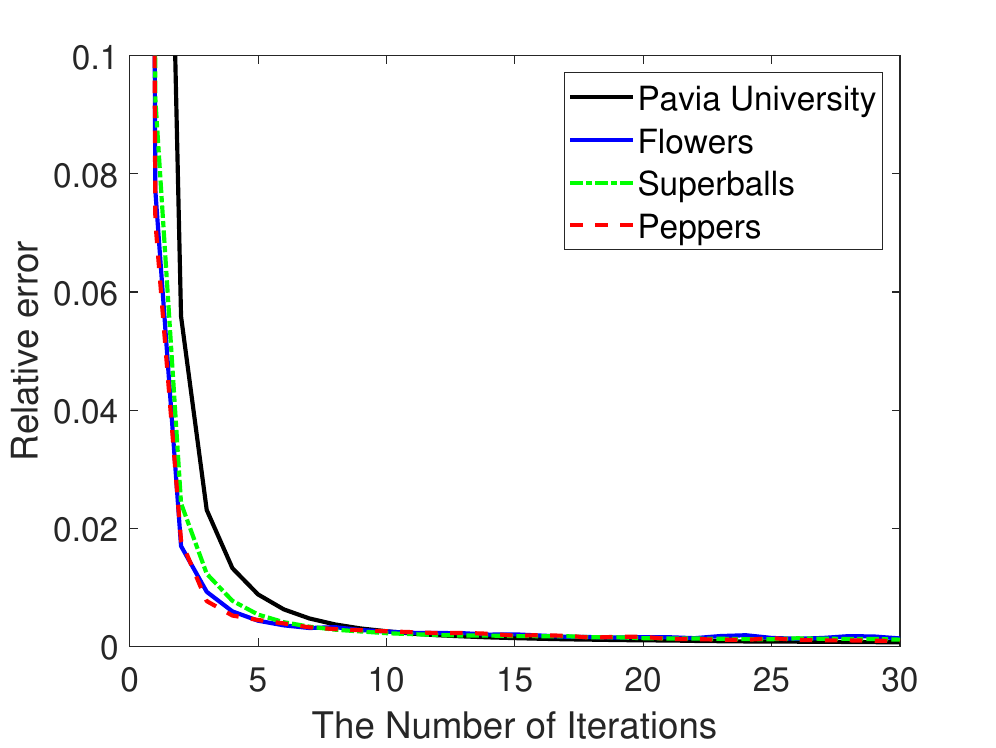}}
\quad
\subfigure[]{
\includegraphics[width=2.25in]{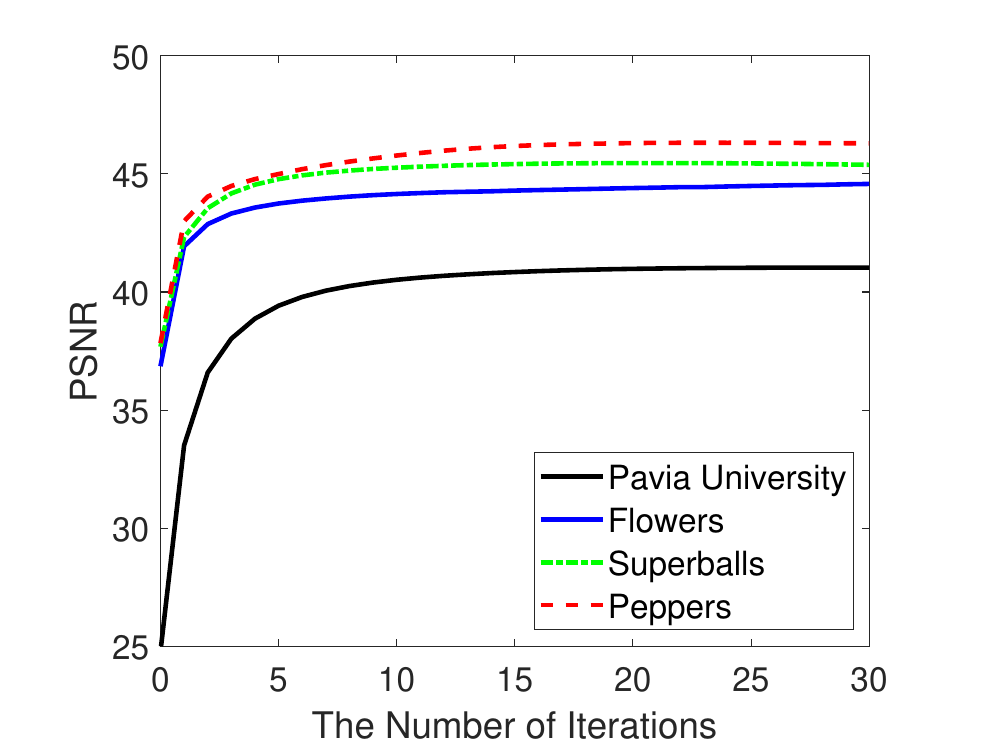}}\vspace{-0.5em}
\caption{The plots of relative error ($\left\|\mathcal{X}^{k}-\mathcal{X}^{k-1}\right\|_{F}/\left\|\mathcal{X}^{k}\right\|_{F}$) (left) and PSNR (right)
against the number of iterations for four images.}\vspace{-1.5em}
\label{fig7}
\end{figure}

Finally, we empirically demonstrate the convergence of the newly developed algorithm.
Specifically, all four images are tested by Algorithm \ref{algorithm1}.
The plots of relative error and PSNR versus the number of iterations are displayed in Fig. \ref{fig7}.
The curves depicted in Fig. \ref{fig7}(a) reveal a consistent trend of the relative errors converging towards zero.
Furthermore, the PSNR values grow quickly as the number of iterations increases, eventually reaching a stable state, as demonstrated in Fig. \ref{fig7}(b). These observations collectively constitute robust evidence of the algorithm's convergence.

%%%%%%%%%%%%%%%%%%%%%%%%%%%%%%%%%%%%%%%%%%%%%%%%%%%%%%%%%%%%%%%%%%

\setcounter{equation}{0}
\section{Conclusion}\label{Conclusion}
In this paper, we introduced a newly developed hyperspectral image fusion model that integrates logarithmic tensor nuclear norm (LTNN) with weighted TV of TR factors.
To better preserve the correlation in the spatial-spectral domain,
the TR decomposition was utilized to represent the HR-HSI.
Using this TR framework as a foundation, the objective of HSI-MSI fusion is reframed as the estimation of the three TR factors.
To enhance the continuity of TR factors for improved maintenance of the smooth structure, the weighted total variation with weight updates was used.
In addition, we employed the mode-2 LTNN to constrain each TR factor for delving deeper into the low-rank characteristics.
Numerical experiments showed that our proposed approach is superior to existing HSI-MSI fusion methods.
In future work, we intend to develop some adaptive techniques for selecting the appropriate TR rank.

%%%%%%%%%%%%%%%%%%%%%%%%%%%%%%%%%%%%%%%%%%%%%%%%%%%%%%%%%%%%%%%%%%%%%%%%%%%%%
\section*{Acknowledgments}
This work was partly supported by
the Major Program of the National Natural Science Foundation of China (No. 61890962),
the Natural Science Foundation of Jiangxi Province (20232BAB201017),
the Science Foundation for Post Doctorate of China (2020M672484),
%the Natural Science Foundation of China (No. 12201286),
and Jiangxi Postgraduate Innovation Special Fund Project (YC2022-S987).

%%%%%%%%%%%%%%%%%%%%%%%%%%%%%%%%%%%%%%%%%%%%%%%%%%%%%%%%%%%%%%%%%%%%%%

%%%% Bibliography  %%%%%%%%%%
\setcounter{equation}{0}

\end{document}